\documentclass{aa}  
\usepackage{graphicx}
\usepackage{txfonts}
\usepackage{color}
\usepackage{natbib}
\usepackage[flushleft]{threeparttable}
\usepackage{textgreek}

\bibpunct{(}{)}{;}{a}{}{,} 
\newcommand{\beq}{\begin{equation}}
\newcommand{\eeq}{\end{equation}}
\newcommand{\bea}{\begin{eqnarray}}
\newcommand{\eea}{\end{eqnarray}}

\newcommand{\msol}{{\,\rm M}_{\odot}}
\newcommand{\mL}{\mathcal{L}}
\newcommand{\mP}{\mathcal{P}}

\begin{document} 

\title{Modelling radio luminosity functions of radio-loud AGN by the
  cosmological evolution of supermassive black holes.}

\titlerunning{AGN radio luminosity function}

\author{Marco Tucci \inst{1} \and Luigi Toffolatti \inst{2,3}}

\institute{Department of Astronomy, University of Geneva,
  ch. d’\'{E}cogia 16, CH--1290 Versoix, CH \\
  \email{marco.tucci@unige.ch} \and Departamento de F\'{i}sica
  Universidad de Oviedo, C. Federico Garc\'{i}a Lorca 18, E--33007
  Oviedo, Spain \and Instituto
  Universitario de Ciencias y Tecnolog\'{i}as Espaciales de Asturias
  (ICTEA), C. Independencia 13, 33004 Oviedo, Spain \\ \email{ltoffolatti@uniovi.es} }

\date{\today}
   
\abstract{} {We develop a formalism to model the luminosity functions
  (LFs) of radio-loud active galactic nuclei (AGN) at GHz frequencies
  by the cosmological evolution of the supermassive black hole (SMBH)
  hosted in their nuclei. The mass function and Eddington ratio
  distributions of SMBHs computed in a previous work published by one
  of the authors have been taken as the starting point for this analysis.}
{Our approach is based on physical and phenomenological relations that
  allow us to statistically calculate the radio luminosity of AGN cores,
  corrected for beaming effects, by linking it with the SMBH at their
  centre, through the fundamental plane of black hole
  activity. Moreover, radio luminosity from extended jets and lobes is
  also computed through a power-law  relationship that reflects the
  expected correlation between the inner radio core and the extended
  jets and lobes. By following a classification scheme well established in
  the field, radio-loud AGN are further divided into two classes,
  characterized by different accretion modes onto the central BH. If
  the Eddington ratio, $\lambda$, is $\leq 0.01$ they are called 
  low-kinetic (LK) mode AGN; if $\lambda\geq 0.01$, they are called
  high-kinetic (HK) mode AGN,  this critical value roughly
  corresponding to the transition between radiatively inefficient and
  efficient accretion flows. The few free parameters used in the
  present model are determined by fitting two different types of
  observational data sets: local (or low-redshift) LFs   of radio-loud
  AGN at 1.4 GHz and differential number counts of extragalactic radio
  sources at 1.4 and 5 GHz.}
{Our present model fits well almost all
  published data on LFs of LK mode AGN and of the total AGN population
  up to redshifts $z\leq 1.5$ and also in the full range of
  luminosities currently probed by data. On the other hand, it tends
  to underestimate some recent measures of the LF of HK mode AGN at
  low redshifts, but only at low radio luminosities. All in all, the
  good performance of our model in this redshift range is remarkable,
  considering that all the free parameters used but the fraction
  of HK mode AGN are redshift independent. The present model is also
  able to provide a very good fit to almost all data on number counts
  of radio-loud sources at 1.4 and 5 GHz.
}{}


   \maketitle

\section{Introduction}
\label{sec1}

Active galactic nuclei (AGN) have been of particular interest over the
past decade due to the crucial role they play in galaxy
formation and evolution. They are associated with the accretion of
material on supermassive black holes (SMBHs), with masses that vary
from millions to billions of solar masses.

Depending on their mode of accretion, AGN are typically divided into
two main categories \citep[for a review, see e.g.][]{hec14}. The
first category, called quasar-mode or radiative-mode AGN,
consists of objects that   efficiently   convert the potential energy
of the gas accreted by the SMBH in the form of radiation. Radiatively
efficient accretion flows in SMBHs are usually described by a standard
geometrically-thin, optically-thick accretion disc
\citep{sha73}. Powerful jets are present in a small fraction
  of radiative-mode AGN. The second category (called radio-mode or
jet-mode AGN) is apparently associated with a mode of accretion
where most of the released energy is in kinetic form. The
geometrically-thin accretion disc is replaced by a geometrically-thick
structure in which the inflow time is much shorter than the radiative
cooling time. These are called advection-dominated or radiatively
inefficient accretion flows \citep[ADAFs or RIAFs;
e.g.,][]{nar94,nar95,qua99}. One of their characteristic properties  is that they are  capable of launching two-sided jets.

The transition between the two modes of accretion seems to occur at a
critical value of the Eddington ratio of $\lambda_{cr}\approx0.01$
\citep{mac03,jes05}. The observed dichotomy between \citet{fan74}
class I (FR\,I) and II (FR\,II) radio galaxies confirms this
hypothesis: the line separating low-power FR\,I and high-power
FR\,II galaxies in a plot of AGN radio luminosity against host galaxy
optical luminosity is interpreted as a threshold in accretion rate and
corresponds to $\sim0.01$ \citep{ghi01,xu09,ghi09}. However, this
  interpretation has been  questioned in some recent analyses
  \citep{gen13,var20}. A similar dichotomy is observed for
Galactic black holes in X-ray binaries, where the transition between
spectral states is observed at the same $\lambda_{cr}$, and radio
emission in high-accreting systems is substantially suppressed
\citep[e.g.][]{mac03}.

Radio galaxies can be also classified based on emission line strengths
and line flux ratios, as low-excitation and
high-excitation. The former population of radio galaxies has been
historically associated with jet-mode AGN, while the latter    
forms a small sub-population of radiative-mode AGN that produce
powerful jets. \citet{bes12} showed that, in the nearby Universe, the
low-excitation population  typically has a low accretion rate, below
one per\,cent of the Eddington rate, whereas high-excitation radio
galaxies predominately accrete above this value.

In this paper we follow the nomenclature used in \citet{mer08} to
distinguish AGN with different accretion modes (based on their
analogies with Galactic black holes in X-ray binaries). If the
accretion rate is below the critical ratio
$\lambda_{cr}\simeq10^{-2}$, AGN are said to be in low-kinetic (LK)
mode (corresponding to jet-mode AGN defined above). LK mode AGN
are typically radio-loud objects and can be associated with FR\,I
radio galaxies, with the only possible exception of LINERs, which
are likely members of the radio-mode population but with rather
modest radio luminosities. On the other hand, AGN accreting above
this critical value ($\lambda_{cr}\approx0.01$) can be observed in two
different states: high-kinetic (HK) mode 
AGN, or high-radiative (HR) mode 
AGN, depending on the presence or lack of powerful jets. AGN in HR mode,
which are the majority of high-accreting SMBHs, are radio-quiet and
will be not considered in the following analysis. HK mode AGN are
instead radio-loud objects and can be roughly associated with FR\,II
radio galaxies \citep[see e.g.][]{hec14}.

Table\,\ref{t0} shows the expected correspondence of the
Merloni\,\&\,Heinz classification with common AGN classes, and summarizes
the main properties. We recall that HK mode AGN are
characterized by radiatively efficient accretion flows and by the
presence of powerful radio jets, which are probably produced by
rapidly-spinning black holes \citep{yua14}. 
Type-1 and type-2 AGN are also included in the table. These two
classes are defined according to their optical spectra: type-1 AGN
show both broad and narrow lines, while type-2 AGN only narrow
lines. The lack of broad emission lines is typically associated with the
presence of obscuration along the line of sight. However, at least a
fraction of type-2 AGN appear to be unobscured and intrinsically
lacking a broad-line region. They appear to be low-luminosity and
low accretion rate AGN \citep[e.g.][]{tru11, hic18}.

The observed emission from AGN-powered radio sources includes
radiation from classical extended jets and double lobes, and 
from the compact radio component (jet core) that are more directly
associated with the energy generation and collimation near central
SMBHs. According to the unification model \citep[see
e.g.][]{urr95}, the appearance of radio sources, including their
spectra, depends primarly on their axis orientation relative to the
observer. A line of sight close to the source  jet-axis offers a
view of the compact base of the approaching jet. Its radio emission is
Doppler-boosted and presents the typical flat spectrum\footnote{Radio
  sources are typically divided according to their spectral index
  $\alpha$ between 1.4 and 5\,GHz (we adopt the convention
  $S\propto\nu^{-\alpha}$). They are classified as flat-spectrum sources
  if $\alpha\le0.5$ and steep-spectrum otherwise.} associated with
optically-thick sources; they are generally called blazars.
In the case of a side-on view, the observed emission is dominated by
the extended optically-thin components with a steeper spectrum. In
general, however, at an intermediate angle of view, radio sources (i.e.
classical steep-spectrum radio sources, usually associated with giant
elliptical galaxies and, at lower luminosities, with starburst or less
active galaxies) can have a significant contribution from both
components.

The radio luminosity function (LF) is an important and common tool for
understanding the evolution of AGN over cosmic time. In recent
years, a large effort has been made on studying and modelling the
evolution of radio LFs, by exploiting large-area multi-wavelength
surveys \citep[e.g.][]{mas10,rig11,smo17,yua17,mal19}, including the
use of cosmological simulations \citep[e.g.][]{bon19,tho20}. The
evolution of the radio AGN population is also expected to be strictly
connected to the SMBH accretion history, being the radio emission of
AGN core directly dependent on the central black hole mass, spin, and
accretion rate \citep[see e.g.][]{hec14}. Recently, following this
idea, empirical or semi-analitic models of the radio AGN LF were
developed based on the growth of SMBHs across cosmic time
\citep{sax17,wei17,gri20}.

In this work, we attemp to model the LF of the radio-loud AGN
population at GHz frequencies based on the SMBHs evolution. The
starting point of the work is the SMBH mass functions (MFs)  and Eddington
ratio distributions computed by \citet{tv17}. In Section\,\ref{sec2}
we develop the formalism to link these quantities to the radio LF of
radio-loud AGN populations.
AGN in LK and HK mode are studied separately, due to the different
accretion rate behaviour and radio properties. In Section\,\ref{sec3}
we present the observational data used to determine free parameters of
the model, which are computed in Section\,\ref{sec4} where model
predictions are also compared to observations. Finally, in
Section\,\ref{sec5} we summarize our main findings and possible future
developments.

Throughout this paper we adopt a flat Lambda cold dark matter
(\textLambda CDM) cosmology with $\Omega_{\Lambda} = 0.7$ and
H$_0 = 70\,{\rm km\,s}^{-1}\,{\rm Mpc}^{-1}$.

\begin{table*}
  \centering 
  \caption{Common classes of the AGN population.}
  \begin{tabular}{llcccccc}
  \hline
 \multicolumn{2}{c}{Nomenclature}& Eddington & Radio &
   morphology & \multicolumn{2}{c}{emission lines} & blazars \\
\multicolumn{2}{c}{ }& ratio & emission &
   & \multicolumn{2}{c}{ } & ($\alpha_{1.4}^5<0.5$) \\
  \hline
 {\bf Low Kinetic} & jet mode & $\la10^{-2}$ & radio 
  loud & FRI\,\&\,FRII & low excitation & type-2 & BL\,Lac\\
 {\bf High Kinetic} & radiative mode & $\ga10^{-2}$ & radio 
  loud & FRII & high excitation & type-1\,\&\,2 & FSRQ \\
 {\bf High Radiative} & radiative mode & $\ga10^{-2}$ & radio 
  quiet & -- &  & type-1\,\&\,2 & -- \\
  \hline
\end{tabular}
\label{t0}
\end{table*}

\section{From mass functions of active SMBHs to radio
  luminosity functions of AGN}
\label{sec2}

In this section we describe how to link the radio emission of AGN,
both from jet cores and from extended jets and lobes, with SMBH physical
properties such as mass and accretion rate. Based on the evolution of
the mass function and the accretion rate of active SMBHs predicted by
\citet{tv17}, we estimate the luminosity function of radio-loud
AGN at GHz frequencies from redshift 0 to 4. We proceed as follows:

\begin{itemize}

\item The intrinsic radio emission from the AGN jet core is linked to
  the SMBH mass and X-ray luminosity through the fundamental
  plane relation of BH activity (Sect.\,\ref{s2s1}).

\item Beaming effects of relativistic jets are taken into account
  in order to calculate the observed core luminosity (Sect.\,\ref{s2s2}).

\item The extended radio emission from AGN jets and lobes is obtained
  from the intrinsic core luminosity using an empirical power-law
  relation. Both are  expected to be related to the bulk
  kinetic power of jets (Sect.\,\ref{s2s3}).

\item We finally estimate the radio luminosity function and number
  counts of radio-loud AGN, including the contributions to the total
  radio power coming from the core and the extended emissions
  (Sects.\,\ref{s2s4}--\ref{s2s5}).
  
\end{itemize}

Hereafter we adopt a simple power-law spectrum for AGN-powered
radio sources in the 1--5\,GHz frequency range (i.e.
$S\propto S^{-\alpha}$) with $\alpha_f=0.0$ for flat-spectrum sources
and $\alpha_s=0.75$ for steep-spectrum ones, respectively.

\subsection{Intrinsic luminosity function of radio jet cores}
\label{s2s1}

Theoretical arguments have shown that the amount of synchrotron
radiation emitted from a scale-invariant jet depends both on the black
hole mass and the accretion rate \citep{fal95,hei03}. The first
observational evidence of this dependency was the discovery of the
fundamental plane (FP) of black hole activity, which is the  relationship
between X-ray luminosity, radio luminosity, and black hole mass for
low accretion rate black holes, connecting Galactic black holes in
X-ray binaries and AGN \citep{mer03,fal04}. In the FP relation the
X-ray emission is considered to be  a tracer of the accretion rate, while
the radio luminosity is used as a probe for the AGN jet. Many authors
have since confirmed the existence of a black hole FP relationship
\citep{kor06,mer07,gul09,plo12,sai15}.

According to the FP relation, the radio luminosity of AGN cores is
related to the X-ray luminosity and the black hole (BH) mass, $M$, by
\citep[e.g.][]{mer03,fal04,plo12}
\beq
\log{\mL_c}=\xi_X\log{L_X}+\xi_M\log{M}+\beta_R\,,
\label{s2e1}
\eeq
where $\mL_c$ is the {intrinsic} (i.e. unbeamed) core luminosity
at 5\,GHz; $L_X$ is  the (unabsorbed) X-ray luminosity in
the 2--10\,keV band (both in unit of erg\,s$^{-1}$);
and $\xi_X,~\xi_M$, and $\beta_R$ are the correlation
coefficients of the FP relation.

The observed correlation coefficients can be compared with the values
predicted by theoretical models. These coefficients depend roughly on
the accretion flow (if it is radiatively efficient or inefficient) and
on the dominant physical process that originates the observed
X-rays. Several components can potentially contribute to the X-ray
emission: synchrotron and synchrotron self-Compton radiation from the
jet; emission from the accretion flow; and  inverse Compton scattering of
soft disc photons on hot electrons of the corona. \citet{mer03} show
how different theoretical models for emission processes can be
directly translated into predictions of the FP coefficients in the
cases where X-rays are predominantly originated by optically thin
synchrotron jet emission or by inverse Compton scattering.

For sources at  low accretion rates, the FP relation is  well
established. Different authors \citep{mer03,fal04,gul09,plo12} have
determined it with  reasonable agreement of the correlation
parameters \citep[with some exceptions; e.g.][]{bon12} and in
agreement with theoretical expectations for radiatively inefficient
accretion flow models \citep[in particular with the advection
dominated accretion flow, ADAF, solution;][]{mer03}. Regardless of the
physical explanation, this scaling relation implies that the same
mechanism governing the accretion and ejection processes from black
holes holds over approximately nine orders of magnitude in mass.
For low accretion rate (or LK mode) AGN we  adopt the
correlation coefficients obtained by \citet{mer07,mer08} based on the
\citet{mer03} data: $\xi_X=0.62$, $\xi_M=0.55$, and $\beta_R=8.6$, with
an intrinsic scatter of 0.6\,dex.

When high accretion rate sources are included in the FP relation, the
inferred coefficients change and the intrinsic scatter
increases. \citet{li08} and \citet{don14} explored the FP for samples
of radiatively efficient black holes. \citet{don14} considered X-ray
binaries and radio-quiet type-1 AGN with Eddington ratio higher than
1\%. The correlation coefficients they found are different with
respect to  low-accretion AGN and compatible with those expected from
radiatively efficient accretion disc models: the slope with the X-ray
luminosity is steeper ($\xi_X\simeq1.6$) and a small anti-correlation
with the black hole mass ($\xi_M\sim-0.2$) is found. Very similar
results are also found in \citet{li08}, but only for radio-loud
broad-line AGN; their sample also includes   radio-quiet broad-line
AGN that provide results in agreement with the FP of LK mode
sources.

Because of the larger uncertainty in the FP relation for HK mode AGN,
the correlation coefficients are not fixed in the model as for
LK mode AGN, but considered as  {free parameters}.



It is more convenient for us to write the FP relation
(Eq.\,\ref{s2e1}) in terms of the Eddington ratio (i.e. the ratio of
the bolometric luminosity $L_B$ to the Eddington luminosity
$L_{\rm Edd}\simeq1.26\times10^{38}\,M/\msol$\,erg\,s$^{-1}$,
$\lambda\equiv L_B/L_{\rm Edd}$) as
\beq
\log{\mL_c} = \xi_X\log{\lambda}+(\xi_M+\xi_X)\log{M}+\beta'_R\,,
\label{s2e2}
\eeq
where $\beta'_R=\beta_R+\xi_X(38.1+\log{f})-\log(5\times10^9)$, and
$f=L_X/L_B$ is the bolometric correction \citep[from ][]{mar04}. Standard radio monochromatic luminosities are
now considered in Eq.\,\ref{s2e2} (in  units of
erg\,s$^{-1}$\,Hz$^{-1}$).

We impose a continuity condition in the FP relationship at the
transition of the LK and HK mode regimes (i.e.  at the critical
accretion rate of $\lambda_{cr}=0.01$). Under this condition we can write
HK mode FP parameters as a function of LK mode ones:
%
\begin{eqnarray}
\xi_M^{HK} & = & \xi_M^{LK} + \Delta\xi_X \nonumber, \\
\beta_R^{HK} & = & \beta_R^{LK} + \Delta\xi_X(38.1+\log\lambda_{cr}+\log f)\,,
  \label{s2e2a}
\end{eqnarray}
where $\Delta\xi_X=\xi_X^{LK}-\xi_X^{HK}$. The first equation is
obtained by requiring the continuity condition to be valid for each
SMBH mass. Thus, if the FP relation is fixed for LK mode AGN, only the
coefficient $\xi^{HK}_X$ is needed to compute $\mL^{HK}_c$.

Once the FP relationship is established it is easy to extract the
intrinsic luminosity function of AGN cores at 5\,GHz (i.e.
$\Phi(\mL_c)$\footnote{Hereafter luminosity functions, mass
  functions, and Eddington ratio distributions are always defined by
  logarithmic intervals. The redshift dependence is ignored at the
  moment.}) as a function of the mass function $\Phi_{AGN}(M)$ and the
Eddington ratio distribution $P(\lambda|M)$ of active SMBHs. For
sources in the $X$ mode (with $X=LK$ or $HK$) we have
\bea \Phi_X(\mL_c) & = & \int d\log(M)
d\log(\lambda)\,\Phi_X(\mL_c,M,\lambda)=
\nonumber \\
& = & f_{loud}^X\int d\log(M)
d\log(\lambda)\,P(\mL_c|M,\lambda,X)\Phi_X(M,\lambda)
\nonumber \\
& = & f_{loud}^X\int d\log(M)\,\Phi_{AGN}(M)\int_{a_0}^{a_1}
d\log(\lambda)\,P(\lambda|M)\times
\nonumber \\
& \times & \frac{1}{\sqrt{2\pi}\sigma_{FP}}
\exp\bigg\{-\frac{[\log{\mL_c}-\log{\mL_c^{FP}(M,\lambda)}]^2}
{2\sigma_{FP}^2}\bigg\}\,,
\label{s2e3}
\eea
where $f_{loud}^{LK}$ is the fraction of LK mode AGN that are
radio-loud, and $f_{loud}^{HK}$ is the fraction of HK mode AGN of the
total AGN population accreting at above the critical ratio,
$\lambda_{cr}$. For simplicity, we assume the same mass function for
LK and HK mode AGN. The quantity $\mL_c^{FP}(M,\lambda)$ is the
core luminosity estimated from the FP relation for a given SMBH with
mass $M$ and Eddington ratio $\lambda$. The integrals on $\log(M)$ and
$\log(\lambda)$ account for the intrinsic scatter in the FP relation,
which is assumed to have a Gaussian distribution with dispersion
$\sigma_{FP}$. The integral limits $a_0$ and $a_1$ depend on the
accretion mode: $a_0=-4,~a_1=-2$ for sources in the LK mode and
$a_0=-2,~a_1=1$ for those in the HK mode. Below the Eddington ratio of
$10^{-4}$, SMBHs are typically classified as inactive and behave as
radio-quiet AGN. Finally, the minimum mass in the SMBH mass function
is fixed to $10^5\msol$. The contribution of less massive BHs is still
very uncertain and expected to be negligible in the luminosity--flux
density range we are investigating \citep[see e.g.][]{kol17,gre19}.

\subsection{Mass function and Eddington ratio distribution of active
  SMBHs}
\label{s2s1b}

\begin{figure*}
  \centering \includegraphics[width=8cm]{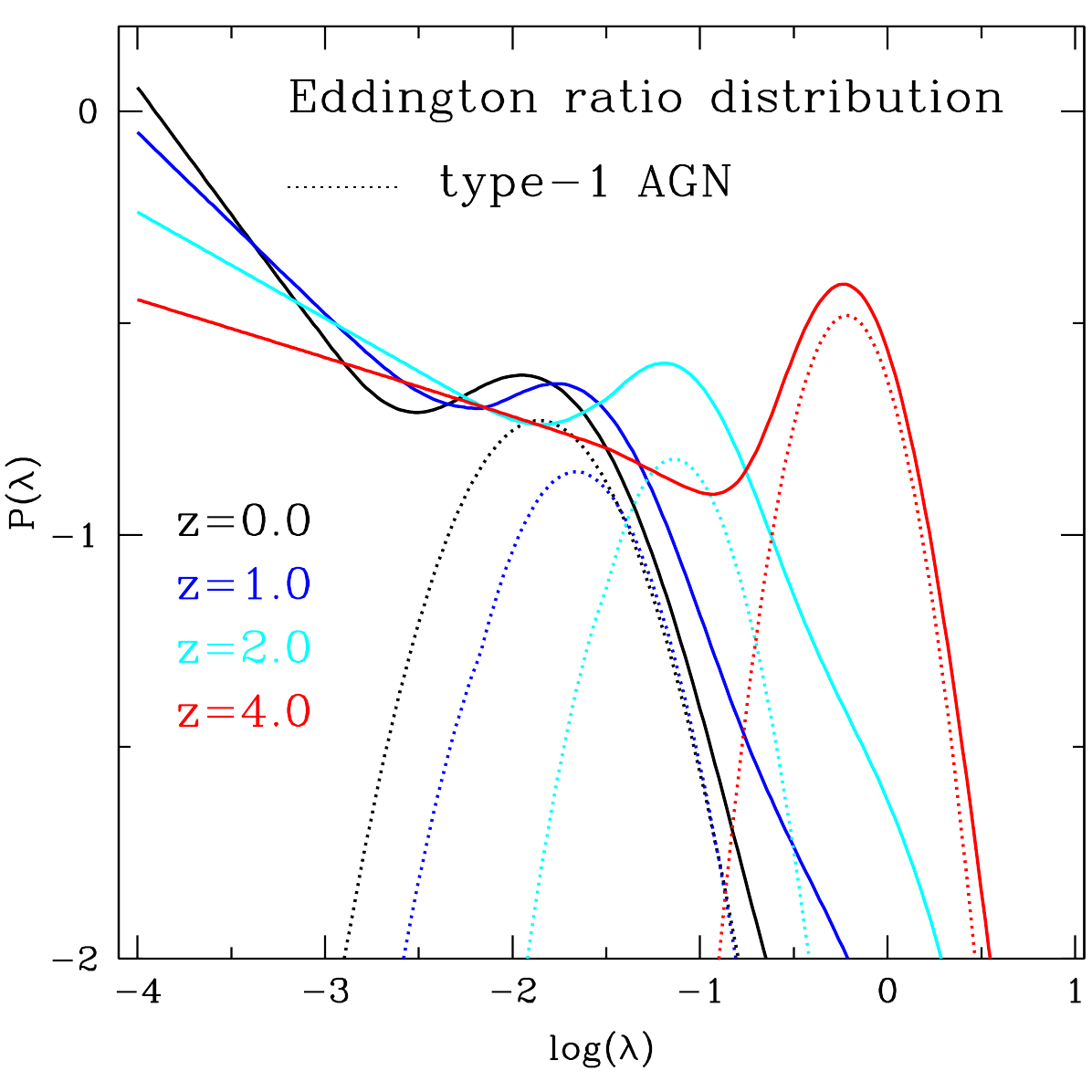}
  \includegraphics[width=8cm]{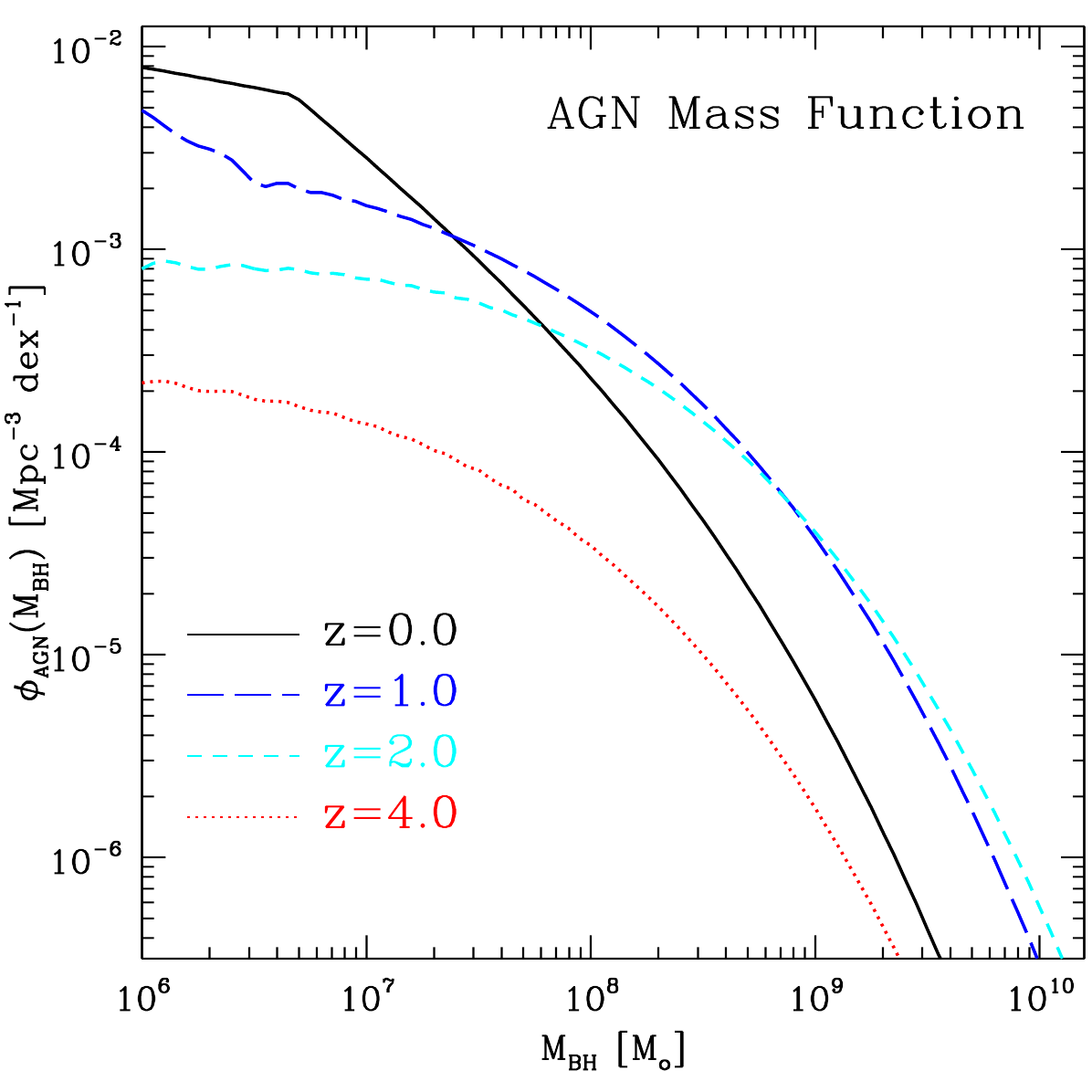}
  \caption{Eddington ratio distribution ({\it left panel}) and AGN
    mass function ({\it right panel}) from \citet{tv17} as a function of
    redshift. The dotted lines in the left panel correspond to the
    Eddington ratio distribution for type-1 AGN.}
  \label{s2f0}
\end{figure*}

\citet{tv17} studied the evolution of the SMBH population, total and
active, via the continuity equation, backwards in time from z = 0 to z
= 4. Type-1 and type-2 AGN are differentiated in that model with
different Eddington ratio distributions, chosen on the basis of
observational estimates; a log-normal distribution is employed for
type-1 AGN, and a power-law distribution with an exponential cut-off
at super-Eddington luminosities is employed for type-2 AGN (see
Fig.\,\ref{s2f0}). \citet{tv17} showed that the evolution of the AGN
 MF is well constrained by the quasar bolometric
luminosity function\footnote{The authors used the function from
  \citet{hop07}.}, with small dependence on the average radiative
efficiency, the Eddington ratio distribution, and the local mass
function of SMBHs. The model contains information on the main physical
quantities that determine the intrinsic properties of SMBHs (i.e.
mass and accretion rate).

In the following analysis we use the AGN MF provided by \citet{tv17}
for an average radiative efficiency of $.epsilon=0.07$, and the
Eddington ratio distribution defined in their Eq.\,12. Their evolution
in redshift is shown in Fig.\,\ref{s2f0}. We can see from the
Eddington ratio distribution that at low redshift most type-2 AGN
accrete at very low rates, while the distribution of type-1 AGN peaks
at $\lambda\sim0.01$. When  the redshift increases, the distribution for
type-2 AGN flattens, and that for type-1 AGN becomes sharper and
peaked at $\lambda\gg0.01$. Looking at the MF, the number density of
low-mass AGN decreases by more than an order of magnitude between
$z=0$ and 4, while for massive AGN ($M>10^8\,M_{\odot}$) the number
density peaks at redshifts 1–2 and then rapidly decreases up to
$z=4$. We have verified that using the AGN MFs that result from
$\epsilon=0.05$ and 0.1 (the other two main cases considered in that
paper) does not have any relevant impact on our final results.

\begin{figure}
  \centering
  \includegraphics[width=9cm]{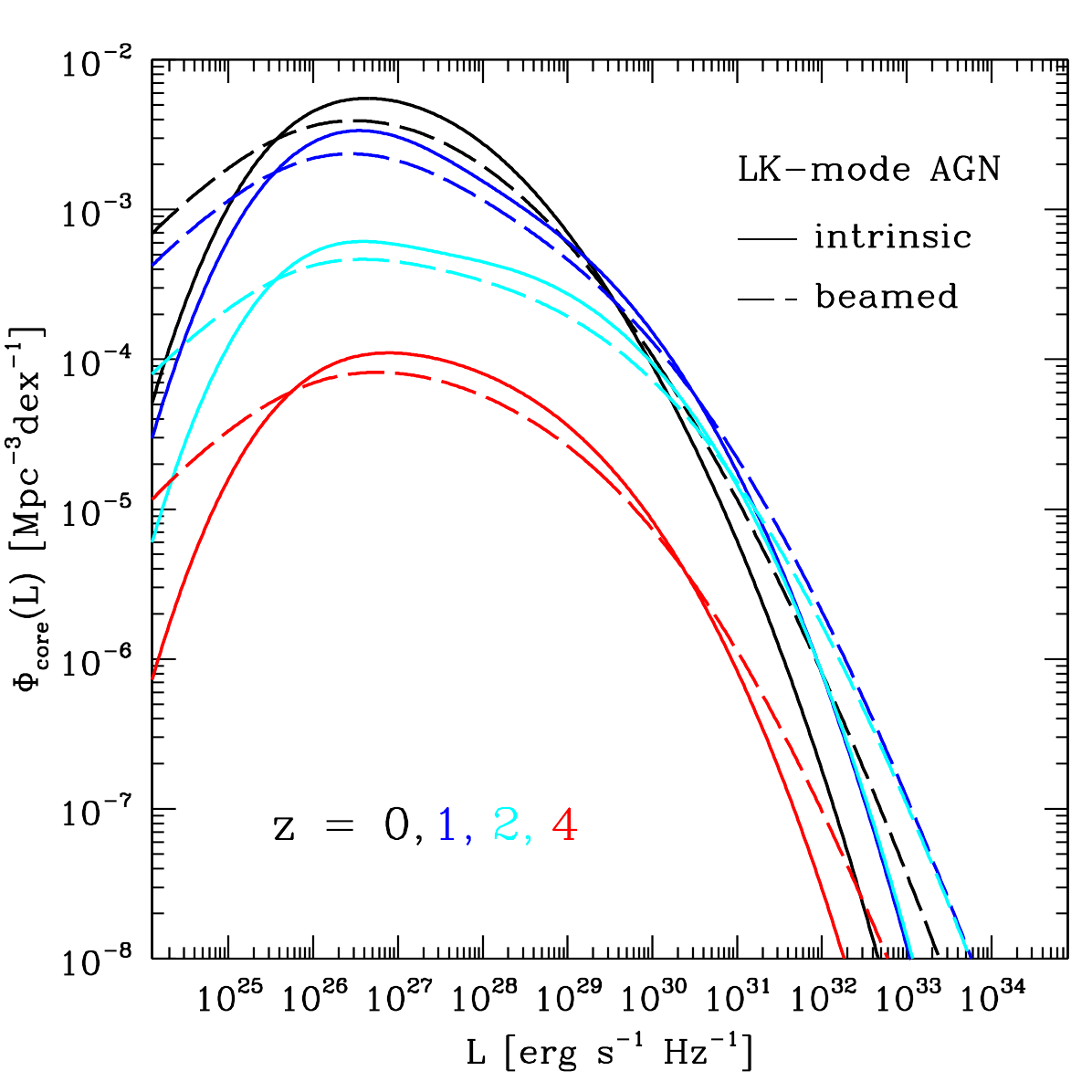}
  \includegraphics[width=9cm]{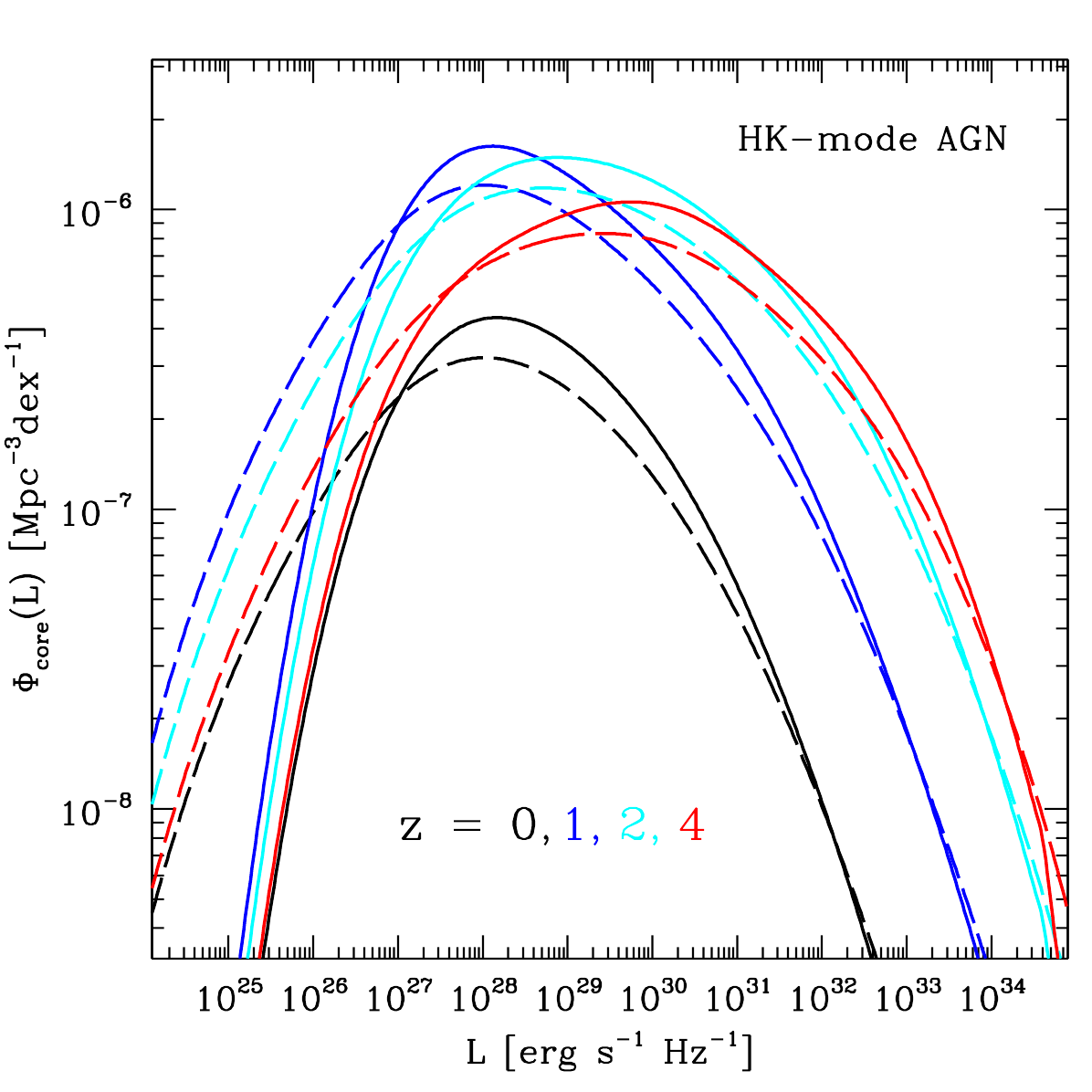}
  \caption{Luminosity function of AGN core jets at 5\,GHz and at
    different redshifts ($z=0,\,1,\,2,\,4$), before (solid lines) and
    after (dashed lines)  including beaming effects for sources in
    the LK mode ({\it top panel}) and HK mode ({\it bottom panel}).}
  \label{s2f1}
\end{figure}

In Fig.\,\ref{s2f1} we show the intrinsic radio luminosity functions
of AGN cores at 5\,GHz in the LK and HK mode as a function of
redshift, obtained by Eq.\,\ref{s2e3}. For LK mode AGN, using the FP
relation of \citet{mer08}, the core LFs present a peak at luminosities
of $10^{26}$--$10^{27}$\,erg\,s$^{-1}$\,Hz$^{-1}$. The rapid decrease
below these luminosities is probably due to the cuts in the AGN MF at
$M=10^5\msol$ and in the Eddington ratio distribution at
$\lambda=10^{-4}$. The core LF also shows   a significant decrease in
amplitude at $z>1$, due to the strong evolution of  the low-mass AGN MF
observed in Fig.\,\ref{s2f0}. At the same time the fraction
of high-lumiosity objects increases with redshift,  due to the
combined effect of a higher average Eddington ratio and a flatter AGN
MF.

A quite different behaviour is instead observed in Fig.\,\ref{s2f1}
for HK mode AGN. The intrinsic core luminosity is computed now using
$\xi^{HK}_X=1.70$, $\sigma_{FP}=0.6$\footnote{These values correspond
  to the best-fit model found in the following analysis.}, and the
continuity condition for the FP relation. As expected, HK mode AGN
are much less numerous than LK mode AGN, but more luminous. The peak
in the LF is found from $10^{28}$ to
$10^{30}$\,erg\,s$^{-1}$\,Hz$^{-1}$, at increasing luminosities with
increasing redshift. We observed an important evolution of the LF.
Going to high redshifts, HK mode AGN become more numerous and
typically brighter by 2--3 orders of magnitude than local ones. Again,
this increase in number and in luminosity with redshift of HK mode AGN
reflects the evolution of the mass density and of the Eddington ratio
distributions for massive (type-1) AGN from the \citet{tv17} model.

\subsection{Beaming effects and observed AGN core luminosities}
\label{s2s2}

Radio-loud  AGN are characterized by the presence of relativistic jets
that emit synchrotron radiation. The observed luminosities of AGN core
jets are affected by Doppler beaming effects and can be significantly
different from their intrinsic values. Here we derive the relation
between intrinsic and observed AGN core LF, given a distribution of
the Lorentz factor $\gamma$. We   assume the simple model in which
each AGN contains two identical,  straight, and oppositely directed jets
with bulk flow velocity $\beta$ in units of the speed of light. Jets
are randomly oriented such that if $\theta$ represents the angle
between the jet axis and the line of sight, $\cos\theta$ is uniformly
distributed between 0 and 1. Finally, according to current models of
the emission in inner cores of AGN, in the optically thick regime the
spectrum of core emission is assumed to be a power law with index
$\alpha_f=0$ \citep[see e.g.][]{bla79}.

The  observed luminosity ($L_c$) of AGN core jets is related to the
intrinsic luminosity ($\mL_c$) via the formula
\beq
L_c\simeq\frac{\mL_c}{[\gamma(1-\beta\cos\theta)]^p}\,.
\label{s2e4}
\eeq
The Lorentz factor $\gamma$ is related to the bulk velocity by
$\gamma=\sqrt{1-\beta^2}$ and $\beta\gamma=(\gamma^2-1)^{1/2}$. The
exponent $p$ depends on the emission model assumed for AGN jets and is
$p=2+\alpha_f$ for continuous jets or $p=3+\alpha_f$ for discrete
moving sources \citep[see e.g.][]{urr95}. We   take as reference
value $p=2+\alpha_f$, which is probably more appropriate for the
continuous jet emission associated with flat-spectrum cores of
AGN. The contribution of opposite-directed jets can be safely ignored
in Eq.\,\ref{s2e4} because it only affects the shape of the very
faint end of the beamed core LF \citep{lis03}.

We take a power-law distribution for the Lorentz factor of AGN jets,
$P(\gamma)\propto\gamma^{-\alpha_{\gamma}}$ with
$\gamma_{min}\le\gamma\le\gamma_{max}$. This distribution provides the
best fit to the apparent speed distribution in AGN jets
\citep{lis97,lis03,liu07}. The AGN population contains  a wide
range of apparent jet speeds, which is inconsistent with a
single-valued intrinsic speed distribution, and extends from 0 to about
35c \citep{kel04,coh07,lis09}. Because the maximum value of the
apparent velocity $\beta_{app}$ sets  the approximate value of the
maximum Lorentz factor $\gamma_{max}$ \citep{ver94}, we adopt
$\gamma_{min}=1$ and $\gamma_{max}=30$. For the slope of the
distribution we take the best-estimated value (i.e. $\alpha_{\gamma}=1.7$) from \citet{liu07}, which is in agreement with the results of
\citet{lis97} and \citet{lis03}. We have verified that changing
$\alpha_{\gamma}$ from 1.3 to 2 (i.e. the interval indicated by
data on the apparent velocities of jets) has little impact on our
results.

These analyses are typically based on complete samples of radio-loud
AGN, including BL\,Lacs, flat-spectrum radio quasars, and radio
galaxies,  objects with very different intrinsic
luminosity values. \citet{lis97} investigated a possible correlation between
$\gamma$ and the intrinsic luminosity of the AGN jet in the form of
$\mL_c\propto\gamma^{-\zeta}$. They found that observational data do
not require such a correlation, and simple models with $\gamma$
independent of $\mL_c$ provide as good fits to the data as $L$--$\gamma$
dependent models. Based on this outcome and for the sake of simplicity, we
consider the Lorentz factor to be independent of the intrinsic
luminosity of the AGN jet, and we use the same distribution for AGN in
LK and HK mode.

Given the Lorentz factor distribution, the observed LF of AGN core
jets, $\Phi(L_c)$, can be estimated from the intrinsic LF by
\beq
\Phi(L_c)=\int d\log\mL_c\,\Phi(L_c,\mL_c)=
\int d\log\mL_c\,\Phi(\mL_c)\,P(L_c|\mL_c)
\label{s2e5}
,\eeq
where the conditional probability, $P(L_c|\mL_c)$, to observe an
object with a core luminosity $L_c$ if the intrinsic core luminosity
is $\mL_c$, is
\bea
P(L_c|\mL_c) & = & \int_{\gamma_0}^{\gamma_1} 
d\gamma\, P(L_c,\gamma|\mL_c)=
\int_{\gamma_0}^{\gamma_1} 
d\gamma\, P(L_c|\gamma,\mL_c)P(\gamma|\mL_c)  \nonumber \\
& = & \int_{\gamma_0}^{\gamma_1} d\gamma\, P(\cos\theta)\,P(\gamma)\,
\bigg(\frac{d\log{L_c}}{d\cos\theta}\bigg)^{-1}  \nonumber \\
& = & \frac{\ln(10)}{p}\bigg(\frac{\mL_c}{L_c}\bigg)^{1/p}
\int_{\gamma_0}^{\gamma_1} d\gamma\,
\frac{P(\gamma)}{\beta\gamma}\,.
\label{s2e6}
\eea
The limits of the integration are fixed by the condition that
$\gamma_{min}\le\gamma\le\gamma_{max}$ and
$0\le\cos\theta=(1-r/\gamma)/\beta\le1$, where
$r=(\mL_c/L_c)^{1/p}$. We find that
$\gamma_0=\max[\gamma_{min},\,r,\,0.5(r+1/r)]$ and
$\gamma_1=\gamma_{max}$. Moreover, the conditional probability is zero
if $r>\gamma_{max}$ or $r+1/r>2\gamma_{max}$. We can then write the
observed luminosity function as
\beq
\Phi(L_c)=\frac{\ln(10)}{p}\int_{\log{\mL_{min}}}^{\log{\mL_{max}}}d\log{\mL_c}\,
\Phi(\mL_c)\,\bigg(\frac{\mL_c}{L_c}\bigg)^{1/p}
\int_{\gamma_0}^{\gamma_1} d\gamma\,\frac{P(\gamma)}{\beta\gamma}\,,
\label{s2e7}
\eeq
where
\beq
\mL_{min}=\bigg[\gamma_{max}-(\beta\gamma)_{max}\bigg]^pL_c
~~~{\rm and}~~~\mL_{max}=\gamma_{max}^p L_c\,.
\label{s2e8}
\eeq

Using the intrinsic core LFs computed in the previous section,
Fig.\,\ref{s2f1} shows  the observed LFs after
 applying the beaming correction. We expect an important Doppler
boosting of the AGN luminosity when the line of sight is close to the
jet axis (i.e. $\theta\la1/\gamma$). This effect is observed as an
increase in  the high-lumiosity tail of the intrinsic core LFs. On
the other hand, when the angle of view is $\gg1/\gamma$ the observed
luminosity can be strongly reduced, as observed in the LFs at the
lowest luminosities. The global effect of beaming is therefore a
broadening of LFs at low and high luminosities.

\subsection{Radio luminosity of AGN from extended jets and lobes}
\label{s2s3}

The luminosity of AGN cores can be strongly suppressed by debeamed
effects when they are observed at angles $\theta\gg1/\gamma$ with
respect to the jet axis. In these cases the emission from extended
jets and lobes is expected to dominate over the core, and AGN will
show the typical steep spectra ($\alpha_s>0.5$).

It is well established that the extended radio luminosity of AGN is
correlated with the bulk kinetic power of the jet, and hence linked to
the AGN central engine \citep[e.g.][]{raw91,wil99}. A
commonly used relation in converting radio luminosity to jet power is
that presented by \citet{wil99}. Based on the minimum energy condition
to estimate the energy stored in the radio lobes,
\citet{wil99} derived a relation between the jet kinetic power $W_j$
and the radio luminosity of FR\,II radio sources of the form
$W_j\propto L^{6/7}$.

Different methods to estimate the jet kinetic power, independently of
the radio lobe luminosity, have been developed and used to test the
\citet{wil99} relation, finding a good agreement with it
\citep{ode09,dal12,god13}.

For FR\,I galaxies the conversion of mechanical power to observed
radio luminosity is more complex due to their strong interactions with
the enviroment. These sources are known to be decelerated to
sub-relativistic speeds during the propagation in the intergalactic
medium. The \citet{wil99} relation has also been calibrated  against
FR\,I sources by studying cavities and bubbles produced by jets in the
intergalactic medium \citep[see
e.g.][]{raf06,har07,bir08,cav10}. These authors found a relation
similar to that of FR\,II sources.

On the other hand, the radio core emission is expected to be a good
tracer of the kinetic jet power. According to the standard model of
the synchrotron jet emission \citep{bla79}, the 
core luminosity should present a dependence on the kinetic jet
power of the form $\mL_c\propto W_j^{17/12}$. \citet{hei03} showed
that in scale-invariant jet models with radiatively inefficient
accretion modes the flat-spectrum core emission follows the relation
$\mL_c\propto W_j^{(17+8\alpha_f)/12}M^{-\alpha_f}$. 
This relation was
observationally confirmed by \citet{mer07} from a sample of 15 nearby
AGN for which the average kinetic power was determined based on the
observed X-ray cavities.

We  therefore expect a direct link between the intrinsic core
luminosity of AGN and the luminosity of their extended
jets and lobes. Based on the above discussion, we assume a power-law
relationship between them (i.e.  in logarithmic scale):
\beq
\log{L_{ex}}=\xi_W\log{\mL_c}+\beta_W\,.
\label{s2e11}
\eeq
According to the expected relations between kinetic jet power and
radio luminosity from jet core and extended jets and lobes \citep{wil99,
  bla79}, the power-law index is estimated to be
$\xi_W\approx0.8$\footnote{This value is calculated by considering the
  above relations: $\mL_c\propto W_j^{17/12}$ and
  $W_j\propto L_{ex}^{6/7}$.}.

Assuming a Gaussian distributed scatter around the previous relation
(with variance $\sigma_W$), the conditional probability to have an
extended jet--lobe luminosity $L_{ex}$ from an AGN with a given core
luminosity $\mL_c$ is
\beq
\mP(L_{ex}|\mL_c) = \frac{1}{\sqrt{2\pi}\sigma_W}\,
e^{-[\log{L_{ex}}-\log{\widetilde{L}_{ex}(\mL_c)}]^2/2\sigma_W^2}\,,
\label{s2e13}
\eeq
where $\widetilde{L}_{ex}(\mL_c)$ is the luminosity estimated from
Eq.\,\ref{s2e11}. The values of $\xi_W$, $\beta_W$, and $\sigma_W$ are considered as free parameters of the model and different for LK
and HK mode AGN.

Given the intrinsic core LF of AGN $\Phi(\mL_c)$, the LF associated with
the extended jet--lobe luminosity, $\Phi(L_{ex})$, can be written as
\beq
\Phi(L_{ex}) = \int d\log({\mL_c})\,\mP(L_{ex}|\mL_c)\,\Phi(\mL_c)\,.
\label{s2e12}
\eeq

\subsection{Radio luminosity function of radio-loud AGN}
\label{s2s4}

At this point we have all we need to estimate the LF of
radio-loud AGN at radio frequencies and in the redshift range of
$z=0$--4, starting from the intrinsic core LF at 5\,GHz. The following
formulas are valid for AGN  in LK and in HK mode.

The observed AGN radio luminosity is the sum of the beamed core
emission ($L_c$) plus the extended emission from jets and lobes
($L_{ex}$):  $L=L_c+L_{ex}$. Both core and extended luminosities
are related to the intrinsic core luminosity. In the previous
sections we   provided these relations at 5\,GHz.

The total LF inferred at a radio frequency $\nu$ can be written as
\beq
\Phi_{\nu}(L)=\int  d\log{\mL_c}\,\mP(L_{\nu}|\mL_c)\,\Phi_5(\mL_c)\,.
\label{s2e14}
\eeq
Assuming that AGN core and extended emissions scale with frequency as
a power law of indices $\alpha_f$ and $\alpha_s$, respectively, we have
that the total luminosity at frequency $\nu$ is related to the 5 GHz
core and extended luminosities as
\beq
L_{\nu}=(\nu/5)^{-\alpha_f}L_c+(\nu/5)^{-\alpha_s}L_{ex}\,,
\eeq
where $L_c$ and $L_{ex}$ are at 5\,GHz. The conditional probability
$\mP(L_{\nu}|\mL_c)$ depends on the conditional probabilities
$\mP_5(L_c|\mL_c)$ and $\mP_5(L_{ex}|\mL_c)$ at 5\,GHz in the following
way:
\bea
\mP(L_{\nu}|\mL_c) & = & 
\int d\log{L_c}\,\mP(L_{\nu}|L_c,\mL_c)\mP_5(L_c|\mL_c) \nonumber \\
& = & \int d\log{L_c}\,\mP_5(L_{ex}|L_c,\mL_c)\mP(L_c|\mL_c)
\bigg(\frac{d\log{L_{\nu}}}{d\log{L_{ex}}}\bigg)^{-1} \nonumber \\
& = & \int 
d\log{L_c}\,\frac{L_{\nu}}{L_{ex,\nu}}\,\mP_5(L_{ex}|\mL_c)\,\mP_5(L_c|\mL_c)\,.
\label{s2e15}
\eea
Inserting Eq.\,\ref{s2e15} in Eq.\,\ref{s2e14} and using
Eq.\,\ref{s2e6} and \ref{s2e13} for $\mP_5(L_c|\mL_c)$ and
$\mP_5(L_{ex}|\mL_c),$ respectively, we find
\bea
\Phi_{\nu}(L) & = & \frac{\ln{(10)}}{p}\frac{1}{\sqrt{2\pi}\sigma_W}
\int d\log{\mL_c}\,\Phi_5(\mL_c) \times \nonumber \\
& & \int_{\widetilde{L}_{c0}}^{\widetilde{L}_{c1}} 
d\log{L_c}\,\frac{L_{\nu}}{L_{ex,\nu}}\,\bigg(\frac{\mL_c}{L_c}\bigg)^{1/p} \times
\nonumber \\
& &
\exp\Bigg\{-\frac{[\log{(L_{ex})}-\log{\widetilde{L}_{ex}(W_j)}]^2}{2\sigma_W^2}\Bigg\}
\times
\nonumber \\
& & 
\int_{\gamma_0}^{\gamma_{max}} d\gamma\,\frac{P(\gamma)}{\beta\gamma}\,,
\label{s2e16}
\eea
where 
\bea
\widetilde{L}_{c0} & = & \log{\mL_c}-p\log{\gamma_{max}} \nonumber, \\
\widetilde{L}_{c1} & = & \log{\mL_c}-p\log({\gamma_{max}-\sqrt{\gamma_{max}^2-1}})\,.
\label{s2e17}
\eea
We can also estimate the LF for flat- and steep-spectrum sources
separately. For flat-spectrum sources (and in an equivalent way for
steep-spectrum sources) the LF is
\beq
\Phi_{\nu}^{flat}(L)=\int d\log{\mL_c}\,\mP({\rm flat}|L_{\nu},\mL_c)\,
\mP(L_{\nu}|\mL_c)\,\Phi_5(\mL_c)\,,
\label{s2e18}
\eeq
where $\mP({\rm flat}|L,\mL_c)$ is the conditional probability that an
object is classified as flat-spectrum having a total luminosity
$L_{\nu}$ at frequency $\nu$ and an intrinsic core luminosity $\mL_c$
at 5\,GHz.

The observational condition for a source to be flat-spectrum is that
the flux density at 1.4\,GHz is
$S_{1.4}\le S_5(1.4/5)^{-0.5}\simeq1.89S_5$. In terms of
luminosity\footnote{The relation between luminosity and observed flux
  density at frequency $\nu$ is
  $S_{\nu}=L_{\nu}/(4\pi d_L^2)\,(1+z)^{\alpha}$, where $d_L$ is the
  luminosity distance and $z$ the source redshift.}, taking into
account the K-correction and the contributions from the
flat-spectrum core and from the extended steep-spectrum lobes, the
condition at 5\,GHz becomes $L_c\ge \rho_{fl}\,L_{ex}$ with
\beq
\rho_{fl}=\bigg(\frac{0.28^{-\alpha_s}-1.89}{1.89-0.28^{-\alpha_f}}\bigg)
(1+z)^{\alpha_f-\alpha_s}\simeq 0.8(1+z)^{-0.75}\,,
\label{s2e20}
\eeq
assuming $\alpha_f=0$ and $\alpha_s=0.75$.

The conditional probability in Eq.\,\ref{s2e18} is therefore
\bea
\mP({\rm flat}|L_{\nu},\mL_c)& \equiv & \mP(L_c\ge \rho_{fl}L_{ex}|L_{\nu},\mL_c)
\nonumber \\
& = & \int d\log{L_c} P(L_c\ge\bar{\rho}_{fl}L_{\nu},L_c|L_{\nu},\mL_c) 
\nonumber \\
& = & \frac{\int_{\log[\bar{\rho}_{fl}L_{\nu}]}^{\log{L_{max}}}
  d\log{L_c}\,\mP(L_c|\mL_c)}
{\int_{\log(\mL_c/\gamma_{max}^p)}^{\log{L_{max}}} d\log{L_c}\,\mP(L_c|\mL_c)}
\,,
\label{s2e21}
\eea
where $\log(L_{max})=\min(\log(L_5),\, \widetilde{L}_{c1})$ (see
Eq.\,\ref{s2e16} and \ref{s2e17}), and
\beq
\bar{\rho}_{fl} = \frac{\rho_{fl}}{(\nu/5)^{1-\alpha_s}+\rho_{fl}(\nu/5)^{1-\alpha_f}}\,.
\eeq
The denominator in the Eq.\,\ref{s2e21} is introduced to have the right
normalization:  $P({\rm flat}|L,\mL_c)+P({\rm steep}|L,\mL_c)=1$.

\subsection{Number counts}
\label{s2s5}

Differential number counts, $n(S)$, defined as the number of sources
per unit area with flux density $S$ within $dS$, are the typical
observables from large-area radio surveys. Observed number counts are
projected counts over a large range of redshifts and can be written as
the integral along redshift of the radio luminosity function
\bea
n(S) & = & \frac{dN}{dS}=\int dz\,\frac{dN}{dSdz}=\int dz\,\frac{dV_c}{dz}\,
\Phi(L,z)\frac{d\log{L}}{dS} \nonumber \\
& = & \frac{1}{\ln(10)S}\int dz\,\frac{dV_c}{dz}\,\Phi[L(S,z),z]\,,
\label{s2e23}
\eea
where $V_c$ is the comoving volume and 
\beq
\frac{dV_c}{dz}=\bigg(\frac{c}{H_0}\bigg)^3
\frac{1}{E(z)}\bigg(\int_0^z\frac{dz'}{E(z')}\bigg)^2\,,
\label{s2e24}
\eeq
with $E(z)=\sqrt{\Omega_m(1+z)^3+\Omega_{\Lambda}}$.

\section{Data sets}
\label{sec3}

In this section we review observational data on LFs and
number counts of radio-loud AGN at frequencies from a few hundred
MHz to a few GHz. Low-redshift LFs at 1.4\,GHz and number counts at 1.4
and 5\,GHz are   used to constrain the free parameters of the model.
Other available data, including LFs at high redshifts ($z>0.5$) and number
counts at $\nu<1.4$, are   taken as a cross-check of the consistency
of the model.

\subsection{Observational data on radio luminosity functions}
\label{s3s1}

\begin{figure}
  \centering
  \includegraphics[width=7cm]{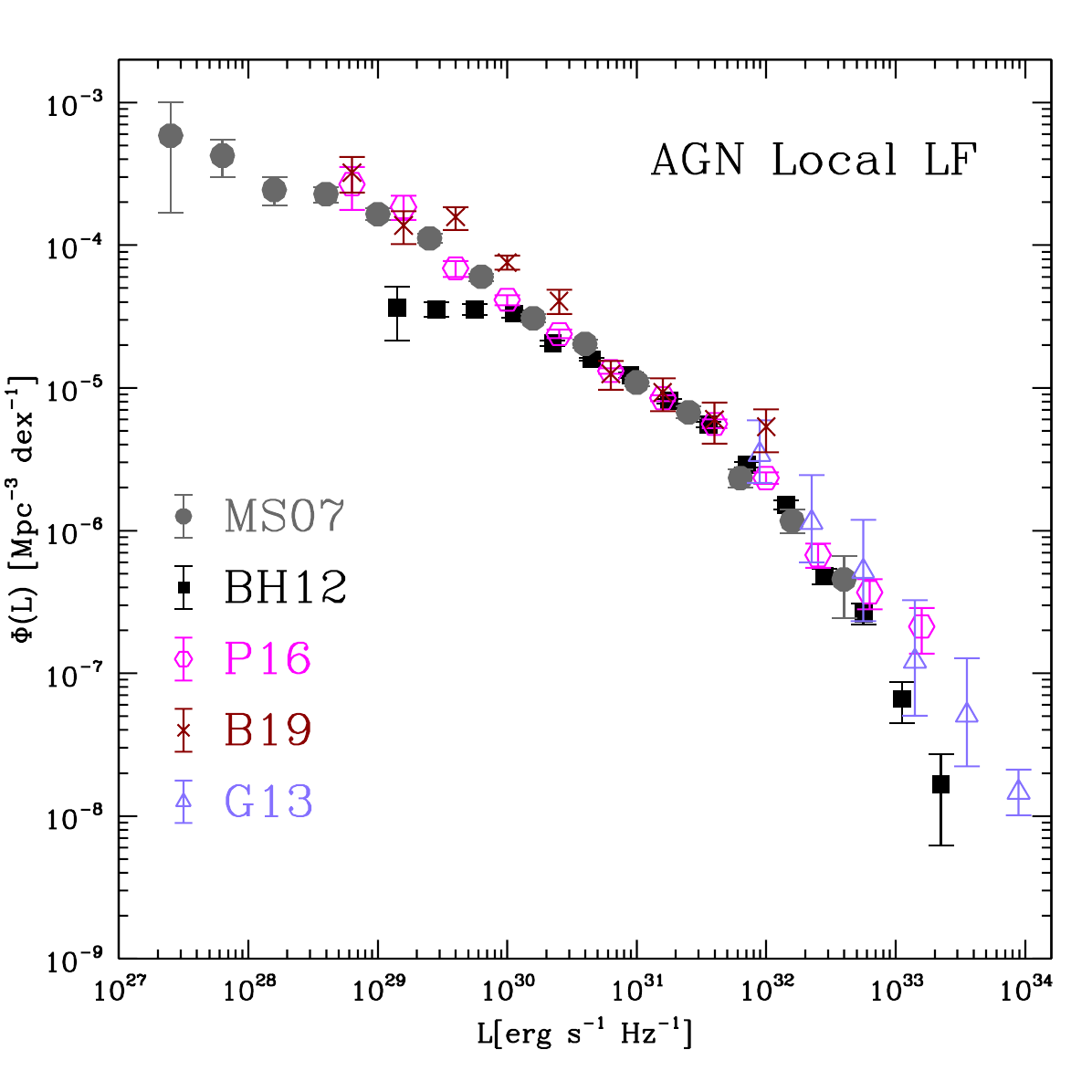}
  \includegraphics[width=7cm]{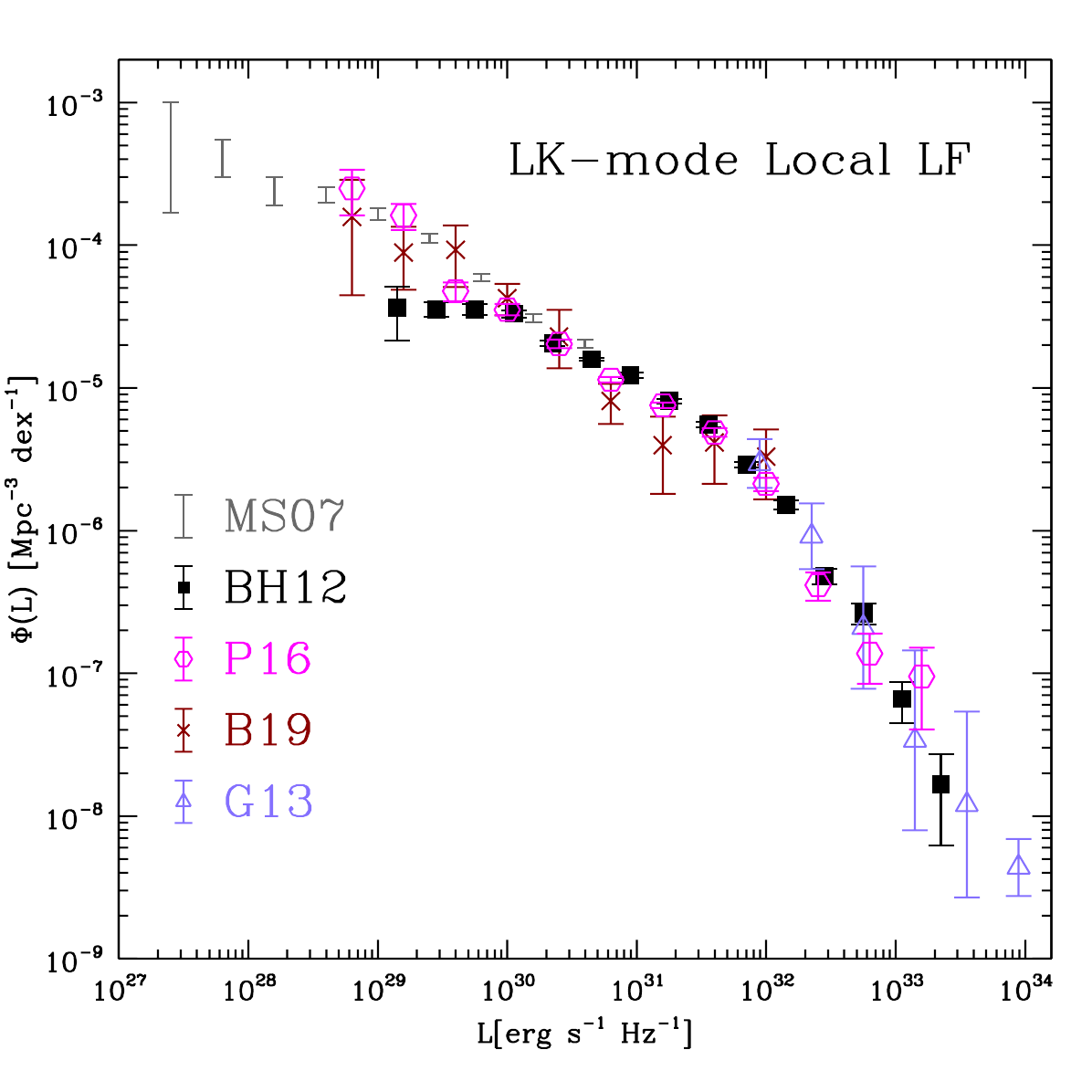}
  \includegraphics[width=7cm]{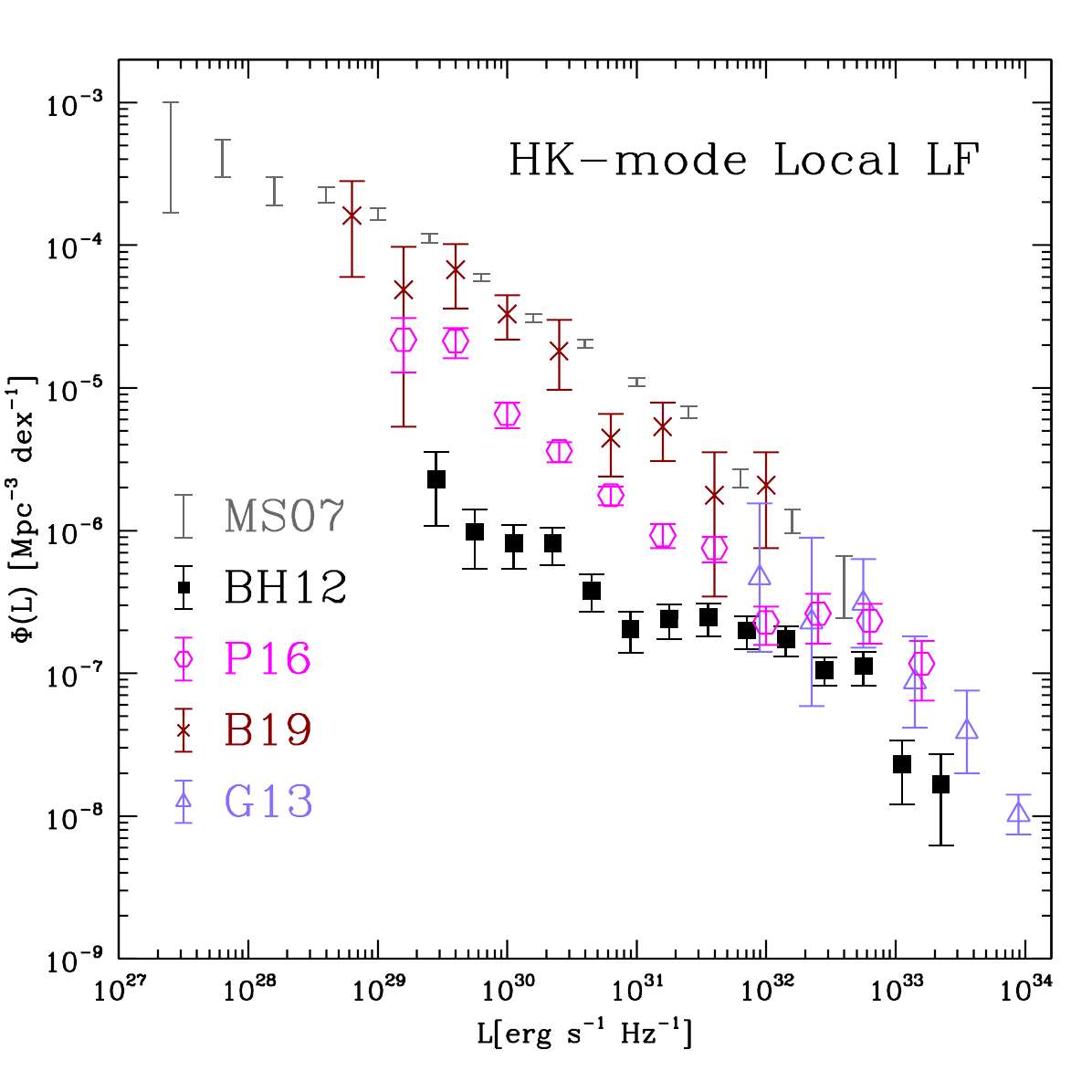}
  \caption{Local luminosity function for all radio-loud ({\it top
      panel}), low-excitation (or LK mode; {\it middle panel}) and
    high-excitation (or HK mode; {\it bottom panel}) AGN at 1.4\,GHz
    from \citet{bes12} (BH12, black solid squares), \citet{pra16}
    (P16, cyan open circles), \citet{but19} (B19, dark red crosses),
    and \citet{gen13} (G13, open violet triangles). The local LF for
    radio-loud AGN from \citet{mau07} (MS07, grey solid points and bars)
    is also reported in all the panels for comparison.}
  \label{s3f1}
\end{figure}

Accurate measurements of radio LFs are mainly restricted to the local
Universe or at low redshifts, $z<1$. The local LF of radio-loud AGN
has been achieved at 1.4\,GHz by the combination of the NRAO VLA Sky
Survey \citep[NVSS;][]{con98} and the Faint Images of the Radio Sky at
Twenty cm \citep[FIRST;][]{bec95} survey  with recent
large-area spectroscopic surveys
\citep[e.g.][]{mac00,sad02,mau07,bes12,van12}. In particular,
\citet{mau07} determined the local LF from a sample of NVSS sources
associated with galaxies brighter than $K=12.75$\,mag in the 6\,degree
Field Galaxy Survey \citep[6dFGS,][]{jon04}. Through visual
examination of 6dF spectra they were able to distinguish between
star-forming galaxies and radio-loud AGN. The latter population is
found to be 40 per\,cent of the sample, corresponding to $\sim2600$
objects. More recently, using a much larger sample of radio-loud AGN
($\sim7000$ objects) constructed by combining the seventh data release
of the Sloan Digital Sky Survey \citep[SDSS,][]{yor00} with the NVSS
and the FIRST survey, \citet[][hereafter BH12]{bes12} found  very good
agreement with the \citet{mau07} local LF at radio luminosities
$L\ga10^{30}$\,erg\,s$^{-1}$Hz$^{-1}$ (see Figure\,\ref{s3f1}). At
lower luminosities their LF flattens below the \citet{mau07}
estimates. This is likely due to the incompleteness of the BH12
sample, whose flux limit is 5\,mJy, corresponding to a radio
luminosity of $\approx10^{30}$\,erg\,s$^{-1}$Hz$^{-1}$ at redshift
$z=0.1$.

Using the wide range of emission line measurements available, BH12
classified radio sources in low-excitation and high-excitation objects
and derived the local LF at 1.4\,GHz for the two populations
separately. More recently, \citet[][hereafter P16]{pra16} also
derived the radio luminosity function for the two populations of radio
galaxies by using an optical spectroscopic survey of radio galaxies
identified from matched FIRST sources and SDSS images. Their local
luminosity function for the radio-loud AGN population is in good
agreement with the BH12 estimates at
$L\ga10^{30}$\,erg\,s$^{-1}$Hz$^{-1}$. Nonetheless, the P16 LF for
high-excitation AGN has a steeper slope than the BH12 value, and
significantly exceeds it at low luminosities (see
Figure\,\ref{s3f1}). P16 explain the discrepancy as being due to the
stricter method employed by BH12 to identify AGN with respect to
star-forming galaxies. \citet{gen13} derived the local radio LF for
high-power low- and high-excitation sources, using the Combined
NVSS-FIRST Galaxy catalogue (CoNFIG), extending measurements up to
$10^{34}$\,erg\,s$^{-1}$Hz$^{-1}$. As displayed in Figure\,\ref{s3f1},
the local LF presented by Gendre et al. (2013) is in very good
agreement with the estimate of P16. On the other hand, it shows an
excess at the highest luminosities in comparison with BH12, especially
for HK mode AGN.

According to previous data, low-excitation (or LK mode) sources
clearly dominate the local LF at low and intermediate luminosities, up
to $10^{32}$\,erg\,s$^{-1}$Hz$^{-1}$, with a fraction of
high-excitation (or HK mode) AGN $\la10\%$. The two
populations become roughly equal in number at
$\sim10^{33}$\,erg\,s$^{-1}$Hz$^{-1}$. Hoever, these results  are not
confirmed by recent findings from \citet[][hereafter B19]{but19}. They use
Australia Telescope Compact Array (ATCA) observations of the XMM
extragalactic survey south field (ATCA XXL-S). Differently to previous
analyses, they first identify high-excitation AGN (both radio-loud
and radio-quiet) based on multi-wavelength information; the remaining
sources are then divided into low-excitation AGN and star-forming
galaxies. B19 found a comparable contribution of high- and
low-excitation AGN to the total AGN LF in the whole luminosity
range. Their high-excitation LF is significantly larger than
estimates from P16 (by a factor $\sim$2--5) and BH12 (by one order of
magnitude; see Fig.\,\ref{s3f1}). These discrepancies are attributed
again to the different classification criteria used to identify
star-forming galaxies and AGN (see the discussion in their
Section\,4.3.2), and to differences in the optical and radio depths
probed by the samples.

The fact that three studies \citep{bes12,pra16,but19} employ different
methods and/or criteria to separate radio-loud AGN and star-forming
galaxies, and that they  find quite different results that are not consistent with each
other, indicates the difficulty and the uncertainties present in the
classification procedure. In the following analysis we decide to use
both the BH12 and P16 measurements for the model fitting. On the other
hand, due to the larger uncertainties associated with them and some
discrepancies in the local AGN LF with previous estimates (a factor of
3--4 between $29.5\la\log(L_{1.4{\rm GHz}})\la30.5$; see
Fig.\,\ref{s3f1}), the B19 data will be considered only for
comparison.

Only a few studies attempted to disentangle the contributions of sources
with different spectral classes to the local LF
\citep{tof87,sti91,dun90,pad07,mao17}. At 1.4\,GHz steep-spectrum
sources are the dominant population, while flat-spectrum sources are
found to be less than 10\,per\,cent. Determinations of the
local and low-redshift LF of flat-spectrum sources are provided by
\citet{tof87} and \citet{pad07}. These are based on small
samples, 
and cover different ranges of luminosities, below and above
$\sim10^{32}$\,erg\,s$^{-1}$Hz$^{-1}$, respectively. Estimates from
\citet{pad07} are obtained from a sample of BL\,Lac objects with
average redshift of 0.26 and assuming no evolution, while
\citet{tof87} used flat-spectrum sources at $z\le0.07$. The two LFs
join almost smoothly with each other (see Fig.\,\ref{s4f4}).

The evolution of the radio LF with redshift has been studied by
different authors
\citep[e.g.][]{sad07,don09,smo09,yua12,mca13,bes14,pad15,smo17,cer18},
typically out to $z\sim1$. For example, \citet{smo09} derived the LF
for low radio power AGN
($L_{1.4\,{\rm GHz}}\la5\times10^{32}$\,erg\,s$^{-1}$Hz$^{-1}$) from a
sample of the VLA--COSMOS survey in four redshift bins in the range
$0.1\le z\le1.3$, finding a modest evolution of the LF. This result
was confirmed by \citet{don09}, using a catalogue of AGN at
$z\sim0.55$ constructed from the cross-correlation of the NVSS/FIRST
survey with the MegaZ--LRG of luminous red galaxies from the
SDSS. While the comoving number density of low-luminosity AGN
increases only by a factor $\sim1.5$ between $z=0.1$ and 0.55, this
factor reaches values of more than 10 at
$L\ga10^{33}$\,erg\,s$^{-1}$Hz$^{-1}$.

\citet{bes14} and \citet{pra16} investigated the evolution of
radio-loud AGN in the radiative and jet mode (or high- and
low-excitation mode) separately out to $z=1$. They showed that the
space density values of the two AGN populations have quite different
evolutions with redshift: the space density of radiative-mode AGN
strongly increases from the local Universe to $z\sim1$ by a factor of
4--7, while jet-mode AGN have a slower evolution or no evolution
depending on radio luminosity. In agreement with this, \citet{but19}
found a weak positive evolution in low-excitation mode AGN, and a
stronger one in high-excitation mode AGN (although weaker than in
previous works).
This is consistent with the mild evolution seen in the total radio
luminosity function at low radio power and the more rapid evolution of
luminous radio galaxies, already observed in previous analyses.

\begin{table}
  \centering 
  \label{t01}
  \begin{threeparttable}
  \caption{Local luminosity functions used in the model fitting procedure.}
  \begin{tabular}{c|cc|cc}
  \hline 
 & \multicolumn{2}{c}{\bf BH12-based} & \multicolumn{2}{c}{\bf P16-based}  \\
 \hline 
 \hline
 AGN & Data$^a$ & Luminosity$^b$ & Data$^a$ & Luminosity$^b$ \\
  \hline 
 & MS07 & $\le30.6$ & MS07 & $\le28.2$ \\
 radio-loud & BH12 & (30.6,\,33) & P16 & (28.2,\,33.5) \\
 & G13 & $\ge33$ & G13 & $\ge33.5$  \\
  \hline 
 & MS07 & $\le30.6$ & MS07 & $\le28.2$ \\
 LK mode & BH12 & (30.6,\,33) & P16 & (28.2,\,33.5) \\
 & G13 & $\ge33$ & G13 & $\ge33.5$  \\
  \hline 
 HK mode & BH12 & [29.45,\,33) & P16 & [28.8,\,33.5) \\
 & G13 & $\ge33$ & G13 & $\ge33.5$  \\
  \hline 
\end{tabular}
\begin{tablenotes}
  \item[a] See Fig.\,\ref{s3f1} for the meaning of acronyms.
  \item[b] Luminosities are given in $\log(L$\,[erg\,s$^{-1}$Hz$^{-1}$]). 
\end{tablenotes}
\end{threeparttable}
\end{table}

\subsection{Luminosity functions used in the model determination}
\label{s3s1s1}

Radio luminosity functions provide essential information for the
determination of free parameters of the model. To this end, we use
the local luminosity function of the total AGN population and of LK
and HK mode  sub-classes, as discussed in detail in the previous
subsection, along with the total LF at $z\simeq0.5$ measured by
\citet[][see Fig.\,\ref{s4f5}]{don09}. We leave instead higher
redshift data for a consistency check of model results.

We execute the model fitting procedure with two different sets of data
for the local LF, one based on the BH12 estimates and the other on the
P16 estimates. In Table\,2 we summarize the data employed with the
corresponding luminosity range. It is important to note that the
total LF from MS07 is also used  for the LK mode LF at very low
luminosities in the hypothesis that at those luminosities the
contribution of HK mode AGN is negligible. Moreover, BH12
measurements for the total and LK mode LF are used only for
$L\ge10^{30}$\,erg\,s$^{-1}$Hz$^{-1}$, due to incompleteness issues of
the sample at lower luminosities (see previous subsection). On the
contrary, we  include in the analysis all the BH12
  estimates for HK mode AGN, taking into account that our model tends
  to underestimate HK mode space densities from BH12 at
  $L<10^{30}$\,erg\,s$^{-1}$Hz$^{-1}$ (see next sections).

In addition, we also consider the following data, all at 1.4\,GHz:

\begin{itemize}

\item{\bf Local LF of flat-spectrum radio sources:} we use the
  estimates from \citet{tof87}, updated according to the 
  adopted cosmology, and \citet{pad07} (see Fig.\,\ref{s4f4}); 

\item {\bf LF of AGN at intermediate redshifts:} we consider results
  from \citet{don09} for the AGN LF at redshifts $z\sim0.55$ (see
  Fig.\,\ref{s4f5}).

\end{itemize}

\subsection{Number counts from large-area surveys}
\label{s3s2}

Relevant information on the
evolution of the space density of radio-loud AGN can be 
provided by observed number counts that gather radio sources in a
large range of redshifts, up to $z\approx4$, with a peak at redshifts
1--2 \citep[see e.g.][]{mas10}.

Thanks to large-area radio surveys, number counts of radio sources
have been accurately estimated at frequencies from hundreds of MHz to
a few GHz and more (for a review, see e.g. \citealt{dez10}; and the
more recent measurements from \citealt{smo17b,but18,ocr19}).

For the model determination we employ number counts at 1.4 and
4.8\,GHz. These are mainly based on the NVSS at 1.4\,GHz, and on the
Northern Green Bank GB6 survey \citep{gre96} and \citet{kue81} at
4.8\,GHz. We use number counts only for flux densities
$S\ge1$\,mJy. At sub-mJy flux densities the relative number
of star-forming galaxies and radio-quiet AGN becomes increasingly
important \citep[see e.g.][and references therein]{mas10}. We do not
include number counts at frequencies larger than 5\,GHz or smaller
than 1.4\,GHz. The assumption of a single-slope power law
spectrum for the core or lobe AGN emission can be considered valid only
in a small range of frequencies (see discussion in Sect.\,\ref{s4s4}).

Finally, we include in the analysis the estimates of number counts for
flat-spectrum sources at 4.8\,GHz from \citet{tuc11}, obtained by
combining different surveys between 1.4 and 20\,GHz, by extrapolating
the spectral energy distribution of flat-spectrum radio sources as
discussed in that paper.

\section{Constraining model parameters}
\label{sec4}

In Sect.\,\ref{sec2} we   describe how to compute the LF of
radio-loud AGN starting from the evolution of the SMBH mass function
and the Eddington ratio distribution. This method implies the use of
phenomenological and theoretical relations that connect SMBH
properties, such as mass and accretion rate, to the AGN luminosity at
radio wavelengths. In most of the cases, these relations are not well
established and free parameters are consequently introduced in the
model. Observational measurements of radio LFs and number counts,
described in the previous section, are exploited to determine these
free parameters and, at the same time, to test the reliability of the
model.

Active galactic nuclei in the LK mode are typically less luminous than HK mode objects,
but are much more numerous, and they are expected to dominate radio LFs at
very low redshifts ($z\la0.5$). On the other hand, AGN in the HK mode
are expected to give the main contribution to number counts observed
from large-area surveys at GHz frequencies at $S\ga10$\,mJy, with a
broad redshift distribution peaking at $z\ga1$ \citep{bro08,mas10}.

We decided to proceed separately for the two populations of
AGN. Firstly, we determined the free parameters for LK mode AGN by
fitting the local LF for this class of objects, which is well measured
by observations. Then, we used the information on number counts and on
HK mode and total LFs to constrain the parameters of HK mode
AGN.

Uncertainties on observational LFs typically include only statistical
errors. As shown in Fig.\,\ref{s3f1}, at some luminosities they are so
small that they are probably underestimated or dominated by systematic
errors \citep{mas10,bes12}. For example, \citet{mas10} adopted an
error of at least 0.2\,dex for the MS07 and \citet{don09} LFs to
determine the parameters of their evolutionary model. In order to
avoid that few observational data with unrealistic small uncertainties
weigh too heavily in the fitting procedure, we decided to fix a minimum
error value for the observational data used in the analysis; we adopted,
conservatively, an error of at least 0.05\,dex for LFs and 0.02\,dex
for number counts.

Hereafter, we mainly focus on results obtained from the BH12-based
set of local LFs, as described in Table\,2. Model predictions are,
however, almost independent of which of the two local LF sets is
employed in the analysis. We   discuss the results
using the P16-based data set in Sect.\,\ref{s4s3}.

\begin{figure}
  \centering
  \includegraphics[width=9cm]{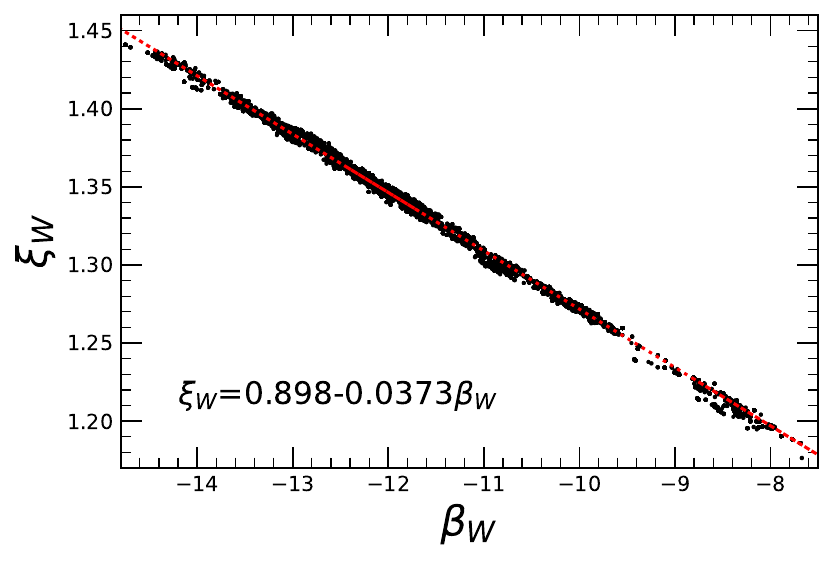}
  \caption{Linear correlation between the model parameters $\xi_W$ and
    $\beta_W$. The solid points are the parameter values found from the
    MCMC chains. The dotted line corresponds to the linear fit reported in
    the panel.}
  \label{s4f3}
\end{figure}

\begin{figure}
  \centering
  \includegraphics[width=9cm]{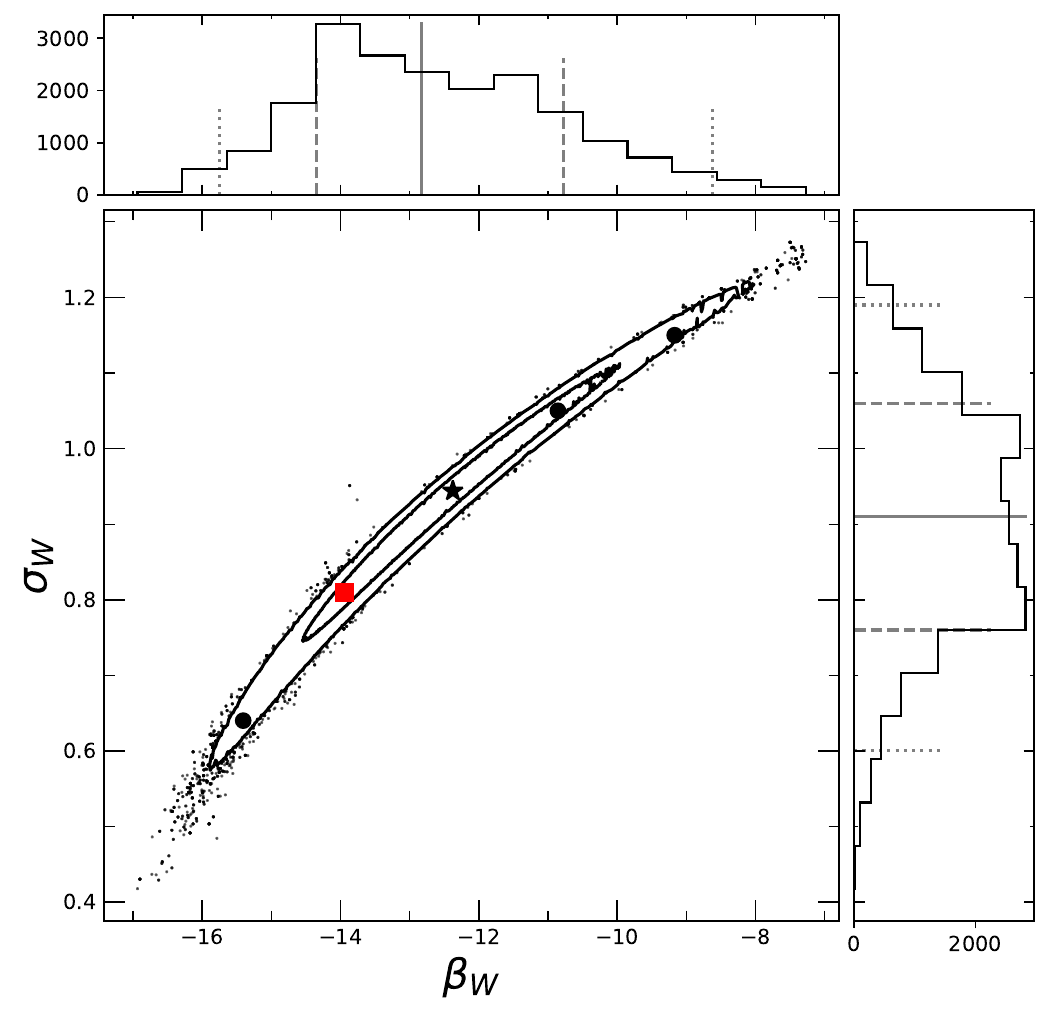}
  \caption{PDF of the model parameters $\beta_W$ and
    $\sigma_W$. Contour plots correspond to the 68.3\%\ and
    95.4\%\,confidence levels. The large solid  points are the model
    sets used in Appendix\,\ref{a1} to fit total LFs and number
    counts: the black star corresponds to the case with the minimum
    $\chi^2$ in the LK mode fit and the red square to the case with the
    minimum $\chi^2$ in the {global} fit. The marginalized
    distributions are also shown, with the median and the 68.3\%\ and
    95.4\% confidence intervals (solid, dashed, and dotted
    lines, respectively).}
  \label{s4f3b}
\end{figure}

\begin{figure*}
  \centering
  \includegraphics[width=9cm]{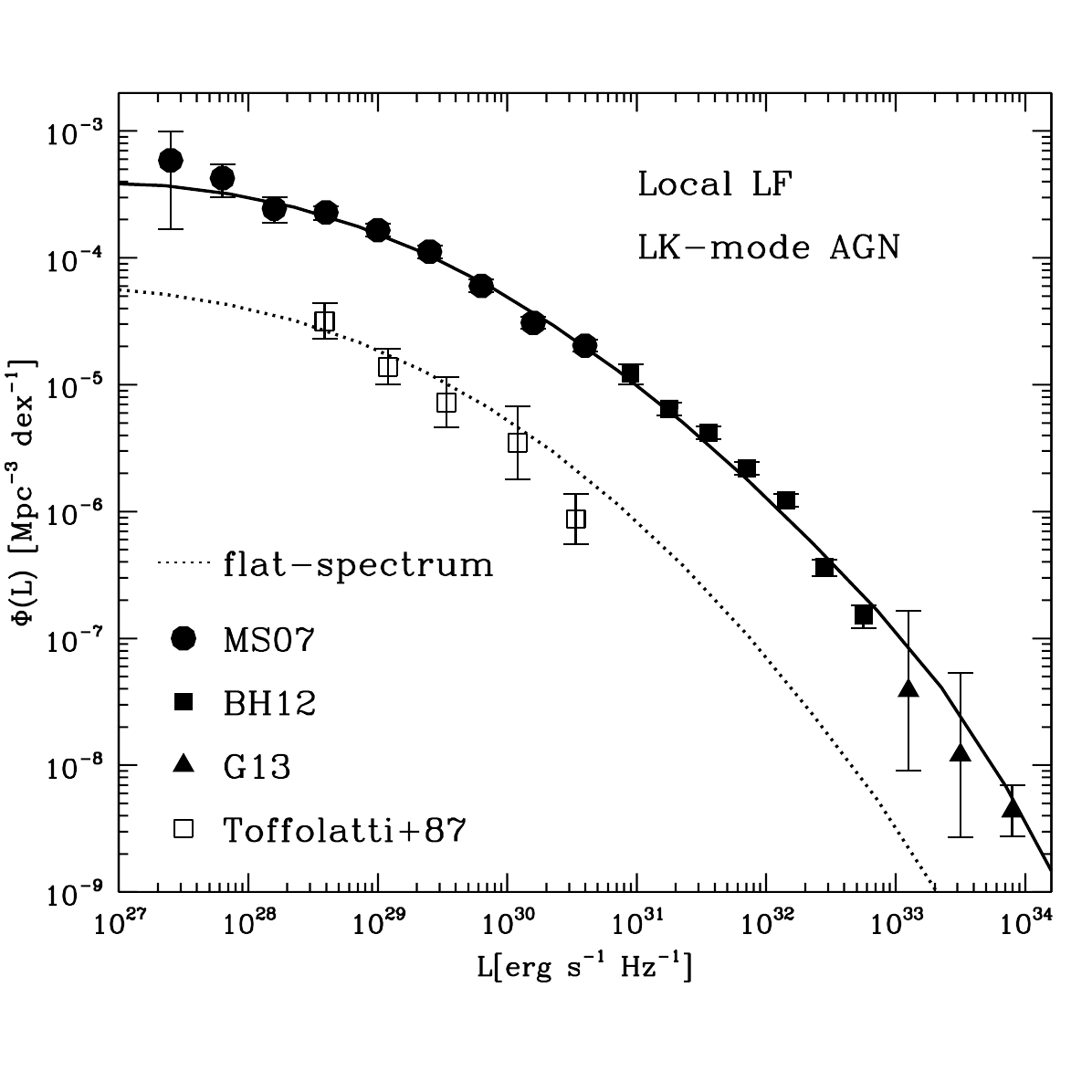}
  \includegraphics[width=9cm]{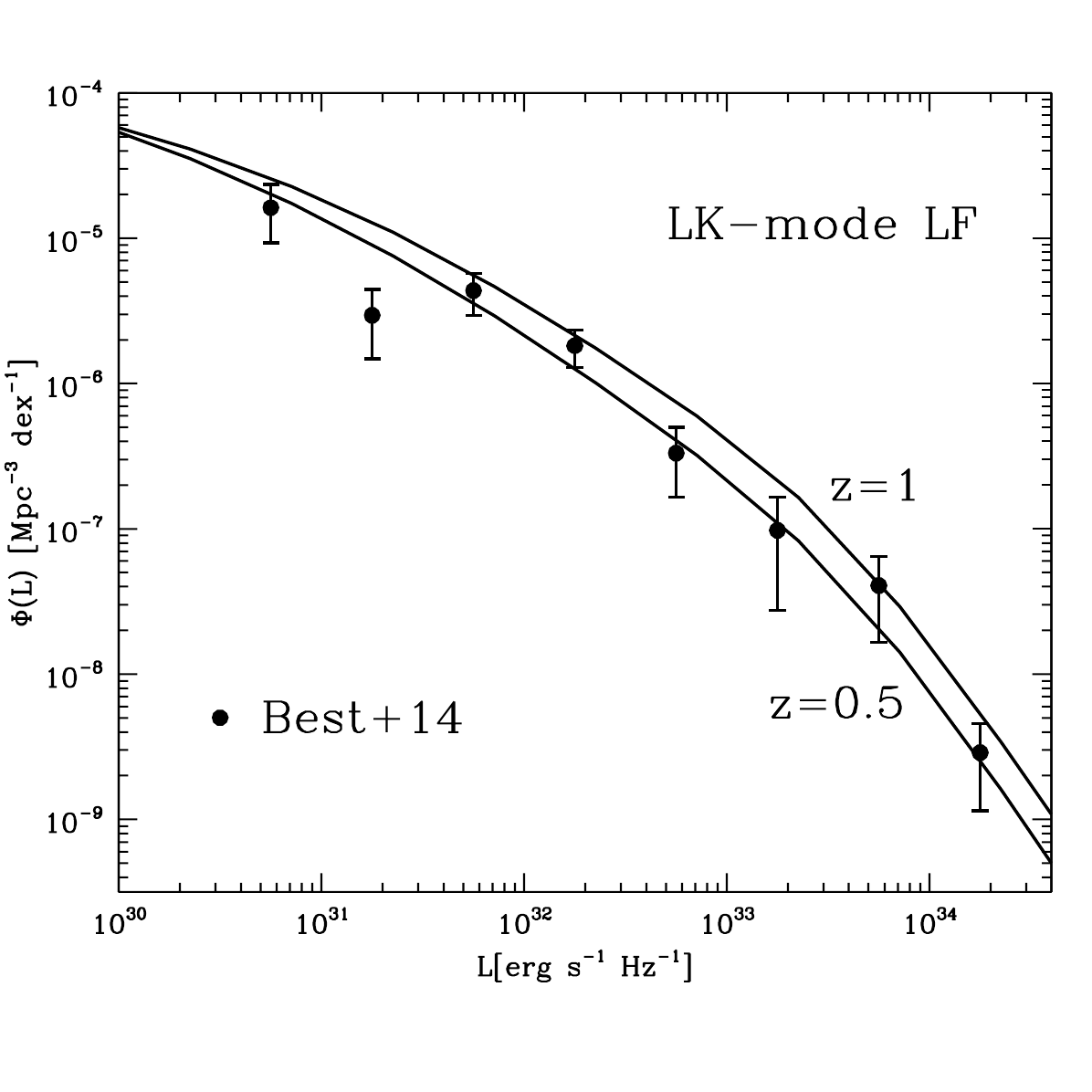}
  \caption{Luminosity function for LK mode AGN ({\it Left panel})
    Model prediction of the local LF for LK mode AGN (solid lines)
    compared to observational estimates for low-excitation AGN
    (\citealt{mau07}, solid points; \citealt{bes12}, solid squares;
    \citealt{gen13}, solid triangles). The dotted lines are the
    predictions for flat-spectrum LK mode AGN, while open squares are
    the local LF for flat-spectrum radio sources estimated by
    \citet{tof87}. ({\it Right panel}) Model predictions of the LK
    mode LF at $z=0.5$ and 1, compared with estimates of the LF of
    jet-mode AGN at $0.5<z<1.0$ from \citet{bes14}.}
  \label{s4f1}
\end{figure*}

\subsection{Results for AGN in the LK mode}
\label{s4s1}

As already stated above, for AGN in the LK mode we adopt the FP
correlation coefficients provided by \citet{mer08}: $\xi_X=0.62$,
$\xi_M=0.55$ and $\beta_R=8.6$, with an intrinsic scatter of
0.6\,dex. We also assume that all LK mode AGN are radio-loud (i.e. $f_{loud}^{LK}=1$). The presence of jets is  commonly
associated with this class of objects \citep{mer08}, in analogy with the
{low, hard} state of black hole X-ray binaries in which radio
emission is always detected.

The only free parameters for this AGN population are the coefficients
$\xi_W$, $\beta_W$, and $\sigma_W$ of Eqs.\,\ref{s2e11} and \ref{s2e13}
that connect the intrinsic core luminosity to the extended emission
from jets and lobes. It is important to keep in mind that these parameters
are independent of redshift, and that the time evolution of the LF for
LK mode AGN is fully regulated by the evolution of the SMBH mass
function and accretion rate (see Sections\,\ref{s2s1b} and \ref{s2s3}).

In order to prevent a too large number of very bright sources, we
impose an exponential cut-off in the values of the intrinsic scatter
$\sigma_{FP}$ and $\sigma_W$ (see Eq.\,\ref{s2e3} and \ref{s2e13}) at
high intrinsic core luminosities, with a minimum value of 0.2:
\beq
\sigma_{FP} = \left\{
  \begin{array}{ll}
    \sigma_{FP} & \log\mL_c\le 34.6\\
    \max(\sigma_{FP}\exp(34.6 - \log\mL_c), 0.2) & \log\mL_c>34.6\,.
  \end{array}
\right.
\nonumber
\eeq
\beq
\sigma_W = \left\{
  \begin{array}{ll}
    \sigma_W & \log\mL_c\le 31.8\\
    \max(\sigma_W\exp(31.8 - \log\mL_c), 0.2) & \log\mL_c>31.8\,.
  \end{array}
\right.
\eeq
The cut-offs on $\sigma_{FP}$ and $\sigma_W$ have an impact on
luminosity functions only at
$L\ga10^{36}$\,and\,$10^{34}$\,erg\,s$^{-1}$\,Hz$^{-1}$, respectively,
and on number counts at the jansky level. We apply the same cut-offs
also to HK mode AGN.

We use a Markov chain Monte Carlo (MCMC) method to determine the three
model parameters by fitting the BH12-based local LF for
low-excitation AGN and the local LF for flat-spectrum sources in the
luminosity range dominated by LK mode AGN
(i.e. $L\la10^{30}$\,erg\,s$^{-1}$\,Hz$^{-1}$).
We find a very tight linear relation between $\xi_W$ and $\beta_W$
(see Fig.\,\ref{s4f3}), meaning that only one of them is a real free
parameter of the model, with $\xi_W$ ranging from 1.3 to 1.45 and
$\beta_W$ from $-15$ to $-8$. We  therefore   directly
determine $\xi_w$ from $\beta_W$, using the linear relation displayed
in Fig.\,\ref{s4f3} and Table\,3.

The fit is then repeated using $\beta_W$ and $\sigma_W$ as the only
free parameters. In Fig.\,\ref{s4f3b} we show the probability density
function (PDF) of the two parameters. We find a significant
correlation between the two parameters. The best-fit model is
obtained for $\beta_W=-12.3$ and $\sigma_W=0.94$,
while the median values (plus the 68\% confidence levels) of
the marginalized PDFs are $\beta_W=-12.83^{+2.06}_{-1.52}$ and
$\sigma_W=0.91\pm0.15$.

As shown in Fig.\,\ref{s4f3b}, the model parameters can be in  a
large range of values, preserving an almost equivalent ability to
fit the data. Changing the parameters at the 1$\sigma$ or 2$\sigma$ level around
the median has little impact on the LK mode local
LF. Differences are more relevant for number counts, as shown in
Appendix\,\ref{a1}. Therefore, we tested  the impact of different
LK mode models when HK mode AGN are also included and total LFs and
number counts are fitted. We consider, in particular, five sets of
parameters (shown as large points in Fig.\,\ref{s4f3b}): the set that
provides the best results will be adopted as our reference model for
LK mode AGN in the following analysis. This corresponds to
$\{\xi_w,\,\beta_w,\,\sigma_W\}=\{1.42,\,-13.95,\,0.81\}$ (see
Appendix\,\ref{a1} and the red point in Fig.\,\ref{s4f3b}).

\begin{figure*}
  \centering
  \includegraphics[width=9cm]{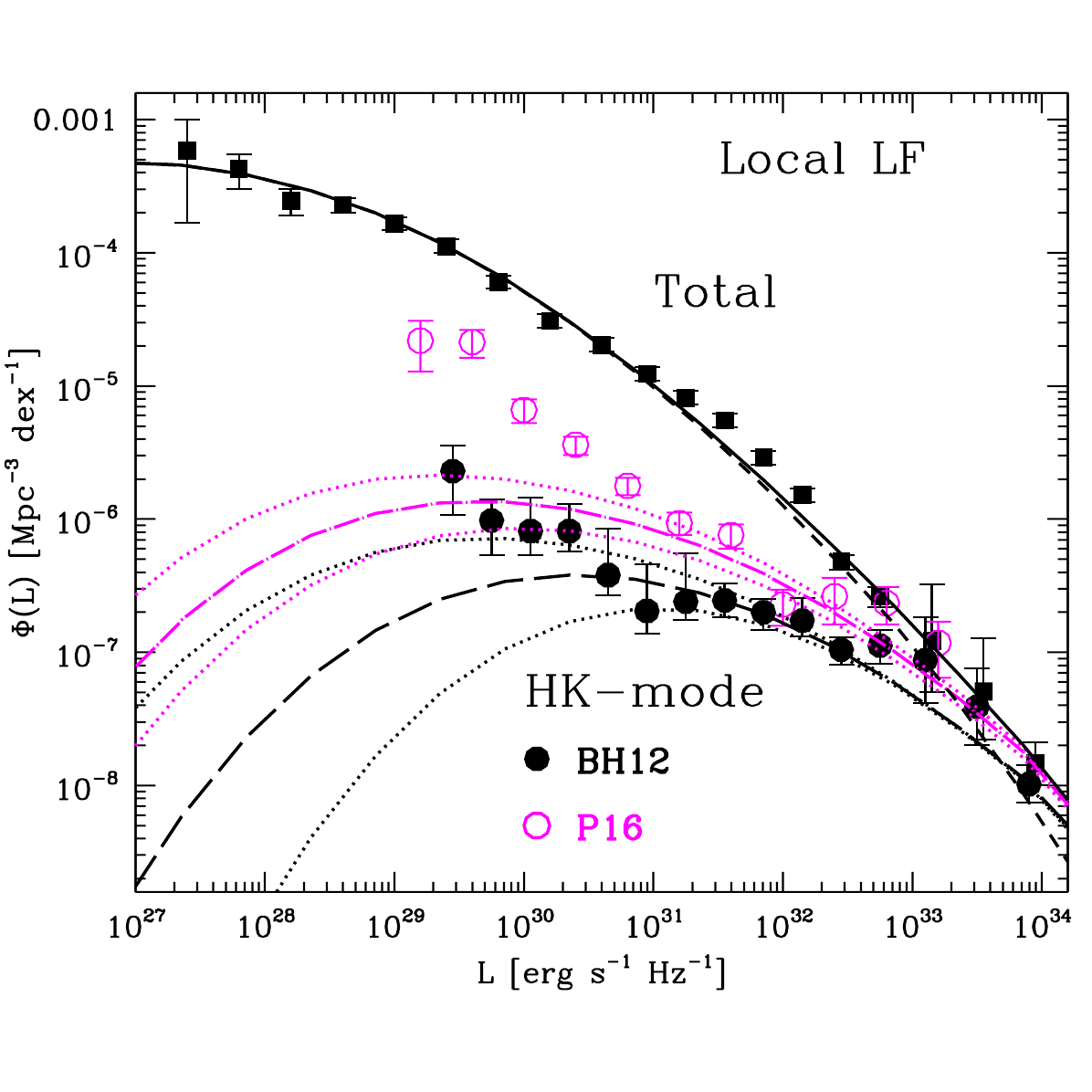}
  \includegraphics[width=9cm]{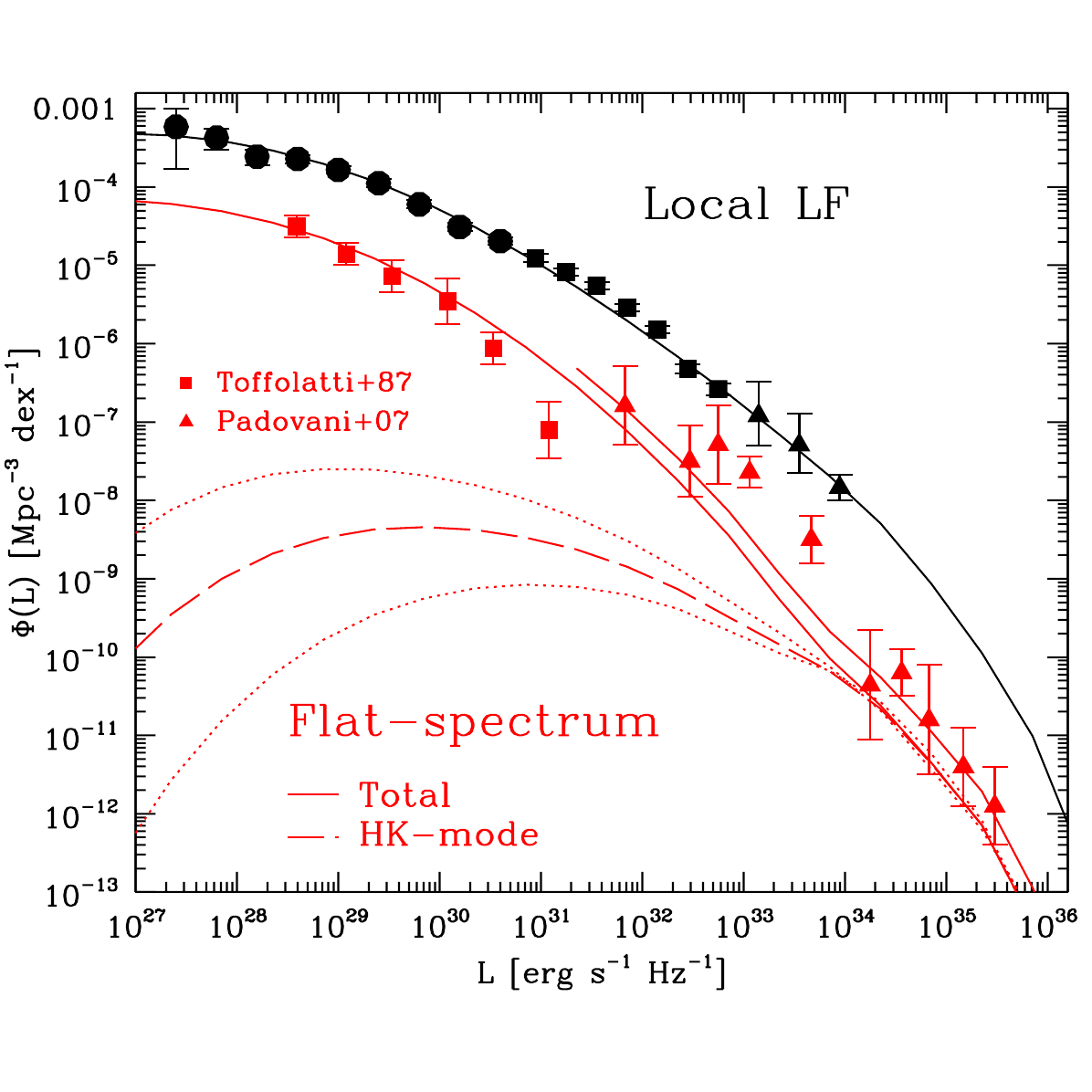}
  \caption{Local luminosity function of radio-loud AGN at 1.4GHz from
    observational data (black solid points, as in Fig.\,\ref{s4f1})
    and from the best-fit model (black solid lines). ({\it Left
      plot}) HK mode LF from \citet[][BH12; solid black
    points]{bes12} and from \citet[][P16; open magenta
    circles]{pra16} are compared with model predictions for HK mode
    AGN (long dashed curves for the best-fit model and dotted lines
    for the upper and lower limits; black lines for the BH12-based model
    and magenta lines for the P16-based model). The predicted LF for
    LK mode AGN is also included (dashed black curve). ({\it Right
      plot}) Local LF of flat-spectrum sources from observations (red
    squares for \citealt{tof87}; red triangles for \citealt{pad07}) is
    compared with model predictions (red lines: solid for the total
    AGN population at $z=0.1$ and 0.3; long dashed and dotted lines
    for the best-fit model and the upper and lower cases of HK mode
    AGN at $z=0.1$).}
  \label{s4f4}
\end{figure*}

As shown in Fig.\,\ref{s4f1}, the model is able to provide a very
good fit to the local luminosity function, even for flat-spectrum
sources (see also Fig.\ref{s4f4}). We also compare our predictions
with observational estimates of the LK mode LF at $0.5<z<1$ from
\citet{bes14}, finding again a good match.


\begin{figure}
  \centering
  \includegraphics[width=9cm]{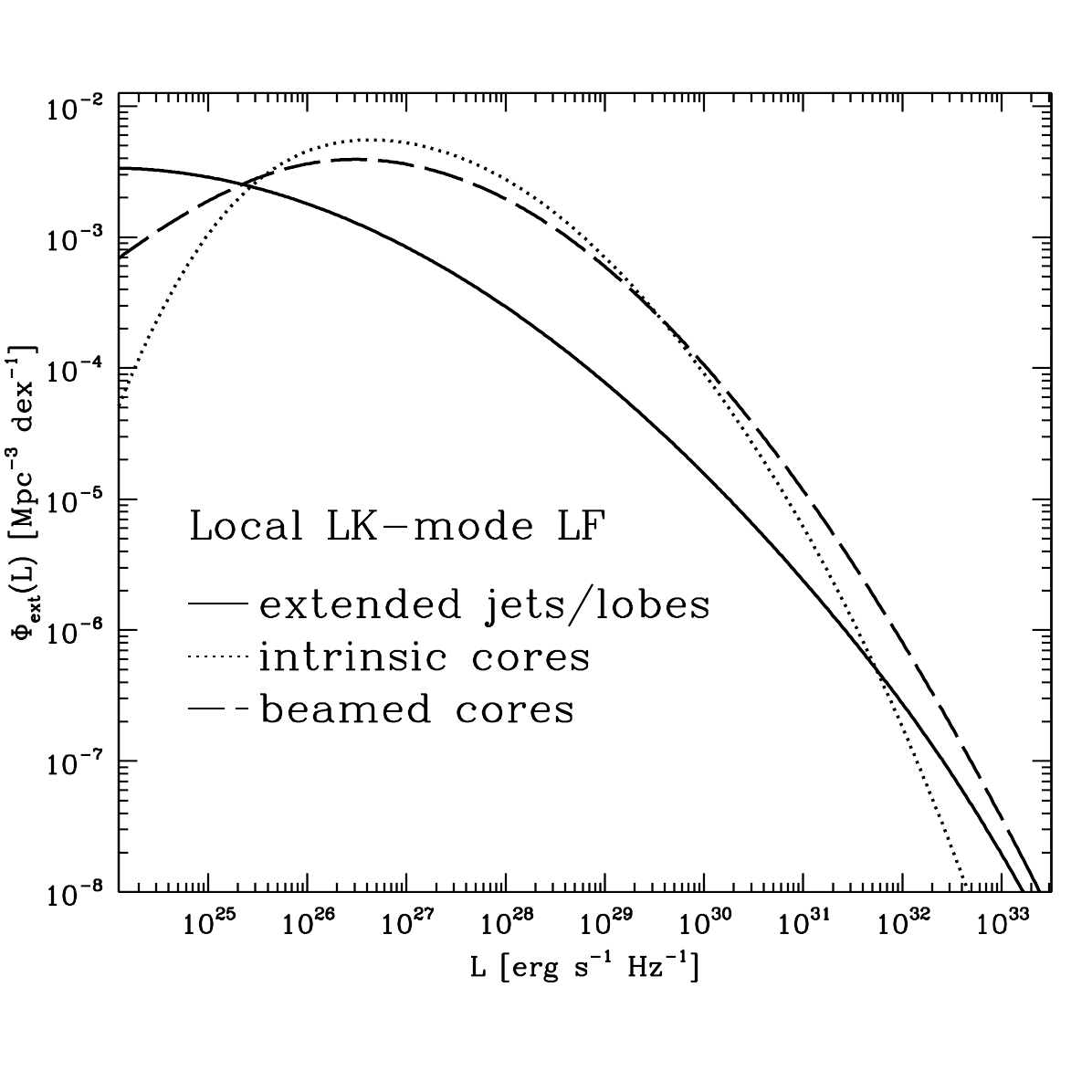}
  \caption{Predicted local LF for the extended jet--lobe AGN emission
    at 5\,GHz (solid line), compared with the corresponding intrinsic
    (dotted line) and beamed (dashed line) core LFs.}
  \label{s4f2b}
\end{figure}

In order to check the hypothesis of $f_{loud}^{LK}=1$, we also
included the fraction of radio-loud LK mode AGN among the free
parameters. We do not find any significant improvement in the fit and
a large fraction of radio-loud AGN ($f_{loud}^{LK}\ga0.3$) is  
required from the data.

Using Eq.\,\ref{s2e12} we can compute the LF associated with the
extended jet--lobe AGN emission, starting from the intrinsic core LF
(Fig.\,\ref{s4f2b}). With respect to the core LF, the peak of the
extended jet--lobe LF is shifted to fainter luminosities by around
two orders of magnitude. Nevertheless, at high luminosities, the
space density for this LF 
is larger than that associated with the intrinsic core emission
and close to the beamed value, as an effect of the large
dispersion in the intrinsic core-extended jet--lobe luminosity
relation.

\subsection{Model parameters for AGN in the HK mode}
\label{s4s2}

The number of free parameters for AGN in the HK mode increases
substantially. First of all, the FP relation 
is not as well established as for LK mode AGN, although there are
indications that a correlation is still present for this population of
sources. We assume that this relation (Eq.\,\ref{s2e1}) is still valid
for HK mode AGN but with unknown coefficients. We impose the
continuity condition at $\lambda_{cr}=0.01$ and, as shown in
Eq.\,\ref{s2e2a}, the only free parameter is $\xi_X$. The dispersion
in the FP relation is fixed to 0.6, as for LK mode AGN. Other free
parameters are the coefficients of the relation between extended and
core luminosity (Eq.\,\ref{s2e11}), $\xi_W$ and $\beta_W$, plus the
scatter in the relation, $\sigma_W$. Finally, differently from
LK mode AGN, the fraction $f_{loud}^{HK}$ of high-accreting SMBHs
that are radio-loud (i.e. in the HK mode) is quite uncertain and
probably redshift dependent. We parametrize this fraction as a
function of redshift as 
\beq
f_{loud}^{HK}(z)=f_0(1+z)^{\alpha_z}\,
\label{s4e1}
,\eeq 
where $f_0$ is the fraction of HK mode AGN at redshift
zero. Both $f_0$ and $\alpha_z$ are free parameters of the
model. We impose the condition that $f_{loud}^{HK}(z)\le0.3$ at
$z=[0,4]$, in agreement with estimates from the literature
\citep{raf09,kra15,kel16}.

To summarize, we end up with six free parameters ($\xi_X$, $\xi_W$,
$\beta_W$, $\sigma_W$, $f_0$, and $\alpha_z$). They can be determined
thanks to the wide and composite amount of data to be fitted. Given
the LFs and number counts of LK mode AGN, we can compare model
predictions with the local LF for the total population of radio-loud
AGN and for HK mode AGN\footnote{Unless expressly stated otherwise, in
  the following subsections we are considering the BH12-based data set
  for the local LFs.}; the LF for flat-spectrum sources at
$z\simeq0.3$ and $L\ga10^{32}$\,erg\,s$^{-1}$\,Hz$^{-1}$; the LF for
radio-loud AGN at redshift $z\simeq0.55$; differential number counts
at 1.4 and 5\,GHz (at $S\ge1$\,mJy); and differential number counts of
flat-spectrum sources at 5\,GHz.

Roughly speaking, redshift-independent parameters from FP and
core-extended luminosity relations determine the shape of LFs at
different redshifts, the shape and the ratio of number counts
at 1--5\,GHz, while their normalizations are fixed mainly by the
fraction of HK mode AGN.

As before, we use a MCMC method to compute the best-fit parameters of
the model and to estimate model uncertainties. Some caveats should  be
taken into account. Firstly, as already seen for LK mode AGN,
the parameters show a large degeneracy and correlation with each other. Consequently, the likelihood of the model parameters presents
not a single minimum but multiple local minima with comparable values
of $\chi^2$. Moreover, predicted LFs and number counts are very
sensitive to small variations in the model parameters. For example,
results are sensitive to variations of $\xi_W$ and $\beta_W$ lower
than 1\%, and to variations of $\sigma_W$ and $\xi_X$ of the order of
few per cent. The MCMC inizialization cannot be chosen randomly, but
the input parameter set has to be firstly tuned, especially for strongly
correlated parameters, otherwise the MCMC algorithm will not be able
to converge to a minimum.

\begin{table*}
  \centering 
  \label{t1}
\begin{threeparttable}
  \caption{Best-fit model parameters for AGN in LK and HK mode
    and their range of values}
  \begin{tabular}{ccccccccccc}
  \hline
  & $\xi_X$ & $\xi_M$ & $\beta_R$ & $\sigma_{FP}$ & $\xi_W$ &
  $\beta_W$  &  $\sigma_W$ & $f_0\,[\times10^3]$ & $\alpha_z$ & $\chi^2$ \\
  \hline 
  BH12-based & \multicolumn{10}{c}{LK mode AGN}\\
 \hline 
  best model & (0.62)$^a$ & (0.55)$^a$ & (8.6)$^a$ & (0.6)$^a$ &
                               (1.42)$^b$  & --13.95 & 0.81 & -- & -- & 33 \\
  median & & & & & & $-12.8^{+1.5}_{-2.1}$ & $0.91\pm0.15$ & &
  & \\
\hline
  P16-based & \multicolumn{10}{c}{ }\\
 \hline 
  best model & (0.62)$^a$ & (0.55)$^a$ & (8.6)$^a$ & (0.6)$^a$ & (1.50)$^b$ & --16.40 & 0.84 & -- &
  -- & 61 \\
  median & & & & & & $-16.4^{+1.6}_{-2.1}$ & $0.83\pm0.17$ & &
  & \\
\hline 
\hline 
   BH12-based & \multicolumn{10}{c}{HK mode AGN} \\
  \hline
  best model & 1.70 & (--0.57)$^c$ & ($\approx$-30)$^c$ & (0.6)$^a$ & (0.78)$^d$ & 7.73
  & 0.79 & 0.61 & 2.08 & 212 \\ 
  median & $1.69\pm0.08$ & & & & & $7.4^{+2.2}_{-2.5}$ & $0.79\pm0.03$
  & $0.6^{+0.3}_{-0.1}$ & $2.1\pm0.1$ & \\
\hline
   P16-based & \multicolumn{10}{c}{ } \\
  \hline
  best model & 1.52 & (--0.46)$^c$ & ($\approx$--30)$^c$ & (0.6)$^a$ & (0.93)$^d$ & 2.76
  & 0.82 & 2.81 & 1.27 & 315 \\ 
  median & $1.51\pm0.04$ & & & & & $2.4\pm1.$ & $0.82\pm0.02$
  & $2.9\pm0.5$ & $1.27\pm0.1$ & \\
\hline
\end{tabular}
\begin{tablenotes}
  \item[a] Fixed parameters.
  \item[b] We use the relation $\xi_W=0.898-0.0373\,\beta_W$
    (BH12-based; see Fig.\,\ref{s4f3}) or
    $\xi_W=0.8875-0.0372\,\beta_W$ (P16-based).
  \item[c] We use the continuity condition at $\lambda=10^{-2}$ (see Eq.\,\ref{s2e2a}).
  \item[d] We use the relation $\xi_W=1.014-0.0305\,\beta_W$
    (BH12-based; see
    Fig.\,\ref{s4f1new}) or $\xi_W=1.010-0.0308\,\beta_W$ (P16-based). 
\end{tablenotes}
\end{threeparttable}
\end{table*}

\begin{figure}
  \centering
  \includegraphics[width=8cm]{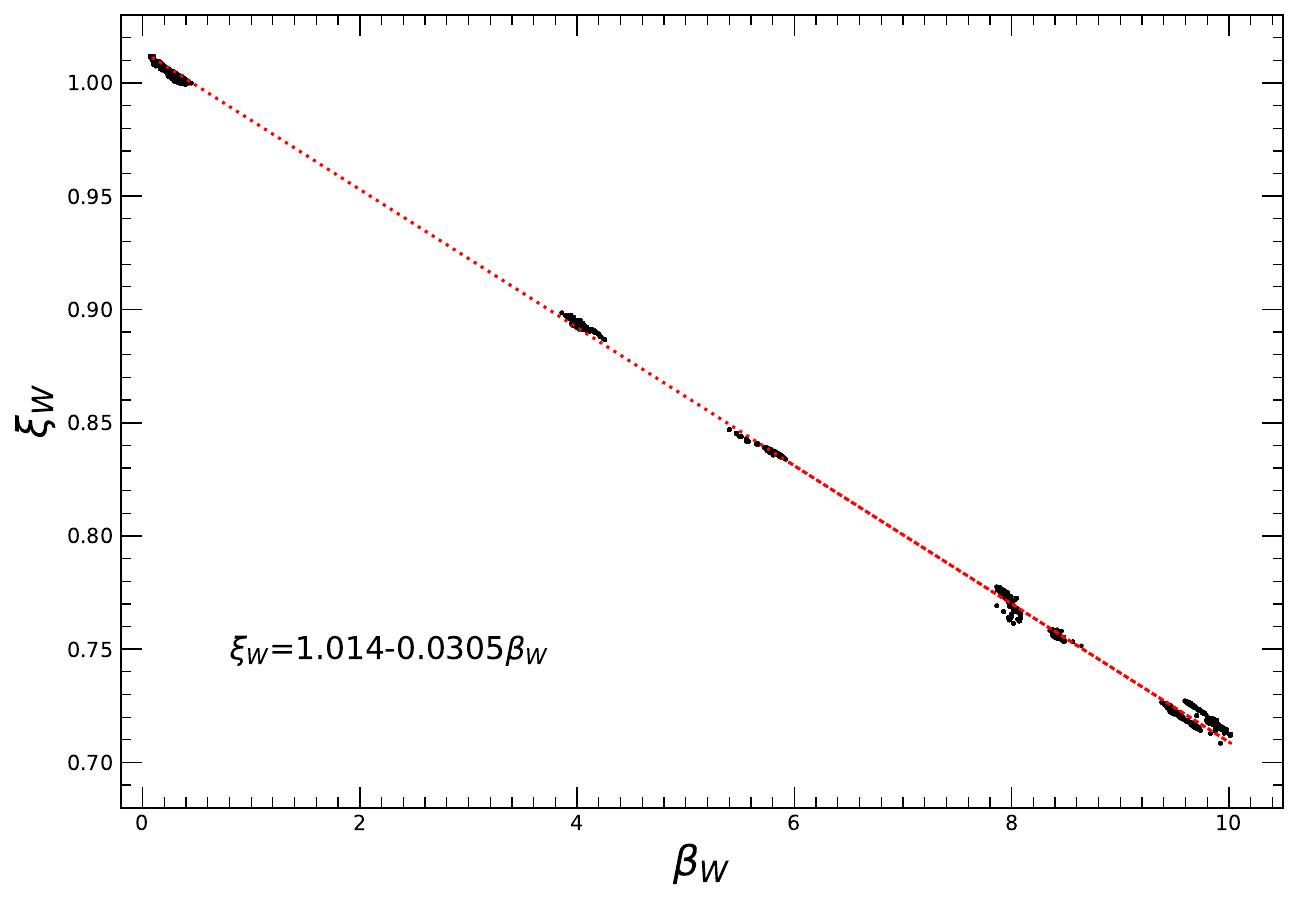}
  \caption{Linear correlation between the model parameters $\xi_W$ and
    $\beta_W$. The solid points are the parameter values found from the
    MCMC chains using different parameter inizializations. The dotted
    line corresponds to the linear fit reported in the panel.}
  \label{s4f1new}
\end{figure}

\begin{figure*}
  \centering
  \includegraphics[width=14.cm]{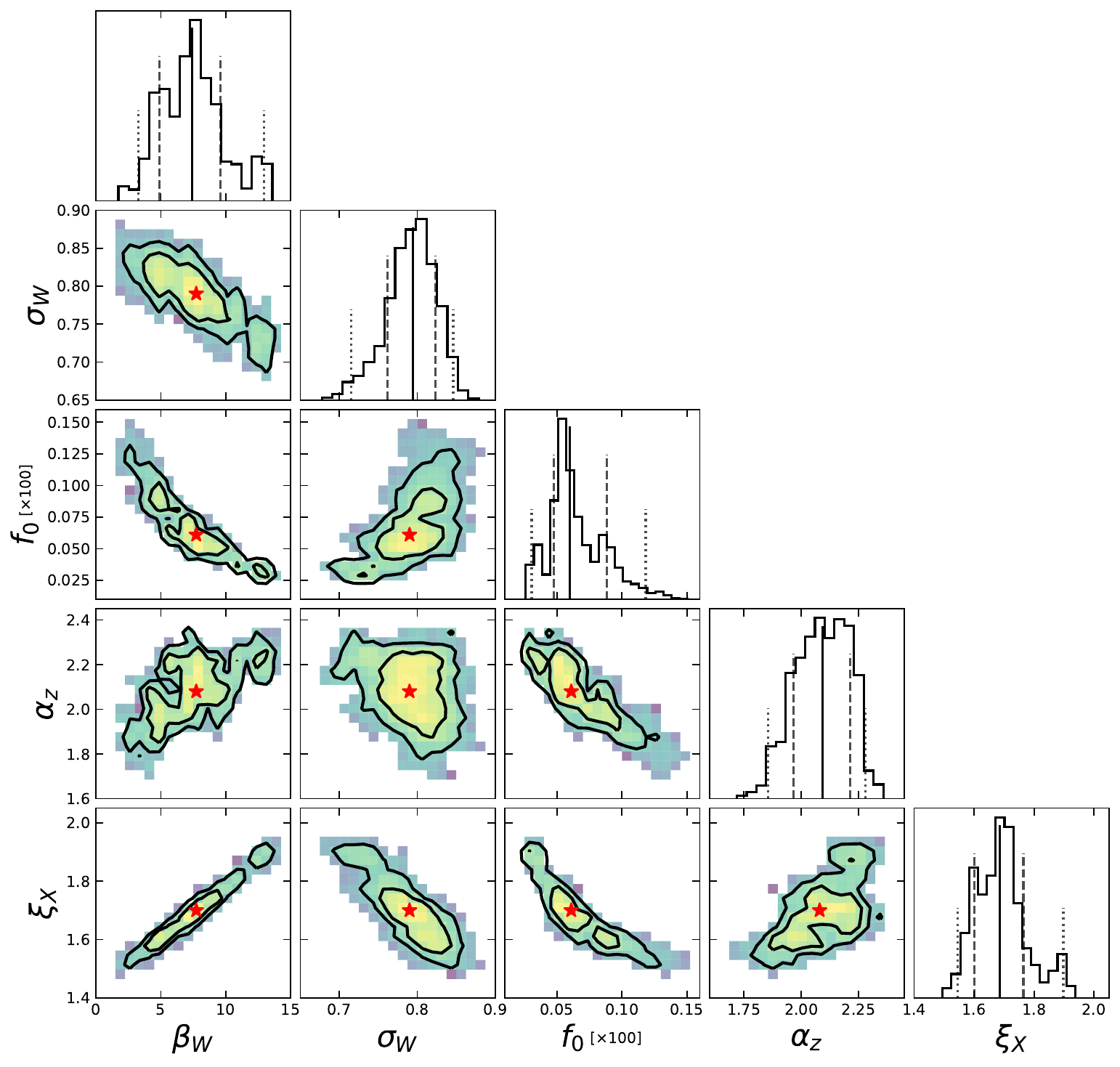}
  \caption{Marginalized PDFs in each 2D sub-space of the model
    parameters. Contour plots correspond to the 68.3\%\ and 95.4\%
    confidence levels. The model set with the lowest
    $\chi^2$ is indicated by the red stars. The 1D marginalized
    distributions are also shown, with the median and the 68.3\%\ and
    95.4\%\ confidence intervals (solid, dashed, and dotted
    lines, respectively).}
  \label{s4f2new}
\end{figure*}

\begin{figure}
  \centering
  \includegraphics[width=8cm]{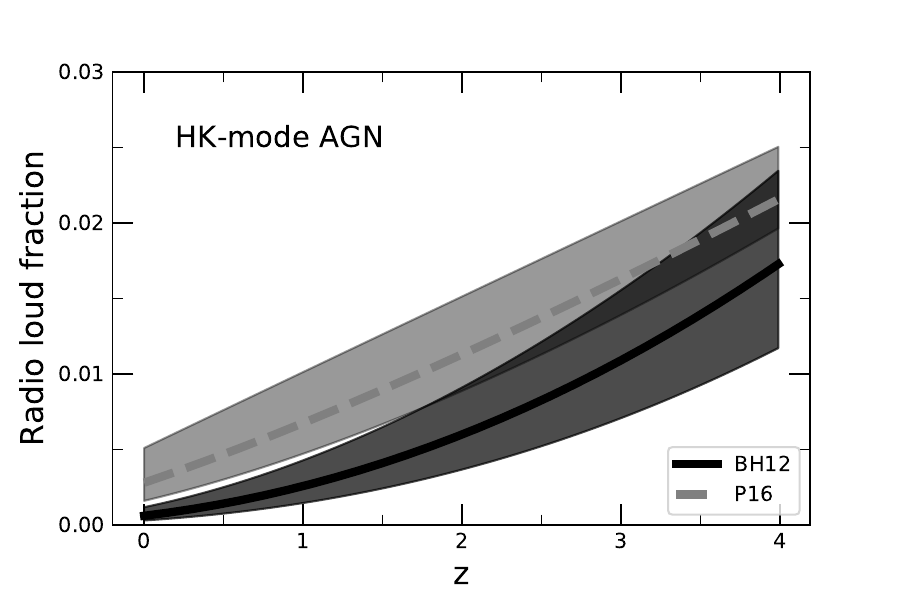}
  \caption{Fraction of high-accreting AGN in HK mode predicted by
    the model as a function of the redshift. The solid (dashed) line
    corresponds to the BH12-based (P16-based) best-fit model and
    the shaded area to models within the 2$\sigma$ confidence level.}
  \label{s4f3new}
\end{figure}

As a starting point, we tried to reduce the number of free parameters
looking for very tight linear correlations between parameters, as
was done for LK mode AGN. We proceeded in the following way. We ran the
MCMC algorithm with different input parameter sets, covering a large
range of possible values for the input parameters.
We find that MCMC chains converge to local minima that depend on the
inizialization of parameters. A strong linear relation is found
between $\xi_w$ and $\beta_W$ (see Fig.\,\ref{s4f1new}), similarly to
LK mode AGN. The correlation is valid over a very large range of
values for $\beta_W$, from 0 to 10. We decided therefore to fix the
parameter $\xi_w$ using the linear fit displayed in
Fig.\,\ref{s4f1new} and Table\,3.

The MCMC method was then run again with the remaining parameters.
The results are now less sensitive to the initialization and the
algorithm is able to converge to the global minimum starting from any
reasonable parameter set. In Fig.\,\ref{s4f2new} we show the
marginalized PDFs of the parameters, while in Table\,3 we report the
best-fit model, the median, and the uncertainties of the parameters,
and in Table\,4 the $\chi^2$ of the fit to the different data sets.
We still find a significant correlation among the parameters, and in
particular between $\xi_X$, $\beta_W$, and $f_0$.
The power-law index $\xi_W$ is found between
0.6 and 0.95,  very close to the theoretical expectation of 0.8
(see discussion in Sect.\,\ref{s2s3}), while the parameter $\beta_W$
covers a  large range of values, between 2 and 14. This
means that the normalization coefficient in the core and extended
luminosity relation can vary by more than ten orders of magnitude. It is
interesting to compare this result with the range of values for
LK mode AGN:  [$-8,\,-16$]. AGN in LK and HK mode 
therefore have  quite a   different relationship between core and extended
luminosity: using the best-fit models, for an intrinsic core
luminosity of $\log\mL_c=27\,(30)\,[33]$ (in erg\,s$^{-1}$\,Hz$^{-1}$)
we get 
$\log L_{ex}=28.8\,(31.1)\,[33.5]$ for HK mode AGN and
$24.4\,(28.6)\,[32.9]$ for LK mode AGN. We conclude therefore that
the model typically predicts a much stronger contribution of extended
jets and lobes for the former AGN class.

The FP parameter $\xi_X$ is found between 1.5 and 1.9, about three times
larger than the value used for LK mode AGN, and (unlike LK mode AGN)
the intrinsic core luminosity is anti-correlated with the SMBH
mass. These results are in good agreement with observational
constraints of the FP relation for high-accretion AGN \citep{don14}.

The fraction of HK mode AGN is between 0.5 and 1 per mil at
$z\simeq0$, and it steeply increases with redshift, with $\alpha_z$
ranging from 1.8 to 2.4 (increasing $f_0$, $\alpha_z$
decreases). Models predict therefore a very low fraction of
high-accreting AGN in the HK mode, especially in the local Universe,
with $f_{loud}^{HK}\ll1$ per cent at $z=0$--1, between 0.4--1 per
cent at $z\simeq2$ and 1--2.4 per cent at $z\simeq4$ (see
Fig.\,\ref{s4f3new}).

\begin{table}
  \centering 
  \caption{$\chi^2$ of model fits to observational data}
  \begin{tabular}{l|cc|cc}
  \hline
   Data set & \multicolumn{2}{c}{BH12-based} & \multicolumn{2}{c}{P16-based} \\
  \hline 
  & $n_{data}$ & $\chi^2$ & $n_{data}$ & $\chi^2$ \\
  \hline 
  \hline 
    \multicolumn{5}{c}{Luminosity Function} \\
  \hline 
    LK mode & & & & \\
    local & 19 & 25 & 17 & 55 \\
    flat-spectrum (z=0.1) & 5 & 8 & 5 & 6 \\
  \hline 
    HK mode  & & & & \\
    local & 15 & 16 & 13 & 76 \\
  \hline 
    Total population  & & & & \\
    local & 19 & 37 & 17 & 42 \\
    $z\simeq0.55$ & 10 & 4 & 10 & 16 \\
    flat-spectrum (z=0.3) & 10 & 14 & 10 & 12 \\
  \hline 
  \hline 
    \multicolumn{5}{c}{Number Counts} \\
  \hline 
    1.4\,GHz & 62 & 102 & 62 & 114 \\
    5\,GHz & 24 & 31 & 24 & 43 \\
    5\,GHz flat-spectrum & 9 & 9 & 9 & 12 \\
  \hline 
  \hline 
    All data & 173 & 244 & 167 & 376 \\
  \hline 
\end{tabular}
\label{tab1}
\end{table}

%

\begin{figure}
  \centering
  \includegraphics[width=9cm]{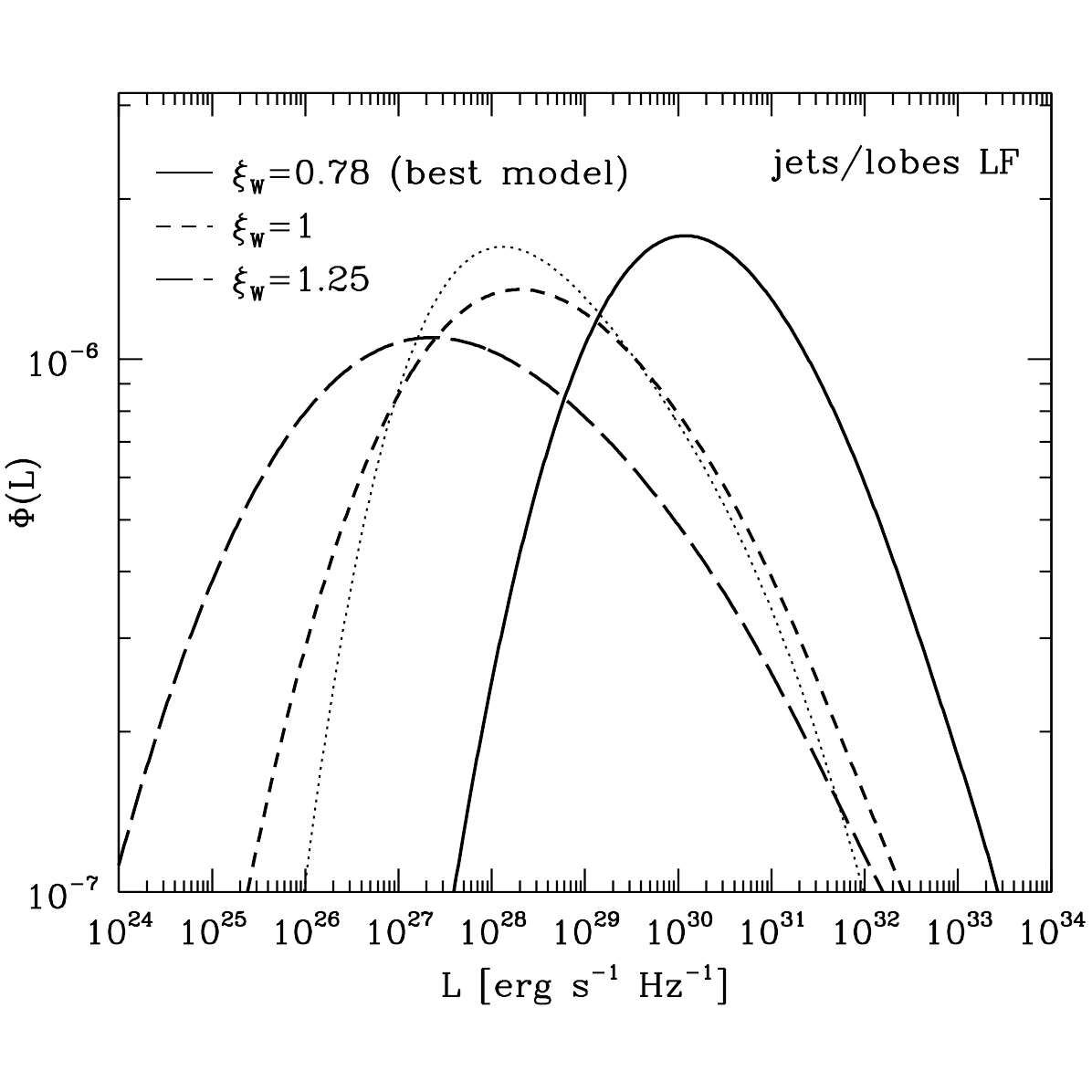}
  \caption{Extended jet--lobe LF at 5\,GHz and $z=1$ as predicted by
    the model for different values of $\xi_W$ (solid and dashed lines)
    compared with the intrinsic core LF (dotted line).}
  \label{s4f7}
\end{figure}

\begin{figure*}
  \centering
  \includegraphics[width=9cm]{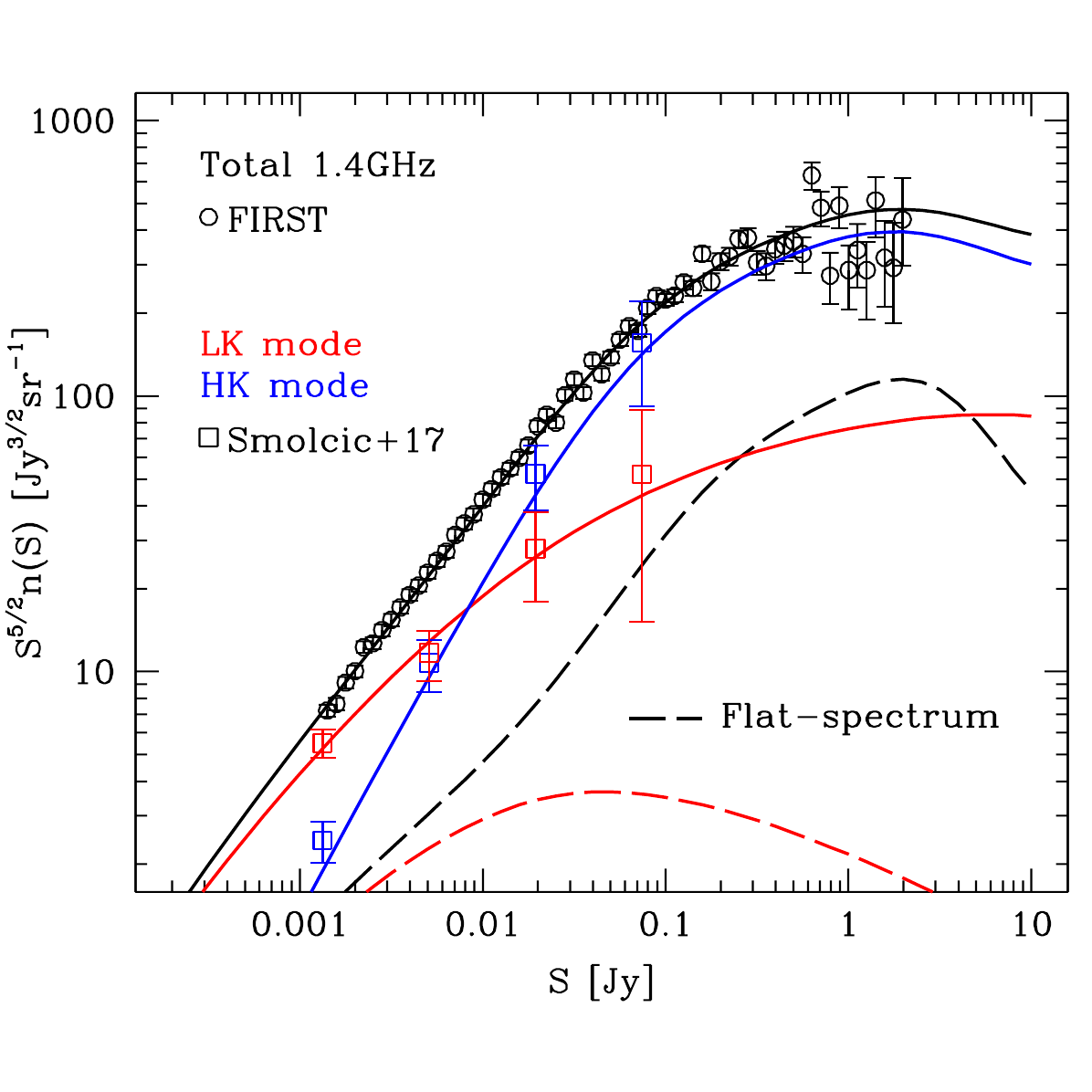}
  \includegraphics[width=9cm]{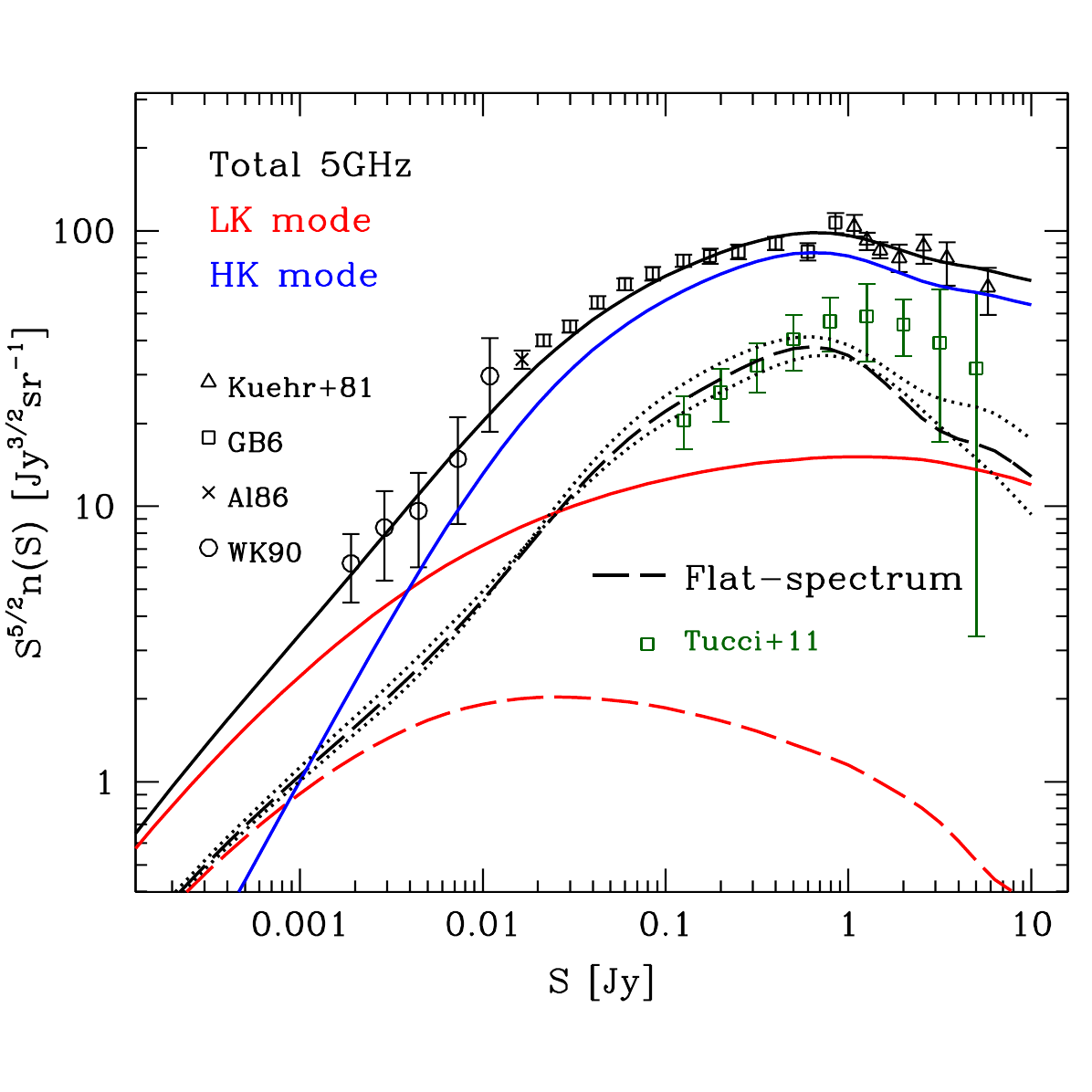}
  \caption{Model predictions of number counts at 1.4\,GHz ({\it left
      panel}) and at 5\,GHz ({\it right panel}) for the total AGN
    population (black lines), for LK mode (red curves) and HK mode
    (blue curves) AGN. Number counts of flat-spectrum sources are
    shown as dashed lines (same colours as for the total population; dotted
    lines in the right panel are for the model upper and lower limits). At
    1.4\,GHz black points are from the FIRST survey \citep{whi97},
    while red and blue open squares are number counts of LK and
    HK mode AGN, respectively, estimated by \citet{smo17b}. At 5\,GHz
    the black points are from a compilation of data collected by
    \citet{dez10}, including the GB6 Catalogue \citep{gre96}, the
    catalogue of Jy sources from \citet{kue81} and data from
    \citet[][WK90]{wro90} and \citet[][Al86]{alt86}. Blue points are estimates for
    flat-spectrum sources from \citet{tuc11}.}
  \label{s4f6}
\end{figure*}

\begin{figure*}
  \centering
  \includegraphics[width=9cm]{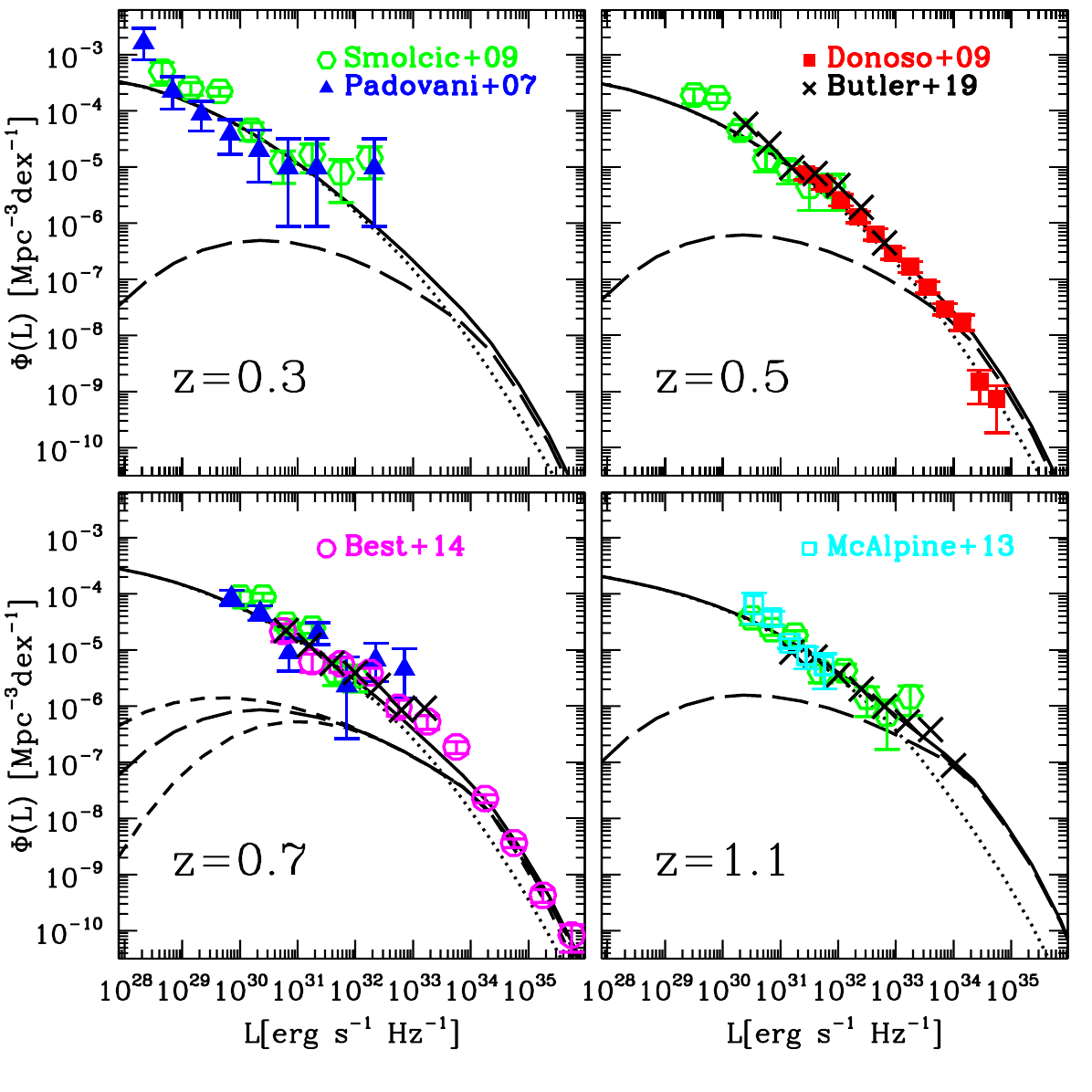}
  \includegraphics[width=9cm]{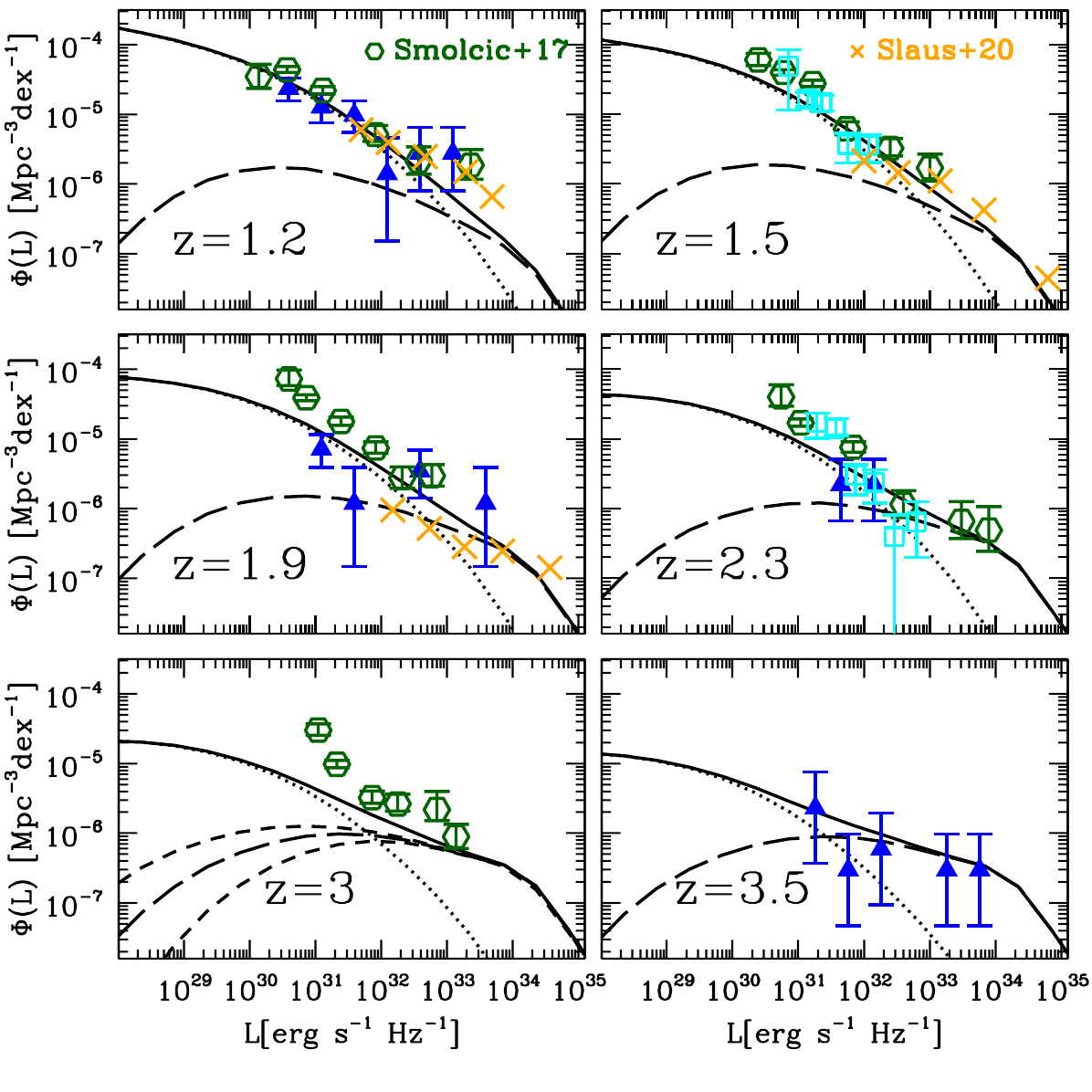}
  \caption{Predictions of the LF for the total (solid lines), LK mode
    (dotted lines), and HK mode (long and short dashed lines for the best
    fit and for the lower and upper limits) AGN population at different
    redshifts. Observational data are from \citet[][blue solid
    triangles]{pad07}, \citet[][green open hexagons]{smo09},
    \citet[][red solid squares]{don09}, \citet[][cyan open
    squares]{mca13}, \citet[][magenta open circles]{bes14},
    \citet[][dark green open hexagon]{smo17}, \citet[][black
    crosses]{but19} and \citet[][orange crosses]{sla20}.}
  \label{s4f5}
\end{figure*}

\begin{figure}
  \centering
  \includegraphics[width=9cm]{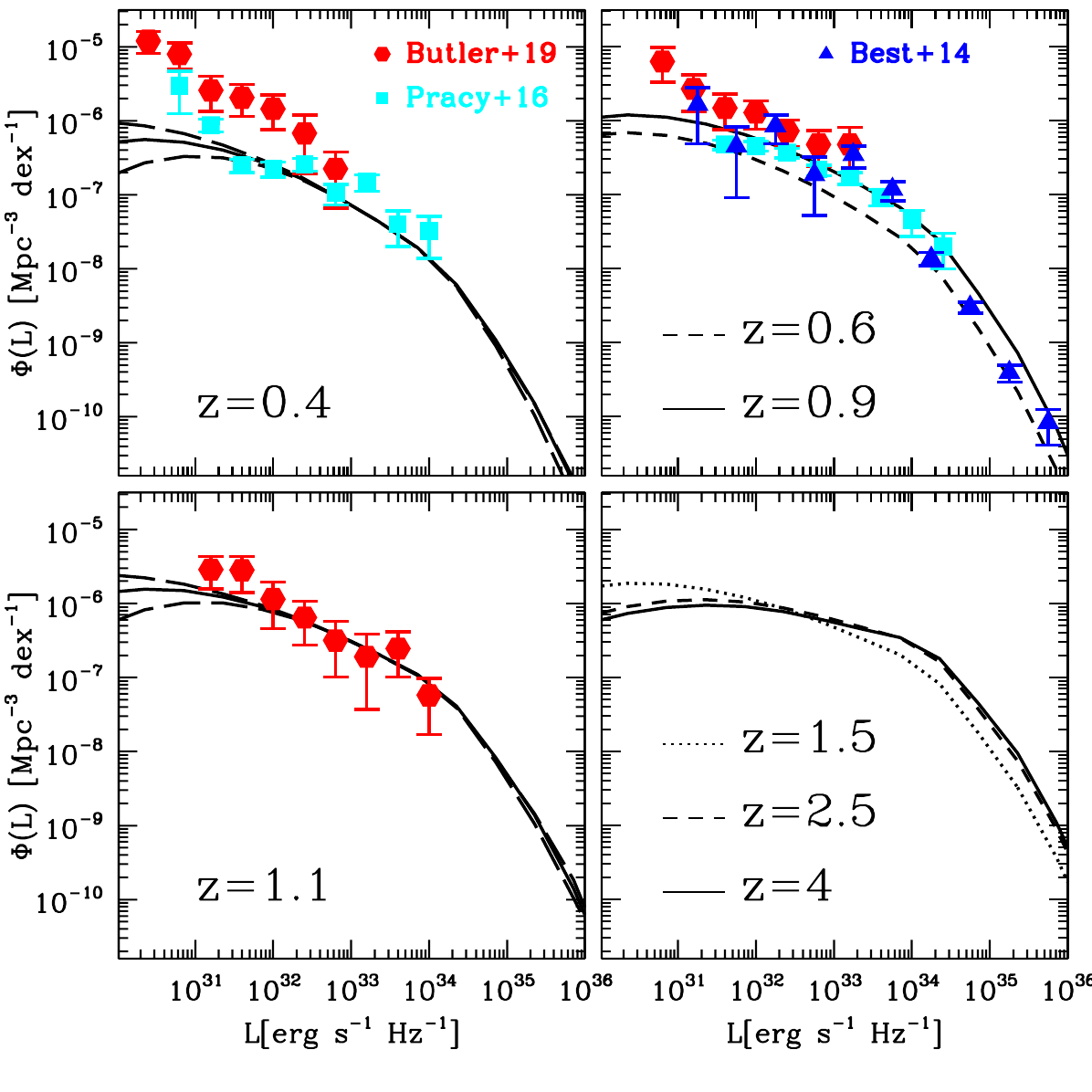}
  \caption{Model predictions for the HK mode LF at the redshifts
    indicated in the panels. In the panels at redshifts 0.4 and 1.1,
    solid lines are for the best-fit model and dashed lines are for the
    lower and upper limits. In the other panels only the best model is
    reported. Observational data are shown as  red hexagons
    from \citet{but19} (in $0.3<z<0.6;~0.6<z<0.9;~0.9<z<1.3$ bins);
    cyan squares from \citet{pra16} (in $0.3<z<0.5;~0.5<z<0.75$ bins); and 
    blue triangles from \citet{bes14} (in $0.5<z<1$ bin).}
  \label{s4f8}
\end{figure}

\begin{figure*}
  \centering
  \includegraphics[width=6cm]{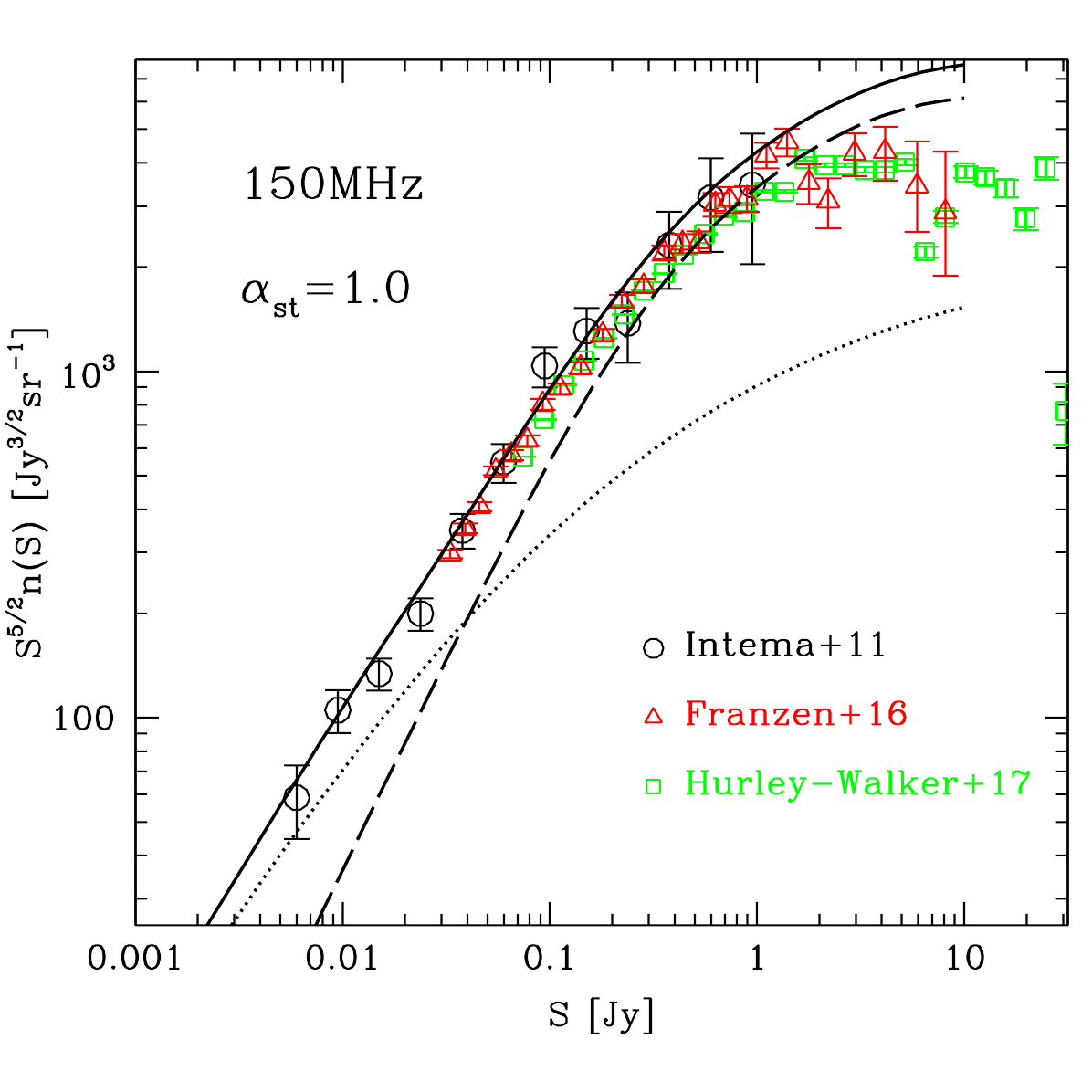}
  \includegraphics[width=6cm]{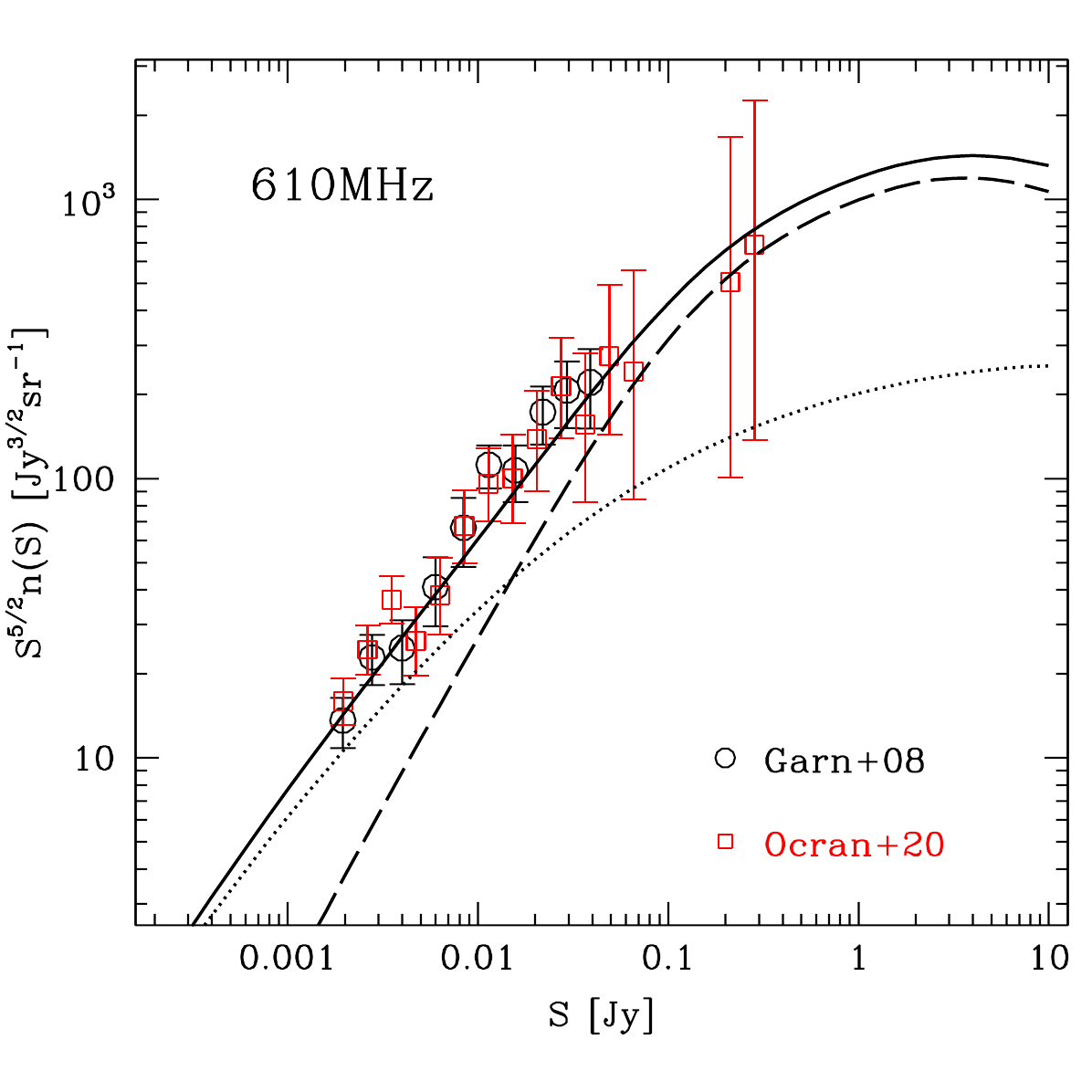}
  \includegraphics[width=6cm]{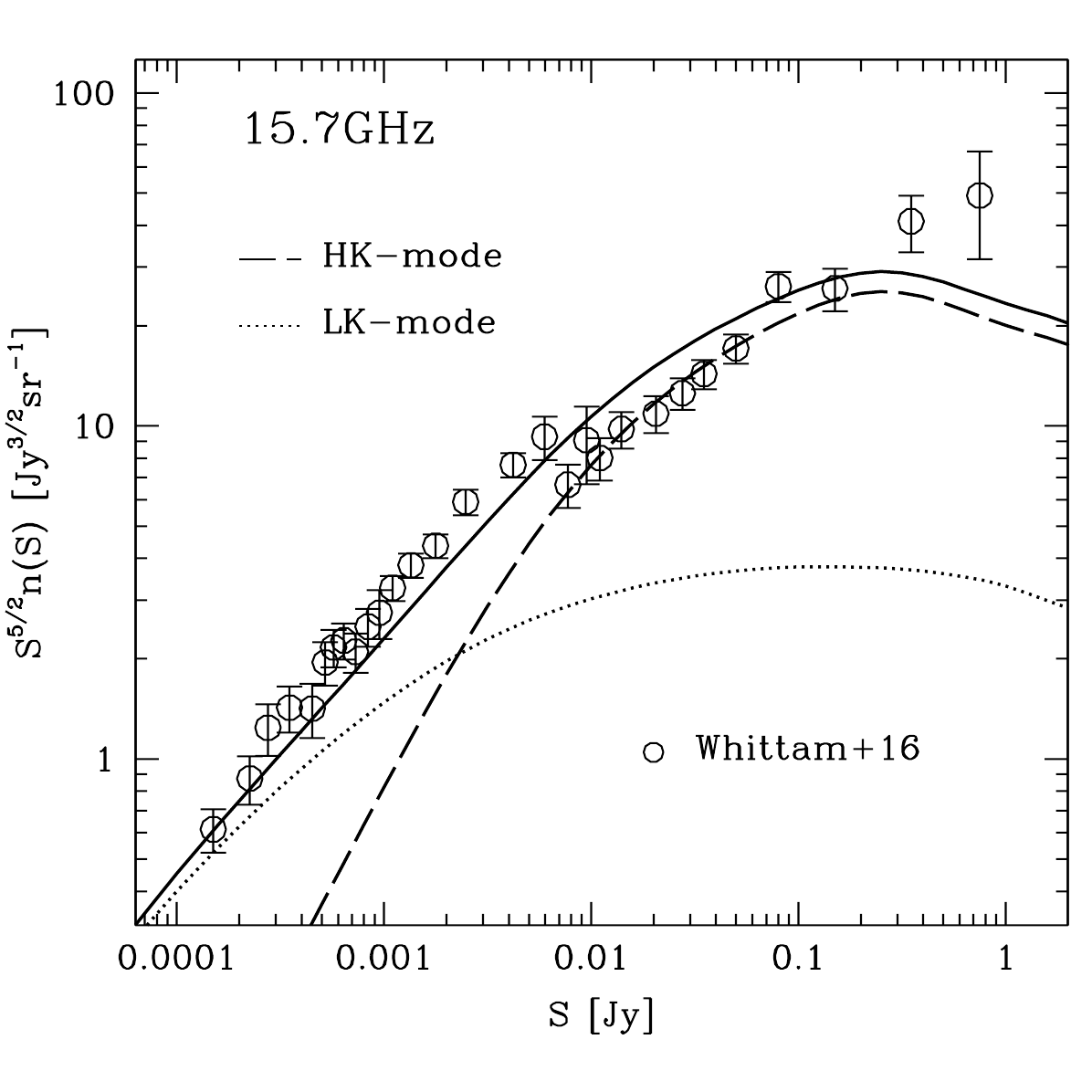}
  \caption{Predictions of number counts at 150 and  610\,MHz and at
    15.7\,GHz for total (solid lines), LK mode (dotted lines), and
    HK mode (dashed lines) AGN. In the predictions at 150\,MHz an
    average spectral index of 1 is used for steep-spectrum sources
    below 1.4\,GHz. The data are for  ({\it 150\,MHz}) \citet[][open black
    circles]{int11}, \citet[][green open squares]{hur17} and
    \citet[][red open triangles]{fra16}; ({\it 610\,MHz})
    \citet[][black open circles]{gar08} and \citet[][red open
    squares]{ocr19}; and  ({\it 15.7\,GHz}) \citet{whi16}.}
  \label{s4f9}
\end{figure*}


\subsection{Model fits to observational data}

Figures\,\ref{s4f4} and \ref{s4f6} provide a visual check of the model
fits to observational data, and in particular to the fits used in the
parameters determination (corresponding to the BH12-based data set
for the local LFs). In these figures, along with the best-fit model,
we also consider {\it lower} and {\it upper limit} cases
that should provide a rough estimate of uncertainties in model
predictions. As the  upper limit we take a model with $f_0\simeq0.14$\%,
$\xi_X\simeq1.5$, and $\beta_W\simeq3$, while as the lower limit a model
with $f_0\simeq0.03$\%, $\xi_X\simeq1.9$, and $\beta_W\simeq16$.

In Fig.\,\ref{s4f4} we consider the local LF at 1.4\,GHz for the total
population of radio-loud AGN and for HK mode AGN. The former is
quite well reproduced by the model in the whole luminosity range in
which data are available.

It is more interesting to focus on HK mode AGN. The best-fit
model provides a good fit to the BH12 HK mode space densities at
$L\ga10^{30}$\,erg\,s$^{-1}$Hz$^{-1}$, whereas it tends to
  underestimate BH12 estimates at lower luminosities. A better
agreement is found for models that predict a relatively large local
fraction of HK mode AGN ($f_0\approx0.1$\%). Model
predictions of the HK mode local LF always show a maximum at
$L>10^{29}$\,erg\,s$^{-1}$Hz$^{-1}$ and a decrease at lower
luminosities. On the contrary, BH12 measurements seem to keep
increasing while moving to low luminosities (we recall that the first points in
the BH12 HK mode LF should be considered as lower limits).

The local LF for flat-spectrum sources is also consistent with
observational data \citep{tof87,pad07} over the seven orders of magnitudes
covered by them. It is interesting to note how model predictions for
the local LF of flat-spectrum HK mode sources can vary at
low to intermediate luminosities according to the parameter set used. At
$\approx10^{29}$\,erg\,s$^{-1}$Hz$^{-1}$, the upper and lower cases differ
by about two orders of magnitude. This is due to the dependence of the
extended jet--lobe LF on the $\xi_W$ (or $\beta_W$) parameter. As
shown in Fig.\,\ref{s4f7}, changing the value of  $\xi_W$  from 0.78 to 1.25,
the peak of the LF moves from $10^{30}$ to
$10^{27}$\,erg\,s$^{-1}$Hz$^{-1}$. With low values of $\xi_W$, therefore,
extended emission from jets and lobes is typically fainter and the number
of flat-spectrum sources increases.

Finally, models provide a very good fit to number counts of the AGN
population  at 1.4\,GHz and at 5\,GHz, and of the flat-spectrum sources
at 5\,GHz, as shown in Fig.\,\ref{s4f6}. Number counts are dominated
by AGN in HK mode starting from flux densities $\ga10$\,mJy. There are
no relevant differences among model predictions using different
parameter sets, mainly due to the strong constraints imposed by number
counts at 1.4\,GHz at $S\la0.1$\,Jy. Some scatter is  observed, but only at the
jansky level for flat-spectrum sources.

\citet{smo17b} provided separate estimates of number counts for
low- to moderate-luminosity AGN (analogous to  LK mode) and moderate- to high-luminosity AGN
(mostly analogous to  HK mode). Very
remarkably, although these data are {not considered} for the fit,
the agreement between these number counts and the  predictions of our
current models is excellent (as displayed in the left panel of
Fig.\,\ref{s4f6}).

\subsection{Model parameters using the P16-based data set}
\label{s4s3}

Model parameters are also determined using the P16-based data set
(see Table\,3). With respect to the BH12-based one, the main change
is the much higher fraction of HK mode AGN found at low to intermediate
luminosities in the local LF. The P16 and BH12 measurements of the
LK mode and total local LF are instead consistent with each other, at
least for $L\ga10^{30}$\,erg\,s$^{-1}$Hz$^{-1}$ (see Fig\,\ref{s3f1}).

For LK mode AGN, the model predicts a quite similar local LF at
intermediate to high luminosities, indepedent of the data set, with
  some differences only at low luminosities
  ($L<10^{29}$\,erg\,s$^{-1}$Hz$^{-1}$).
In terms of parameters, $\beta_W$ from the P16 data set is about
20\,per\,cent smaller with respect to the BH12-based case.

The impact of the different HK mode local LF is quite significant on
the model. As expected, the most affected parameter is the local
fraction of HK mode AGN, whose best-fit value is now about five times
the BH12-based one (i.e. $f_0\sim3$\,per\,mil). The evolution of the
HK mode fraction is softer than before, as shown in
Fig.\,\ref{s4f3new}, and compatible with previous results at high
redshifts. The best-fit values for the parameters $\beta_w$ and
$\xi_X$ are   smaller than the BH12-based results, especially
$\beta_W$, but still compatible with them at the  2$\sigma$ level. The
predicted number counts are instead very similar to the BH12-based
ones due to the strong constraints imposed by observational data. The
  larger $\chi^2$ found for the P16-based best-fit model is
 mainly due to the poor fit to the HK mode local LF (see
Table\,4).

Remarkably, at low luminosities the predicted local LF for HK mode
AGN is still consistent with the BH12 data, and significantly
underestimates the P16 measurements (see Fig.\ref{s4f4}). This can be
understood by the following remarks. The BH12-based model predicts a
number density of HK mode AGN of $10^{-6}$\,Mpc$^{-3}$ at $z=0$,
while, based on the P16 LF, it should be roughly of the order of
$10^{-4}$\,Mpc$^{-3}$ (taking into account only sources with
$L>10^{29}$\,erg\,s$^{-1}$Hz$^{-1}$). The model can increase the local
number density by increasing the local fraction of HK mode AGN
($f_0$). Because the local number density of SMBHs with $\lambda>0.01$
is $10^{-3}$ from the \citet{tv17} mass function, $f_0$ should be of
the order of 0.1, which is  more than 30 times larger than the best-fit
value.  Such a high level of HK mode AGN is probably difficult to
conciliate with the observed number counts, which set quite stringent
constraints on the space density of the HK mode AGN population at
$z\ga0.5$.

We can conclude that our model indeed predicts  a small fraction of
HK mode AGN at low to intermediate luminosities ($\approx0.1$--0.3\%
locally), independently of the observational data considered for the
local HK mode LF. We verify this by (arbitrarily) reducing the
uncertainties in the P16 HK mode estimates in order to give a
high relevance to these data in the fitting procedure. As a result,
not very differently from before, we find a good fit to the P16
  HK mode space densities only at
  $L\ga10^{31}$\,erg\,s$^{-1}$Hz$^{-1}$, but not at lower
  luminosities, where the predicted HK mode LF is still well below
  the P16 estimates.

\subsection{Comparing model predictions with more observational data}
\label{s4s4}

The model allows us to compute the AGN LF at redshifts up to $z=4$,
and number counts at frequencies that are outside the 1--5\,GHz
range. In Figures\,\ref{s4f5}--\ref{s4f9} we compare predictions of
our best-fit (BH12-based) model with observational data that are not
used in the fitting process (with the only exception for the Donoso et
al. data). This is an important check of the reliability of model
forecasts outside the redshift and frequency ranges where the model is
defined.

In Figure\,\ref{s4f5} we can compare the time evolution of the modeled
AGN LF with observations, from redshift $z\simeq0$ up to $z\sim4$. The
model shows quite   good agreement with most of the data at $z\la1.5$,
over a large range of luminosities, from $10^{30}$ to
$10^{35-36}$\,erg\,s$^{-1}$Hz$^{-1}$. At high redshifts, observational
measurements become more difficult to carry out and present larger
uncertainties; however, the general trend of data is  well reproduced by the
model at all redshifts. Some dicrepancies are shown with the data of
\citet{smo17} at $L\la10^{32}$\,erg\,s$^{-1}$Hz$^{-1}$. They steeply
increase as the luminosity decreases, much more than  predicted by
the model. As discussed by \citet{bon19}, however, the LF of
radio-loud AGN estimated by \citet{smo17} might be contaminated by the
presence of radio-quiet AGN whose emission is dominated by the star
formation of host galaxies.

Focusing on HK mode AGN only, the modelled LF increases from redshift
0 to $\sim1.5$ at all (but especially at high) luminosities (see
Fig.\,\ref{s4f5} and \ref{s4f8}). For example, at $z=1.5$ the LF is
larger than the local one by a factor 5 and 20 at $L=10^{30}$ and
$10^{34}$\,erg\,s$^{-1}$Hz$^{-1}$, respectively. This is due to the
strong increase in the fraction of HK mode AGN going back in time and
to the evolution of the \citet{tv17} mass function for high-accreting
SMBHs. At higher redshifts, contrary to LK mode AGN, the HK mode LF
still keeps increasing at high luminosities, while almost no evolution
is found at intermediate to low luminosities, making the shape of the
HK mode LFs quite flat from $10^{30}$ to
$10^{34}$\,erg\,s$^{-1}$Hz$^{-1}$. The relevance of HK mode AGN
become more and more important for the LF with the increase in the
redshift, and they are the dominant mode at $L\ga10^{33}$
($10^{32}$)\,erg\,s$^{-1}$Hz$^{-1}$ at $z=1$ (3).

The evolution of the HK mode LF can be compared with observational
data up to $z=1.1$ (Fig.\,\ref{s4f8}). This is very consistent with
data from \citet{bes14} at $0.5<z<1$, and partially consistent with
\citet{pra16} at $0.3<z<0.75$. On the other hand, as already discussed
for the local LF, the model underestimates estimates of \citet{but19}
at low redshifts. The discrepancy seems to substantially reduce going
to higher redshifts, and we find a good match in the $0.9<z<1.1$ bin.


In principle, luminosity functions and number counts can be estimated
by the model also at frequencies outside the 1--5\,GHz
interval. However, this is a critical operation. The model
assumes a power-law spectrum with constant slope for flat- and
steep-spectrum sources. This approximation is valid in a small range
of frequencies, but clearly fails when spectra are extrapolated over a
large interval. This is particularly true at high frequencies:
flat-spectrum sources become the most relevant population and their
spectral index at $\nu\gg5\,$GHz are typically poorly correlated with
the 1--5\,GHz one \citep[see e.g.][]{sad08, tuc08, mas16}.

Accurate measurements of number counts of radio-loud AGN are
available at 150 and 610\,MHz. At 610\,MHz the model is still able to
provide a good fit of the data, as shown in Fig.\,\ref{s4f9}. On the
contrary, if we consider number counts at 150\,MHz, we need to
introduce a significant steepening of the average spectral index used
for steep-spectrum sources, from $0.75$ to $1$, in order to get the
proper normalization. This steepening
is consistent with the average spectral index found by \citet{ocr19}
for radio-loud AGN that is $\alpha_{1400}^{610}=0.89\pm0.28$. On the
other hand, using the GLEAM 4 Jy sample, \citet{whi20} found a median
spectral index of $0.79$ between 151 and 1400\,MHz, with a steepening
only within the GLEAM band ($\alpha_{72}^{231}=0.83$). This may
indicate that the extrapolation of model predictions to frequencies as
low as 150\,MHz requires a more accurate method than a simple
steepening of the average spectral index.

Finally, in Fig.\,\ref{s4f9} we compare model predictions to number
counts from the Ninth Cambridge (9C) and Tenth Cambridge (10C) radio
surveys at 15.7\,GHz \citep{dav11,whi16}. They cover almost four
orders of magnitude in flux density, from 0.1\,mJy to 1\,Jy (low flux
densities covered by the 10C survey and high flux densities by the 9C
survey). At $S\sim0.01$\,Jy, where the two surveys overlap, number
counts seem to present a break in normalization: the 9C number counts
have  a slope similar to the 10C value, but with a normalization
lower by a factor of $\sim1.5$. The model follows the same overall slope
of the observed number counts, but with a normalization that falls  
between the two different ones determined by the 10C and   9C
surveys. We note, however, that these predictions are
extrapolations using the 1--5\,GHz average spectral indices, which are
probably not suitable at higher frequencies. In addition, HK and
LK mode AGN could have a different spectral behaviour between 5 and
20\,GHz, as suggested by \citet{sad14} and \citet{whi16b}.

\begin{figure}
  \centering
  \includegraphics[width=9cm]{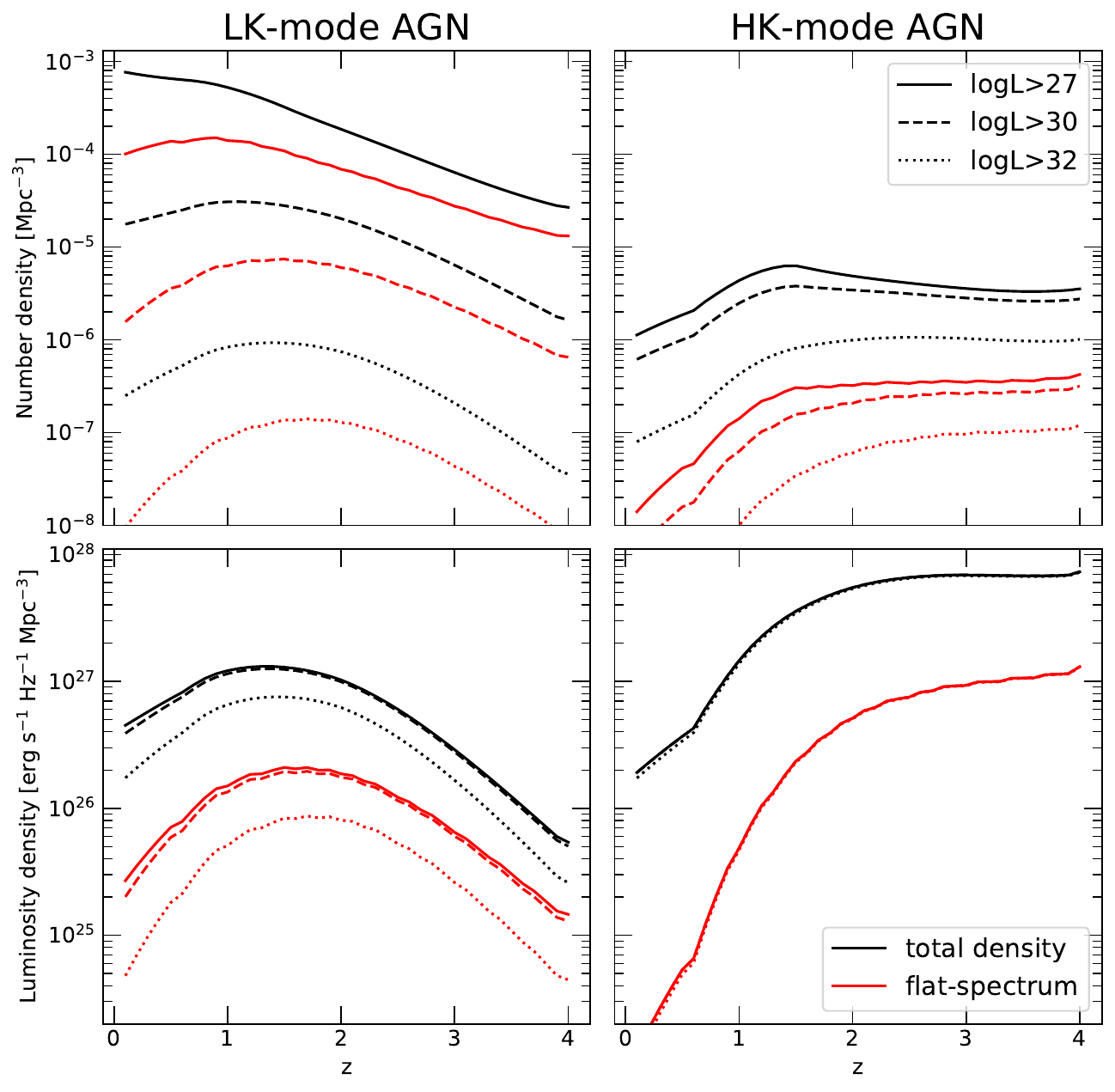}
  \caption{Cosmic evolution of the number density (top panels) and of
    the luminosity density (bottom panels) for LK mode (left panels)
    and HK mode (right panels) AGN. Three different lower limits are
    considered (see inset,  top right panel). Results for
    flat-spectrum sources are also shown (red lines).}
  \label{s4f10}
\end{figure}

\subsection{Predictions on cosmic evolution of radio-loud AGN}

Based on our best-fit model for LK and HK mode AGN, we can
compute the AGN number and luminosity density as a function of redshift,
up to $z=4$, as follows:
\beq
\mathcal{N}=\int\,\phi(L)\,d\log(L);~~~~
\mathcal{L}=\int\,L\,\phi(L)\,d\log(L).
\eeq

In Fig.\,\ref{s4f10} we show the evolution of the number and
luminosity density for the two AGN populations, and for their sub-class
of flat-spectrum sources. We consider three different lower limits in
the integral, $L=10^{27}$, $10^{30}$, and
$10^{32}$\,erg\,s$^{-1}$Hz$^{-1}$ (while the upper limit is fixed to
$10^{39}$\,erg\,s$^{-1}$Hz$^{-1}$). As discussed above and clearly
shown in the figure, LK and HK mode AGN are expected to have a
quite different cosmic evolution. LK mode AGN are much more numerous
at all redshifts, but they are typically faint objects (the fraction
of objects brighter than $10^{32}$\,erg\,s$^{-1}$Hz$^{-1}$ is always
$\ll1$\%). The number density steadily decreases with redshift (except
flat-spectrum sources that show a peak at $z\sim1$--2), while the
luminosity density peaks at $z\sim1$. This implies a
luminosity-dependent evolution for this class of AGN.

On the other hand, HK mode AGN show  strong evolution until redshift
1.5, and  modest or no evolution after that. However, the very flat
slope in the number and luminosity density at high redshifts ($z\ga3$),
observed in Fig.\,\ref{s4f10}, has to be taken with some caution. The
model  becomes less predictive at these redshifts (i.e. when
the contribution to number counts is less important). In addition, the
evolution of the fraction of high-accreting HK mode AGN has been
described by a simple power law of redshift. Although it can be
satisfactory in most of the $z$ range, this parametrization can fail
at the highest redshifts when this fraction may stop growing or even
decrease. Finally, most  HK mode AGN are brighter than
$10^{30}$\,erg\,s$^{-1}$Hz$^{-1}$, and ten per cent or more have
$L>10^{32}$\,erg\,s$^{-1}$Hz$^{-1}$. No significant differences in
the cosmic evolution of flat- and steep-spectrum sources are
observed for HK mode AGN.

\begin{figure}
  \centering
  \includegraphics[width=8cm]{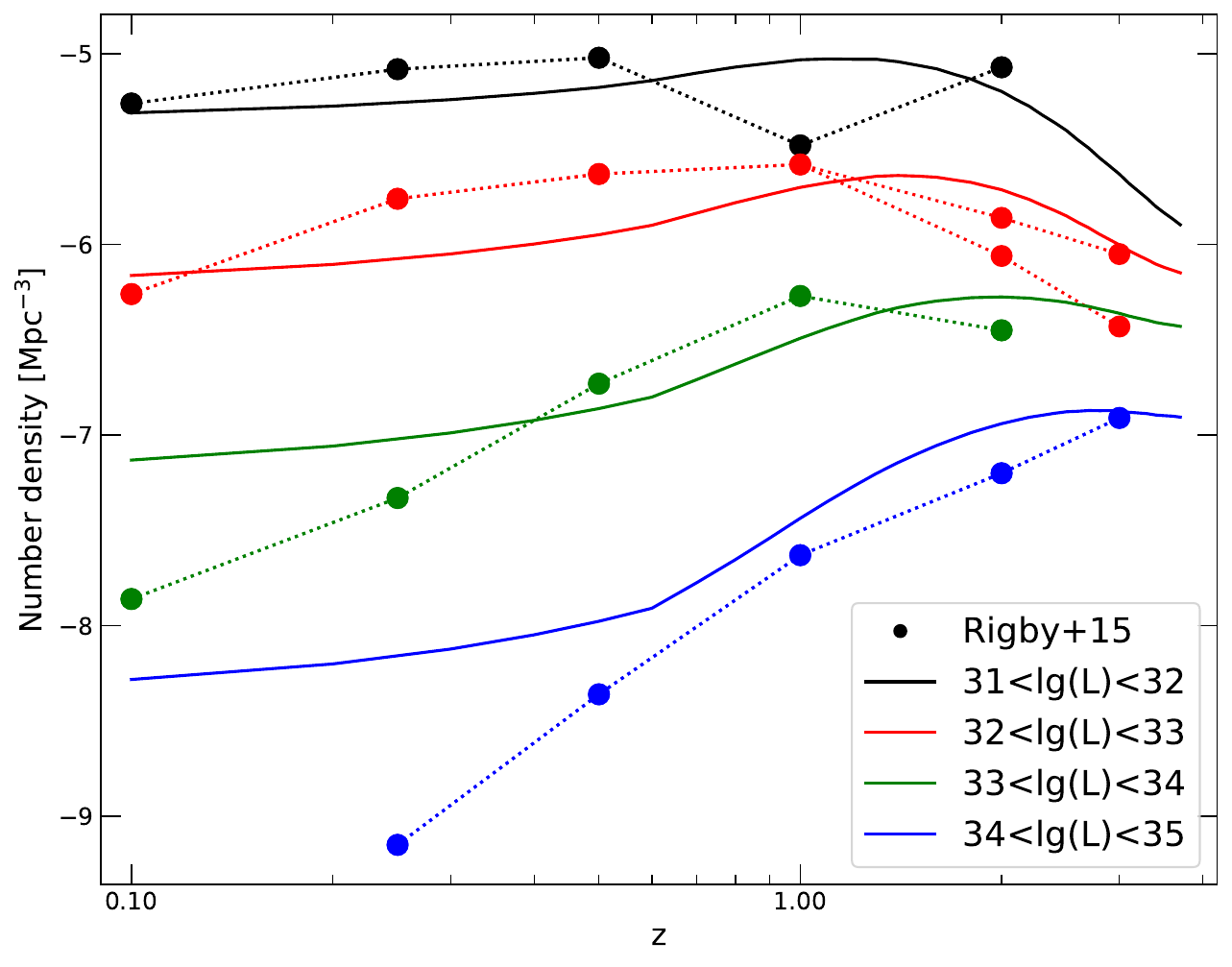}
  \caption{Predicted number densities for steep-spectrum sources in
    different luminosity bins (solid lines) compared to estimates
    from \citet{rig15} (solid points).}
  \label{s4f11}
\end{figure}

Model predictions of the number density evolution of steep-spectrum
sources can be compared with results from \citet{rig15} (see
Fig.\,\ref{s4f11}). They extend the analysis of \citet{rig11}
including fainter objects from the Subaru/XMM-Newton Deep Field, and
covering the luminosity range of
$10^{31}<\log L<10^{35}$\,erg\,s$^{-1}$Hz$^{-1}$. The number density
of steep-spectrum sources was estimated in five luminosity bins,
using a grid-based modelling that takes into account source counts,
redshift samples and local LFs, without any assumptions about the LF
shape and high-redshift behaviour. The model seems to roughly match
the Rigby et al. estimates in all the bins, apart from an excess of
bright ($L>10^{33}$\,erg\,s$^{-1}$Hz$^{-1}$) sources at low redshifts.

\section{Conclusions}
\label{sec5}

In this paper we have developed a formalism to link the luminosity
function of radio-loud AGN at GHz frequencies with the cosmological
evolution of SMBHs. Our approach is based on physical and
phenomenological relations that allow us to connect in a statistical
way radio emission of AGN cores with SMBHs at their centre, through the fundamental plane of   black hole activity, and radio emission from extended jets and lobes, through a   power law relation because radio core and extended emission are both a good
  tracers of the kinetic jet power. 

This formalism allows us to model the evolution of the luminosity
function of radio-loud AGN. They are divided into two classes according
to the measured Eddington ratio, $\lambda$, greater or smaller
than the critical value 0.01
(LK mode and HK mode AGN,
respectively). This critical value should correspond to the transition
between radiatively inefficient and efficient accretion flows, and
explain the observed dichotomies, for example, between FRI and FRII
radio galaxies, low- and high-excitation AGN, and BL\,Lacs and
flat-spectrum radio quasars.

The model takes as input the evolution of mass function and Eddington
ratio distributions of SMBHs. We   used the ones provided by \citet{tv17}
at redshifts from 0 to 4. The model is also characterized by a
remarkably limited number of free parameters, which mainly arise from
uncertainties in the relations discussed above: after fixing tightly
linearly correlated parameters, there are only two free parameters
associated with LK mode AGN and five with HK mode AGN. We adopted a
Markov chain Monte Carlo method to fix the free parameters of the
model by fitting two different types of observational data sets: local
(or low-redshift) LFs at 1.4\,GHz of AGN populations, and differential
number counts at 1.4 and 5\,GHz. The model is able in general to
provide a very good fit to them. This is especially relevant for
1.4\,GHz number counts that are very accurately determined in the flux
density range 1--100\,mJy. The model is also in perfect
agreement with number counts of LK and HK mode AGN estimated by
\citet{smo17b} at that frequency (data that are not used in
constraining the parameters of the model; see Section\,\ref{sec4}).

Moreover, our model predictions have been compared to the recent data
on LFs of radio-loud AGN in the whole redshift range
$0.3<z<4$. Although they were not   used in determining the free
parameters of the model, we find again a very satisfactory fit of
these data. In particular, the modelled AGN evolution seems to
reproduce the observed one in the full range of luminosities covered
by data, at least up to $z\sim1.5$. In addition, the predicted
evolution of the number density of steep-spectrum sources agrees
quite well with the estimates from \citet{rig15}.
The good performance of the model at these redshifts is remarkable,
considering that all the free parameters but the fraction of HK mode
AGN are redshift independent. An important role in driving the correct
AGN evolution is given by the SMBH mass functions and
Eddington ratio distributions provided as input to the model.

The AGN evolution with cosmic time obtained from the model is quite
different for the two classes of AGN and dependent on the luminosity. The
space density of AGN in LK mode typically declines with redshift, with
the only exception of bright objects that positively evolves at
$z\la1$. This produces a total luminosity density that peaks at
$z\sim1$--1.5. On the other hand, HK mode AGN show a positive
evolution up to $z=1.5$--2, weaker at low luminosities and stronger at
high luminosities. At higher redshifts the evolution becomes very
modest or null. The total number and luminosity density consequently keep
increasing up to $z\ga1.5$ and then remain almost constant up to
$z=4$.

The main uncertainty in the model   comes from the LF of HK mode
AGN at low redshifts. The local LF for this class was recently
estimated by three works \citep{bes12,pra16,but19} using different
criteria to distinguish AGN and star-forming radio galaxies. Their
results show a quite big discrepancy in the space density of HK mode
AGN at low/intermediate luminositites
($L<10^{31}$\,erg\,s$^{-1}$Hz$^{-1}$): from a few per\,cent of the
total  \citep{bes12} to ten per\,cent \citep{pra16} or more
\citep{but19}. Such disagreement is probably due to
misclassifications of faint sources. Independently of which
observational data are taken into account, however, our model predicts
a very low space density of HK mode AGN at low to intermediate
luminosities, in agreement with the \citet{bes12} measurements. The local
fraction of high-accreting AGN in HK mode is expected  to be
very small, of the order of a few per\,mil or less, with a strong
evolution with redshift. It is interesting to note (and in support of
the goodness of the model) that at redshift $z\simeq1$, where the
contribution of star-forming galaxies is negligible, model
predictions closely agree with observational estimates. This is true
even for the \citet{but19} measurements that locally display the
greatest differences from our model. Moreover, locally,
a larger fraction of faint AGN in HK mode is probably incompatible
with the positive evolution observed for this class of AGN and with
the constraints imposed by the measured number counts.

Coefficients in the fundamental plane relation for HK mode AGN has
been considered unknown by the model and determined by the fitting
procedure. The model finds a positive correlation between the intrinsic core
radio luminosity and the X-ray luminosity, steeper than the
one observed for LK mode AGN, and in agreement with theoretical
expectations for radiatively efficient accretion flows
\citep[e.g.][]{mer03}. On the other hand, an anti-correlation is
found between the intrinsic core radio luminosity and the SMBH
mass that is  consistent with estimates of \citet{don14}.

Observational data are not yet able to provide strong constraints
on the coefficients of the relation between the luminosity of AGN
intrinsic cores and of extended jets and lobes. Assuming a power-law
relationship between them, we find a large degeneracy between the
normalization and the power-law index of the relation. In addition,
the normalization parameter can vary by orders of magnitude in the
model, providing an equally good fit to the data (both for HK
and LK mode AGN). In order to remove this degeneracy and to narrow
the allowed parameters range, more information about different AGN
sub-populations would be required; for example, accurate estimates of
the local luminosity function of HK mode AGN at
$L<10^{30}$\,erg\,s$^{-1}$Hz$^{-1}$ could provide a better
determination of their fraction, and significantly reduce the
uncertainty on the core and extended emission relation. Similarly,
estimates of the local luminosity function for flat-spectrum radio
quasars and BL\,Lacs separately could give a relevant improvement in
the parameter determination.

Finally, we have extrapolated model predictions outside the frequency
range of 1--5\,GHz. The comparison with the currently determined
number counts at a few hundred MHz and 15\,GHz show again the good
performance of the model, although at 150\,MHz the average
spectral index of steep-spectrum sources has to be increased to fit
the observed number counts. As a consequence, we expect that a more
accurate modelling of AGN spectra for compact and extended
emissions is required when a large range of frequencies is considered.

Predictions of the best-fit models for LFs and number counts of LK
and HK mode AGN are available at the following webpage:
\underline{https://www.astro.unige.ch/\textasciitilde tucci/AGNradioLF/}.

\begin{acknowledgements}

  The authors thank the anonymous referee for the constructive
  remarks, which have improved the quality of the manuscript. MT
  warmly thanks M. Volonteri for the long discussions about SMBHs and
  AGN that allowed this work to be carried out, and M. Spinelli for
  the helpful discussion on MCMC methods. LT acknowledges the Spanish
  Ministerio de Ciencia, Innovaci\'{o}n y Universidades for partial
  financial support under projects PGC2018-101814-B-I00 and
  PGC2018-101948-B-I00. Part of the analysis was performed using the
  cluster Yggdrasil of the University of Geneva.
  
\end{acknowledgements}

\bibliographystyle{aa} 
\bibliography{mybib}

\begin{appendix}

\section{Model predictions for LK mode AGN from
    different parameter sets}
\label{a1}  
  
\begin{table}
  \centering 
  \caption{Parameter values of the five selected models for LK mode AGN}
  \begin{tabular}{cccccc}
  \hline
  model & $\beta_W$ & $\xi_W$  &  $\sigma_W$ & $\chi^2_{LK}$ & $\chi^2_{global}$ \\
  \hline
  1 & -9.17 & 1.24 & 1.15 & 35 & 240 \\ 
  2 & -10.86 & 1.30 & 1.05 & 32 & 230 \\ 
  3 & -12.37 & 1.36 & 0.94 & 31 & 228 \\ 
  4 & -13.95 & 1.42 & 0.81 & 33 & 212 \\ 
  5 & -15.41 & 1.47 & 0.64 & 36 & 224 \\ 
\hline
\end{tabular}
\label{ta1}
\end{table}

\begin{figure}
  \centering
  \includegraphics[width=9cm]{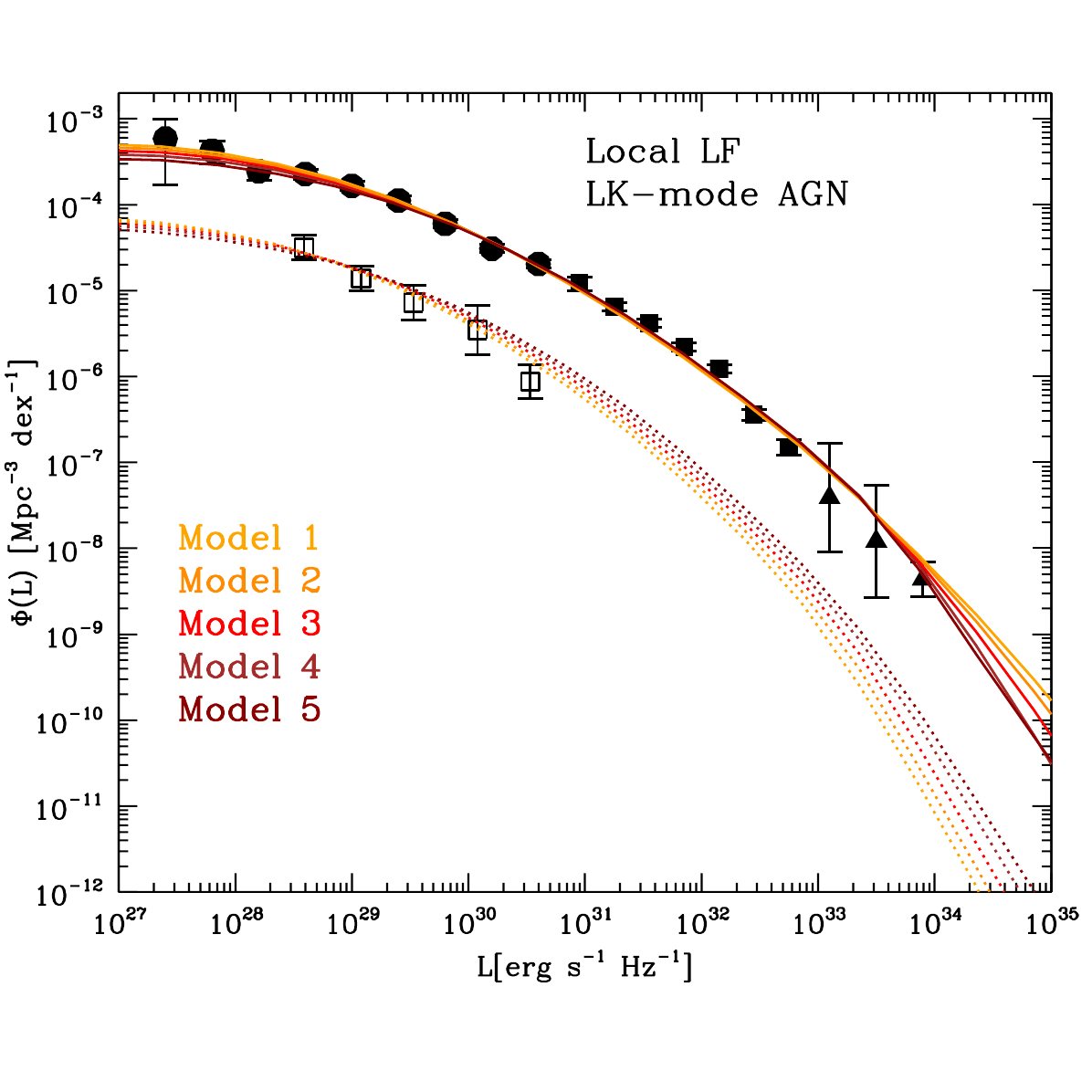}
  \includegraphics[width=9cm]{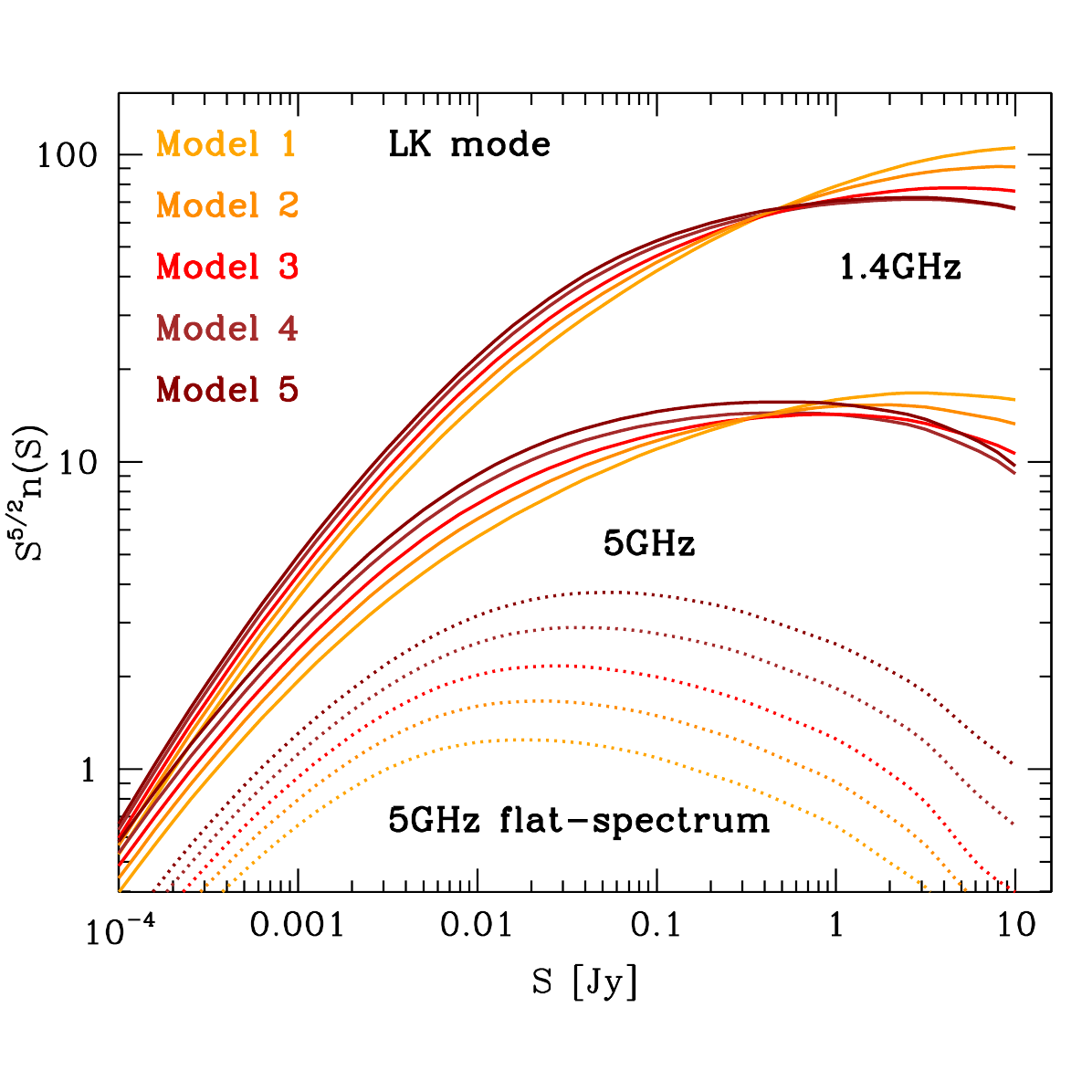}
  \caption{({\it Top panel}) Local LF for LK mode AGN from the model
    parameter sets reported in Table\,\ref{ta1} compared with
    observational data as in Fig.\,\ref{s4f1}. ({\it Bottom panel})
    Predicted number counts of LK mode AGN at 1.4\,GHz (upper curves) and
    5\,GHz (lower curves) from the five parameter sets. The dotted lines show
    the predicted number counts of flat-spectrum LK mode sources at
    5\,GHz.}
  \label{s4f4new}
\end{figure}

As discussed in Sect.\,\ref{s4s1}, model parameters for LK mode AGN
are strongly correlated, and fixing one of the two between $\beta_W$
or $\xi_W$ is almost sufficient to define the model for that AGN
population. The range of possible values for these two parameters is
quite large, and very good fits to observational data are found for
$-16<\beta\la-8$ and $1.25<\xi_W<1.5$. In this Appendix, we discuss
how a different set of parameters can quantitatively affect the
predicted LFs and number counts of LK mode AGN.

Along with the best-fit model (corresponding to Model 3 of
Table\,\ref{ta1}), we select four other  sets of parameters that cover
the full range of possible parameter values
(see Fig.\,\ref{s4f3} and Table\,\ref{ta1}). In
Fig.\,\ref{s4f4new} we show the results for the five different models
in terms of the local LF and number counts. The local LFs (both total and
flat-spectrum) are very similar at the luminosities covered by
observations, with some differences only at higher luminosities.

The choice of the parameter set is more relevant for number
counts. The shape of number counts is  slightly dependent on
the parameters: models with lower values of $\xi_W$ tend to have
steeper number counts at  sub-Jy--Jy levels and lower counts at mJy
levels. Moreover, they predict significantly fewer flat-spectrum
sources.

Differences in number counts can have a significant impact on the
model performance when the full radio-loud AGN population is
considered (i.e. LK mode plus HK mode AGN). Although they are not
large, we find some changes in the accuracy of the global fit
when using the five parameter sets for LK mode sources: the global
best fit is obtained using Model 4 with $\chi^2\simeq212$
that is taken as our reference model for LK mode AGN.

\end{appendix}

\end{document}